\documentclass[a4paper,11pt]{article}

\usepackage{subcaption}
\usepackage{standalone}
\usepackage{pgfplots}
\pgfplotsset{compat=1.16}

\usepackage{jheppub} %
\usepackage{derivative}
\DeclareDifferential{\Dd}{\mathrm{D}}[style-notation=multiple, style-notation-*=single
]

\newcommand{\DdImag}[1]{\Dd[order=(\I)]{#1}}

\usepackage{amssymb}
\usepackage{slashed}

\usepackage[utf8]{inputenc} %
\DeclareUnicodeCharacter{2212}{-} %
\usepackage[LGR,T1]{fontenc} %
\usepackage{textgreek}

\usepackage{lineno}
\usepackage{csquotes}
\usepackage{float}
\usepackage{xcolor}
\usepackage{cancel}
\usepackage{bm}
\usepackage{enumitem} %
\usepackage{mathtools}
\usepackage{cleveref}
\usepackage{ytableau}
\usepackage{comment}
\usepackage{amstext} %
\usepackage{nicematrix}
\usepackage{xspace} %

\newcommand{\normord}[1]{:\mathrel{#1}:}
\newcommand{\dotwo}{\tfrac{d}{2}}
\newcommand{\dd}{\mathrm{d}}
\newcommand{\ONpdf}{\texorpdfstring{$\gO(N)$}{O(N)}}
\newcommand{\psibar}{\bar{\psi}}

\newcommand{\I}{{\rm I}}

\newcommand{\UV}{{\rm UV}}

\def\det{\mathop{\rm det}\nolimits}

\def\tr{\mathop{\rm tr}\nolimits}
\def\str{\mathop{\rm str}\nolimits}

\def\Tr{\mathop{\rm Tr}\nolimits}
\def\bbra{{\langle\kern-2.5pt\langle}}
\def\kket{{\rangle\kern-2.5pt\rangle}}
\def\Bbra{{\Big\langle\kern-3.5pt\Big\langle}}
\def\Kket{{\Big\rangle\kern-3.5pt\Big\rangle}}

\DeclarePairedDelimiter\abs{\lvert}{\rvert}

\DeclarePairedDelimiterX\braket[2]{\langle}{\rangle}{#1\,\delimsize\vert\,\mathopen{}#2}
\DeclarePairedDelimiter\expval{\langle}{\rangle}
\DeclarePairedDelimiter\floor{\lfloor}{\rfloor}

\newcommand   \half{\frac 1 2}
\newcommand   \lptl{\raise .8ex\hbox{$^\leftarrow$} \hspace{-9pt} \partial}
\newcommand   \lrptl{\raise .8ex\hbox{$^\leftrightarrow$} \hspace{-9pt} \partial}

\newcommand   \Z  {\mathbb{Z}}

\newcommand   \SO    {\mathrm{SO}}
\newcommand   \gO    {\mathrm{O}}

\DeclareMathOperator{\vol}{vol}
\DeclareMathOperator{\Str}{Str}

\newcommand   \cD {\mathcal{D}}

\newcommand   \cF {\mathcal{F}}

\newcommand   \cN {\mathcal{N}}

\newcommand   \cR {\mathcal{R}}

\newcommand{\be}{
  \begin{equation}
  \begin{aligned}
}
\newcommand{\ee}{
  \end{aligned}
  \end{equation}
}

\makeatletter
\DeclareRobustCommand\bea{\@ifnextchar[{\@@bea}{\@bea}}
\def\@@bea[#1]#2\eea{\begin{subequations}\begin{align}#2\end{align}\label{#1}\end{subequations}}
\def\@bea#1\eea{\begin{subequations}\begin{align}#1\end{align}\end{subequations}}
\makeatother

\usepackage{tikz}
\usepackage[compat=1.1.0]{tikz-feynman}
\tikzfeynmanset{warn luatex=false}

\usepackage{array}   %
\newcolumntype{L}{>{$}l<{$}} %
\makeatletter
\edef\savedcodes{\catcode`\noexpand\_=\the\catcode`\_}
\edef\@tempa{\csname opt@newtxmath.sty\endcsname}
\def\@tempb{{subscriptcorrection}}
\expandafter\expandafter\expandafter
  \in@\expandafter\@tempb\expandafter{\@tempa}
\ifin@
  \catcode`\_=12 %
\fi
\makeatother
\usepackage{tensor}
\savedcodes

\title{The sphere free energy of the vector models to order $1/N$}

\author[\Delta]{Ludo Fraser-Taliente}%

\affiliation[\Delta]{Rudolf Peierls Centre for Theoretical Physics, University of Oxford, Oxford, OX1 3PU, UK}

\emailAdd{ludovic.fraser-taliente@physics.ox.ac.uk}

\abstract{%
We calculate the large-$N$ expansion of the sphere free energy $F=-\log Z_{S^d}$ of the $\gO(N)$ $\phi^4$ and the Gross-Neveu $(\bar{\psi} \psi)^2$ CFTs to order $1/N$. 
Analytic regularization of these theories requires consistently shifting the UV scaling dimension of the auxiliary field: this can only be done by modifying its kinetic term.
This modification combines with the counterterms to give the result that matches the $\epsilon$-expansion, resolving a puzzle raised by Tarnopolsky in \cite{Tarnopolsky:2016vvd}.
These $F$s can be written compactly in terms of the anomalous dimensions, for both the short-range and the long-range versions of these CFTs.
We also provide various technical results including a computation of the counterterms on the sphere and a neat derivation of the sphere free energy of a free conformal field.
Finally, we observe that the long-range CFT becomes the short-range CFT at exactly the point where its $\Ft=-\sin \tfrac{\pi d}{2} F$ is maximized as a function of the vector's scaling dimension.
}

\newcommand{\id}{\mathbb{I}}
\newcommand{\spinid}{\mathbb{I}_s}
\newcommand{\thalf}{\tfrac{1}{2}}
{%
}%

\newcommand{\Ft}{\tilde{F}}
\newcommand{\normCor}{A}

\newcommand{\FtextOrPDF}{\texorpdfstring{$F$}{F}\xspace}

\newcommand{\NPDF}{\texorpdfstring{$N$}{N}\xspace}

\usepackage{empheq} %

\pgfmathsetmacro\MathAxis{height("$\vcenter{}$")} %
\tikzset{
  sigma/.style={dashed, thick},
  phi/.style={solid, line width=1pt},
}

\begin{document}
\maketitle
\flushbottom

\section{Introduction}

We study the vector models. The $\gO(N)$ $\phi^4$ vector model is defined by
\begin{equation}\label{eq:ONaction}
Z_{\gO(N)}\equiv\int \Dd{\phi} \, e^{-S_{\gO(N),\phi^4}}, \quad S_{\gO(N),\phi^4} \equiv \int_x \half \phi_i C^{-1} \phi_i + \frac{\lambda_0}{8N} (\phi_i \phi_i)^2
\end{equation}
and its complex fermionic cousin, the Gross-Neveu (GN) model, is defined by
\begin{equation}\label{eq:GNaction}
Z_\mathrm{GN} \equiv\int \Dd{\psi} \Dd{\psibar} \, e^{-S_{\mathrm{GN},\psi^4}}, \quad S_{\mathrm{GN},\psi^4} \equiv \int_x -\bar\psi_i C^{-1} \psi^i + \frac{\lambda_0}{2N} (\bar\psi_i \psi^i)^2, %
\end{equation}
where $i$ now runs from $1$ to $n$, and $N \equiv n \tr\spinid$. 
In each action, $C(x,y)$ is the bare position-space propagator of the free field and $\int_x = \int \odif[order=d]{x}\, \sqrt{g}$.
Both of these QFTs are known to provide access to an interacting Wilson-Fisher-like $d$-dimensional Euclidean conformal field theory (CFT) as fixed points of their RG flow: for $2<d<4$, the $\gO(N)$ vector CFT lies in the IR of \eqref{eq:ONaction} and the GN vector CFT lies in the UV of \eqref{eq:GNaction}.

In this paper, we compute the universal part of these CFTs' sphere free energies: we just do the functional integrals
\begin{equation}
F\equiv - \log Z\rvert_{S^d},
\end{equation}
in the limit of a large number $N$ of fields $\{\phi_i, \psi^i\}$, working with a particular regularization and tuning the renormalized couplings to ensure that we are working at the Wilson-Fisher-like fixed point. 
By \enquote*{the universal part}, we mean the part that does not depend on the regularization; \textit{henceforth this caveat will be dropped, and we will freely refer to the $F$s that we compute as the (sphere) free energy}.
The sphere free energy has been of great interest of late, as $\Ft \equiv -\sin(\tfrac{\pi d}{2}) F$ provides a measure of the number of degrees of freedom of an arbitrary CFT \cite{Giombi:2014xxa,Fei:2015oha,Giombi:2024zrt}, generalizing the two-dimensional $c$ to higher-dimensional CFTs \cite{Zamolodchikov:1986c, Cardy:1988cwa}.
Additionally, it was recently determined that the leading large-$N$ correction to the scaling dimensions of $\phi$ and $\psi$ are determined by extremization of $\Ft$ over a family of CFTs \cite{Fraser-Taliente:2024hzv}.

Our $F$ results, given in \cref{sec:Fresults}, are to order $1/N$ and in arbitrary dimension $d$, and confirm and generalize the values conjectured by matching to the $\epsilon$-expansion in \cite{Tarnopolsky:2016vvd}.
They prove to have a simple form in terms of the field scaling dimensions.
\cite{Tarnopolsky:2016vvd} was not able to confirm their conjecture by a direct computation, due to mysterious factors of $3$ and $\tfrac{5}{2}$. 
We resolve this issue by realizing that the analytic regularization, a shift of the scaling dimension of the auxiliary field by $\delta$, must be implemented by adding terms to the action. 
These terms combine with the counterterms $\propto 1/\delta$ to cancel the mysterious factors. %

Following the long tradition in physics of varying parameters that do not appear to be variable, there exist generalizations of these CFTs which have an explicitly non-local action and are called the long-range vector CFTs.
They are parametrised by the scaling dimension of the vector $\Delta_{\phi/\psi}\equiv \tfrac{d-s}{2}$; the short-range CFTs are then specific points on these lines of non-local CFTs.
It is straightforward to work in an arbitrary-$s$ formalism, and so we present our results for both the long-range and short-range CFTs.
Notably, we observe (but do not prove) that the short-range CFTs coincide with the long-range CFTs that have extremal free energy, as parametrised by $s$; if this could be proven \cite{Fraser-Taliente2026prep}, it would serve as an extension to all orders in $N$ of the $\Ft$-extremization noted above. 
This is particularly nice, since the long-range expressions are typically much nicer and more compact than the short-range ones.

\subsection{Paper structure}

We present the long-range and short-range results together, as the analyses are almost identical.
The structure of this paper is the following.
\begin{enumerate}
\item In the remainder of this section, we first describe the short- and long-range CFTs and the crossover between them; then we consider their regularization; and then we compactly summarise our $F$s.
\Cref{sec:discussion} then contains our discussion of these results.
\item In \cref{sec:Fresults} we immediately present the free energy results for the $\gO(N)$, Gross-Neveu, and supersymmetric CFTs, and compare them to each other.
\item In \cref{sec:ONsetup} we describe the correct construction and regulation of the $\gO(N)$ model (the same is done for the GN model in \cref{app:GNsetup}).
\item In \cref{sec:calculatingF} we compute the universal part of the free energy for these models, and comment on why it takes the form it does. 
We provide a compact derivation of $F_\Phi(\Delta)$, the free energy of an arbitrary generalized free conformal field of dimension $\Delta$.
\end{enumerate}
Various details are relegated to the other appendices.
\begin{enumerate}
\item In \cref{app:conventions}, we review our conventions for Gaussian integrals and free fields, as well as position space integrals.
\item We describe how to find the values of the counterterm in \cref{app:counterterms}.
\item We compute our diagrams on the sphere using standard Mellin-Barnes integrals, which are discussed in \cref{app:sphereInts}. 
For completeness, we also discuss how we can use the method of uniqueness (summarised in \cref{app:uniqueness}) in \cref{sec:applyingUniqueness}.
\end{enumerate}
\clearpage

\subsection{The large-\NPDF expansion and the \texorpdfstring{$\epsilon$}{epsilon} expansion}

Working in the large-$N$ limit is made simpler by introducing a real auxiliary field $\sigma\sim \normord{\phi_i\phi_i}$. 
The partition function can then be rewritten exactly using a new action,
\begin{align}
  Z_{\gO(N)} &\stackrel{\text{\S \ref{sec:ONsetup}}}{=} \int \DdImag{\sigma} \Dd{\phi} \, e^{-S_{\gO(N)}}, \quad &S_{\gO(N)} \equiv \int_x \half \phi_i \left(C^{-1} - \frac{\sigma}{\sqrt{N}} \right) \phi_i - \frac{\sigma^2}{2\lambda_0}, \label{eq:ONactionWithSig}\\
  Z_\mathrm{GN} &\stackrel{\text{\S \ref{app:GNsetup}}}{=} \int  \DdImag{\sigma}  \Dd{\psi} \Dd{\psibar} \, e^{-S_\mathrm{GN}}, \quad 
&S_\mathrm{GN} \equiv \int_x - \psibar_i \left(C^{-1} - \frac{\sigma}{\sqrt{N}} \right) \psi^i - \frac{\sigma^2}{2\lambda_0} \label{eq:GNactionwithSig},
\end{align}
where in slightly more generality we can take $C(x,y)$ to be the propagator of a free conformal field of dimension $\Delta_{\phi/\psi}^\UV$ (whether on the sphere or on flat space -- see \cref{app:conventions} for our conventions).
Our strategy for computing $F$ is to work at the critical point by taking $\lambda_0 \to \infty$, and integrate out the vector field $\phi_i$.
This leaves a non-local effective action for $\sigma$ in which $1/N$ appears only as a coupling constant (a continuous parameter), allowing for a perturbative solution for $F$ in $1/N$.

The standard conformally coupled free scalar and fermion have $\Delta_\phi^\UV=\tfrac{d-2}{2}$ and $\Delta_\psi^\UV=\tfrac{d-1}{2}$; these are implemented in curved space by
\begin{equation}
  C^{-1}_{\gO(N)} = -\nabla^2 + \frac{d-2}{4(d-1)} \cR \quad \text{ and } \quad C^{-1}_\mathrm{GN} = +\slashed{\nabla}  %
\end{equation}
respectively (for flat space, these are just $C^{-1}=-\partial^2$ and $+\slashed{\partial}$, which yield the canonical position-space correlators $C(x,y)$ given in \cref{app:space}).
Now, in the IR for $2<d<4$, we can then ignore the irrelevant operator $\sigma^2$ in the actions above (schematically, we send the bare coupling $\lambda_0 \to \infty$); this then gives the interacting CFT with action
\begin{equation}\label{eq:moreGeneralON}
S = \int_x \half \phi_i\left(C^{-1} - \frac{\sigma}{\sqrt{N}}\right)\phi_i + \text{counterterms}
\end{equation} 
and conformal scaling dimensions
\begin{equation}
\Delta_\phi = \Delta_\phi^\UV + O(1/N), \quad \Delta_\sigma = \Delta_\sigma^\UV + O(1/N),
\end{equation}
where $\Delta_\sigma = d-2\Delta_\phi^\UV$, for which we would like to compute $F$.

To properly define any QFT we must specify a regularization.
However, this auxiliary theory is unpleasant to regulate when working at the critical point: using a hard cutoff is too difficult, and DREG alone does not work since the interaction $\sigma \phi_i \phi_i$ is marginal in any $d$.
The computationally easiest way to access the interacting CFT is to combine DREG with an analytic shift in the scaling dimension of the field $\sigma$ in the UV, a regularization which we call DREG$+\delta$. 
This shift is not trivial to implement, particularly because the leading kinetic term for $\sigma$ is dynamically generated by $\phi$ loops. 
It can only be implemented by modifying the action of the theory, which makes $\sigma$ an interacting field even in the UV.
Technically, then, this is an IR duality, since we are considering \textbf{a different QFT that flows to the same Wilson-Fisher-like CFT in the IR}: we illustrate this in \cref{fig:duality}.
Away from the IR, these do not describe the same QFT; this is manifest because the UV that includes a $\sigma$ field is explicitly non-local, as we shall see.
Accounting for this modification to the action, including the counterterms that are required for the correlators to be finite, then leads to the additional contribution to $F$ that matches the $\epsilon$ expansion.

The $1/N$ expansion works because in the limit that the small parameter is taken to zero, the UV and IR CFTs coincide (including the presence of a free field $\sigma$); hence the interacting IR theory can be solved by perturbing around the UV theory. 
This is identical to the case of the $\epsilon$ expansion, as demonstrated in \cref{fig:duality} -- except for the fact that the UV theory is not the CFT of $N$ free scalars.
However, unlike in the $\epsilon$-expansion case, finite $1/N$ does not also provide a regularization of the loop integrals.
Hence, an additional regularization $\delta$ is required -- this is the source of the technical complexity of the $1/N$ expansion compared to the $\epsilon$ expansion.

\begin{figure}[h!]
  \centering
  \begin{tikzpicture}[node distance=1.5cm and 2cm, >=stealth, decoration={markings, mark=at position 0.5 with {\arrow[scale=2.1]{>}}}]
    \node (uv1) [draw, text width=3cm, align=center] {$N$ free scalars};
    \node (uv2) [draw, text width=4cm, align=center, right=of uv1] {$N$ free scalars + a free $\sigma$ field ($\delta$-regulated)};
    \node (ir) [draw, text width=3.5cm, align=center, below=of uv1, yshift=-1.5cm, xshift=2.5cm] {$\gO(N)$ vector CFT};
  
    \draw[red, thick, postaction={decorate}] (uv1.south) to[out=-90, in=135] node[midway, left, xshift=-4pt] {$\lambda_0(\phi_i \phi_i)^2$} (ir.north);
    \draw[thick, postaction={decorate}] (uv2.south) to[out=-90, in=45] node[midway, right, xshift=4pt] {$g_0 \sigma \phi_i \phi_i$} (ir.north);
  
    \draw[<->, red, thick] ([xshift=-0.5cm]uv1.west) -- ([xshift=-0.5cm]uv1.west |- ir.north) node[midway, left] {$\propto \epsilon$};
    \draw[<->, thick] ([xshift=0.5cm]uv2.east) -- ([xshift=0.5cm]uv2.east |- ir.north) node[midway, right] {$\propto 1/N$};
  \end{tikzpicture}
  \caption{A schematic of our IR duality in theory space: there are two different QFTs that flow to the $\gO(N)$ vector CFT.
  The standard UV completion of $N$ free scalars is on the left. The WF CFT found by perturbing this theory by $(\phi_i \phi_i)^2$, \eqref{eq:ONaction}, can be solved for finite $N$ by using DREG, working perturbatively in $\epsilon=4-d$.
  In the large-$N$ limit, DREG no longer suffices: we need a QFT  \eqref{eq:completeAction} with a new regulator $\delta$ that improves the UV behaviour -- this is shown on the right. 
  Now $\sigma \phi_i \phi_i$ is the perturbing relevant operator, and we can solve the IR CFT by working perturbatively in $1/N$.
  Computationally, we can solve for the $\gO(N)$ theory by perturbing about either of these theories because in each case the \enquote*{length} of the flow (i.e. the size of the distance between the UV and IR CFTs, as measured by their conformal data) is proportional to that small parameter $\epsilon$ or $1/N$.
  We ignore the subtlety that in some $d$s the roles of free UV and interacting IR might be reversed \cite{Fei:2014yja}.
  }
  \label{fig:duality}
  \end{figure}  

\subsection{Which CFTs?}

We will also compute $F$ for the CFTs defined by \eqref{eq:moreGeneralON} but with $C$ a conformal propagator with generic scaling dimension $\Delta_\phi^\UV$.
These are a more general family of CFTs that exist in arbitrary dimension, though they may be non-unitary and/or non-local, and reaching them requires careful tuning of counterterms \cite{Gubser:2017vgc,Goykhman:2019kcj,Chai:2021wac,Giombi:2024zrt}.
\textit{We therefore remain entirely agnostic about whether our RG flow from the free theory is towards the UV or the IR, and simply compute $F$ for the Wilson-Fisher-like CFT in each case, working directly at the fixed point.}

For the $\gO(N)$ models, we can choose $C$ to give:
\begin{enumerate}
  \item $\Delta_\phi^\UV = \tfrac{d}{2}-k$ for integer $k$. 
  \eqref{eq:moreGeneralON} then describes a local CFT (albeit manifestly non-unitary for $k>1$).
  We refer to these as the short-range (SR) CFTs, though they are also called the $\Box^k$ CFTs \cite{Gracey:2017erc} or Lifshitz points \cite{Gubser:2017vgc}. For these $C^{-1} \sim (-\partial^2)^k = \Box^k$, as described in \cref{app:sphereKineticTerms}.
  \item $\Delta_\phi^\UV = \tfrac{d-s}{2}$ for arbitrary $s$. 
 \eqref{eq:moreGeneralON} then describes non-local Wilson-Fisher-like CFTs: these are called the long-range (LR) CFTs \cite{Fisher:1972zz,Giombi:2019enr,Paulos:2015jfa,Behan:2017dwr,Slade:2016yer,Behan:2018hfx,Benedetti:2020rrq,Benedetti:2024mqx,Chai:2021arp,Giombi:2022gjj}.
  For these, we have a fractional Laplacian $C^{-1} \sim (-\partial^2)^{s/2} = \Box^{s/2}$, as described in \cref{app:sphereKineticTerms}.
\end{enumerate}
On the sphere, the operators $C^{-1} \sim \Box^{s/2}$ that define a free scalar of the given dimension are in reality more complicated, and will be discussed later.

As was discussed in \cite{Gubser:2017vgc}, the $\Box^k$ $\gO(N)$ CFT is found by flowing to the IR of \eqref{eq:ONaction} with $C^{-1}\sim \Box^k$ for $2k < d < 4k$.
There are likely issues with canonical quantization and Ostrogradsky instabilities in all such higher derivative theories, but \cite{Gubser:2017vgc} also commented that they are probably well-defined as Euclidean theories constructed via the path integral.

\subsubsection*{Crossover from long- to short-range: an order of limits}

In the SR case, the interacting CFT has $\Delta_\phi = \Delta_\phi^\UV + \hat{\gamma}_\phi$, with some anomalous dimension $\hat{\gamma}_\phi$; in the LR case, the interacting CFT has $\Delta_\phi = \Delta_\phi^\UV=\tfrac{d-s}{2}$ exactly: this is because quantum corrections can only generate local divergences, and hence local counterterms -- therefore the non-local kinetic term for $\phi$ cannot be renormalized to any order in $N$\footnote{
This was proven to all orders in $\epsilon \equiv s-d/2$ in \cite{Lohmann:2017qyq}. 
It also follows from defining the LR theory as a defect in a higher-dimensional bulk \cite{Behan:2023ile}.}. 
Somewhat surprisingly, then, these two CFTs coincide at the crossover value \begin{equation}
s=s_\mathrm{SR}=2k-2\hat{\gamma}_\phi
\end{equation}
(possibly up to the addition of $N$ decoupled free fields  \cite{Behan:2017emf,Behan:2018hfx}, although our computations of $\Ft$, along with those of \cite{Giombi:2024zrt}, suggest that this interpretation is not quite correct).
That is, the CFT data of the short-range models and the long-range models are identical at the crossover points \cite{Chai:2021arp}.
In this paper, we show that the same applies for $F$.

The details of how to reach these CFTs are complex.
For any $s > s_\mathrm{SR}|_{k=1}$, the CFT defined by \eqref{eq:moreGeneralON} will generically flow to the $k=1$ short-range CFT, as described by \cite{Behan:2017emf,Chai:2021arp}. %
We can understand this qualitatively in the following way. 
Due to the local divergences coming from the Wilsonian RG, the field theory is always trying to generate a $k=1$ short-range kinetic term for the scalar.
This would give $\phi$ the scaling dimension of $\tfrac{d-s_\mathrm{SR}}{2}$.
However, for $s<s_\mathrm{SR}$ the explicit kinetic term of dimension $\tfrac{d-s}{2} > \tfrac{d-s_\mathrm{SR}}{2}$ \enquote*{wins}, so $\phi$ gets stuck with scaling dimension $\tfrac{d-s}{2}$.
For $s>s_\mathrm{SR}$, unless further counterterms are tuned to cancel those local divergences, we will just find the short-range CFT.

However, in the following \textit{we will assume that further counterterms are tuned to ensure that we can flow to a CFT with $\Delta_\phi = \tfrac{d-s}{2}$ even for $s>s_\mathrm{SR}$}, as was done in \cite{Gubser:2017vgc}.
This is similar to how, despite the fact that the $\phi^4$ theory flows to a mean field theory for any $d>4$, we can reach the $\gO(N)$ CFT for $d>4$ by a judicious choice of action \cite{Fei:2014yja,Giombi:2019upv}.

To forestall a potential point of confusion: there are now two different ways of reaching the $\gO(N)$ CFT.
The reason for this is a difference in the order of limits: the limit $s \to 2k$ does not commute with the limit in which the regulator of this theory ($\delta$) is removed. 
Thus, as we show explicitly in \cref{app:nonminimal}, the short-range CFT can be reached by either
\begin{enumerate}
\item \eqref{eq:moreGeneralON}, where we fix $s\to 2k$ and then take $\delta \to 0$;
\item or \eqref{eq:moreGeneralON}, where we first take $\delta \to 0$ and then take $s \to 2k-2\hat{\gamma}_\phi$.
\end{enumerate}

The same comments apply to the analogous generalizations of the Gross-Neveu model, where $\Delta_\psi^\UV = \tfrac{d}{2}-k$ (for $k-\thalf$ integer) and $\Delta_\psi^\UV = \tfrac{d-s}{2}$: the short- and long-range CFTs coincide at $s_\mathrm{SR} = 2k -2\hat{\gamma}_\psi$.

\subsubsection*{The short-range models lie at the extremal $F$ for the long-range models}

Given that the short-range fixed points lie on the curve parametrised by $s$ (equivalently, parametrised by $\Delta_\phi$), it is natural to ask whether there is anything special about the short-range fixed points from the long-range perspective.
Put another way, if we only knew about the long-range CFTs, parametrised by $s$, how would we notice that $s_\mathrm{SR}$ was special?
The answer is\footnote{There are other distinguishing features: for example, it is only for $s_\mathrm{SR}$ that the theory is local. %
Hence, it is only for $s_\mathrm{SR}$ that the stress-energy tensor is a local operator, appearing in the OPEs of the fundamental fields \cite{Fraser-Taliente:2026gdh} -- this is just as for the long-range melonic CFTs \cite{Benedetti:2019ikb,Benedetti:2019rja,Fraser-Taliente:2025spectrum}.} that, at least to order $1/N^2$, the short-range models lie at the extrema of $F^\mathrm{LR}_i(s)$:
\begin{equation}\label{eq:extremalF}
\odv{F^\mathrm{LR}_i(s)}{s} \Big\rvert_{s=s_\mathrm{SR}} = 0,
\end{equation}
for any $k$ (a similar story holds in the $\epsilon$ expansion \cite{Fraser-Taliente2026prep}). 
In particular, for the standard $k=1$ $\gO(N)$ model and $k=\half$ GN model, $\Ft^\mathrm{LR}_i(s)$ is maximized by $s_\mathrm{SR}$ in $2<d<4$.
Since we calculate $F$ up to order $1/N$, this means that $\Delta_\phi^\mathrm{SR} = \tfrac{d-s_\mathrm{SR}}{2}$ can be found to order $1/N^2$.
This gives a compact encoding of the short-range vector model's conformal data, as the extremum of a particular function.

\begin{figure}[ht!]
\centering
\includegraphics[alt={Two curves are shown, black and blue, with the blue one, corresponding to the interacting theory, having a maximum at a lower F and s.},width=0.7\textwidth]{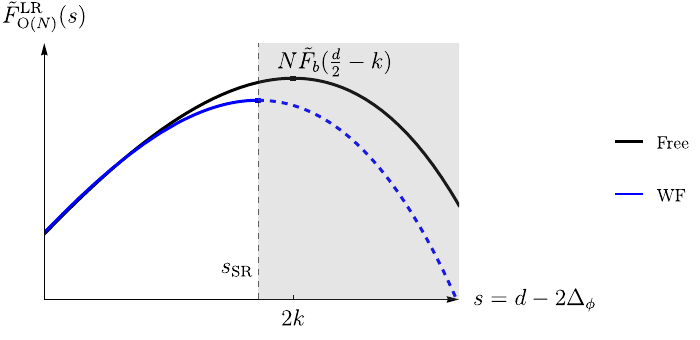}
\caption{A schematic of the long-range free energy $\Ft^\mathrm{LR}(s)$ for the $\gO(N)$ model.
We have taken odd $k$, and assume $d>2k$, meaning that $\Ft$ is maximized to leading order in $N$; this is the situation for the standard $\phi^4$ $\gO(N)$ CFT ($k=1$).
The free CFT of $N$ scalars (black line) has $\Ft=N\Ft_b(\tfrac{d-s}{2})$, which is maximal at $s=2k$, marked with a box. 
The WF interacting CFT (blue line) has a maximum at $s=s_\mathrm{SR}=2k-2\hat{\gamma}_{\phi,1}/N + O(1/N^2)$, marked with a box.
Both maxima are the locations of the short-range CFTs, and we have chosen a value of $d$ such that the free local theory has a larger $\Ft$.
Unless counterterms are tuned, for $s> s_\mathrm{SR}$ we flow to the short-range model on perturbing the free scalars by $\phi^4$, and so that regime of the interacting model is dashed.}
\label{fig:maximum}
\end{figure}

For a free long-range field, it is trivial that the extrema of the long-range $\Ft_{b,f}(\Delta)$ lie at the short-range scaling dimensions $\Delta = \dotwo - k$. 
Therefore, the situation here is \textit{identical} to the case of a free field of dimension $\tfrac{d-s}{2}$: that is a one-parameter family of (generically non-local) CFTs for which the extrema of $\Ft$ lies at the local points $s=2k$, if $d>2k$.
Thus, the interaction just shifts that extremum by $s_\mathrm{SR}-2k = -2\hat{\gamma} \sim O(1/N)$. 
Both lines of CFTs are illustrated in \cref{fig:maximum} for the $\gO(N)$ model, where we have taken:
\begin{itemize}
\item odd $k$ to show the case where $\Ft$ is maximized. %
\item $d>2k$ to ensure that $\Delta_\phi^\mathrm{SR}>0$, and so the CFT makes sense.
\item $4k>d$; this is the condition for the relevance of $\phi^4$ as a perturbation of the free theory, and defines the upper critical dimension \cite{Gubser:2017vgc}. 
\end{itemize} 
For some values of $d$ (but not all) in this range, we have $\Ft(\Delta_\sigma^\mathrm{SR}) < 0$, and so the flow from the $\Box^k$ free theory to the $\Box^k$ WF theory satisfies the $\Ft$-theorem $ \Ft^\mathrm{SR}_\mathrm{free}>\Ft^\mathrm{SR}_\mathrm{WF}$ to leading order in $N$.
This is the situation shown in \cref{fig:maximum}: in the $k=1$ theory, it is true for all $2<d<4$; in the $k=2$ theory, it is true for $5.35\lesssim d<8$.

\subsection{The shortest summary of our results}

The $F$s are presented in \cref{sec:Fresults}, but we wish now to emphasise their compact nature in terms of the scaling dimensions.
The interacting CFT of a bosonic ($x=b$) or fermionic ($x=f$) vector with full scaling dimension $\Delta_x \equiv \tfrac{d-s}{2}$ has %
\begin{equation}\label{eq:FsimpleSummary}
F_\mathrm{CFT} = N F_x(\Delta_x) + F_b(s) + \frac{F_b'(s)}{N} \left(\half \Gamma_\mathrm{tet} +\frac{1}{3}\Gamma_\mathrm{pr} \right) + O(1/N^2),
\end{equation}
regardless of whether it is short- or long-range.
Here, $F_x(\Delta_x)$ is a standard quantity, the sphere free energy of a free boson/fermion of scaling dimension $\Delta_x$; $\Gamma_{\mathrm{tet}}$ and $\Gamma_{\mathrm{pr}}$ are just separate $s$-dependent contributions to the scaling dimension of the auxiliary field $\sigma$, 
\begin{equation}
  \Delta_\sigma=  d-2\Delta_x+\tfrac{1}{N}(\Gamma_\mathrm{tet} +\Gamma_\mathrm{pr})+O(1/N^2),
\end{equation}
and will be defined in \eqref{eq:ONtetpr} and \eqref{eq:GNtet} for the $\gO(N)$ and GN models respectively. 
$s_\mathrm{SR}$ and the corresponding 
$\Delta_{x}=\tfrac{d}{2}-k + \frac{\hat{\gamma}_{x,1}}{N} + \frac{\hat{\gamma}_{x,2}}{N^2}+ O(1/N^3)$ in the short-range version is then picked out by the extremization condition \eqref{eq:extremalF}.
The full expressions for $\hat{\gamma}_{x,2}$ in particular are quite cumbersome (to say nothing of $\hat{\gamma}_{x,3}$), so this is a pleasing encapsulation.

We now discuss the correct way to regulate these theories.

\subsection{Proper regularization}

Focusing now on the $\gO(N)$ model, we attack the action \eqref{eq:ONactionWithSig} by performing the quadratic $\phi$ integral and expanding the logarithm in powers of $N$
\begin{subequations}\label{eq:sigmaEffectiveAction}
\begin{align}
Z_{\gO(N)} &= Z_{\phi, \mathrm{free}}^N \int \DdImag{\sigma} \exp(- S_\sigma)\\
S_\sigma &\equiv \frac{N}{2} \tr \log\left(\mathbb{I}-\frac{M}{\sqrt{N}}\right) -\frac{1}{2\lambda_0} \int_x \sigma^2 \\
&=  -\frac{N}{2} \sum_{q=1}^\infty \frac{1}{N^\frac{q}{2}}\frac{1}{q} \tr M^q -\frac{1}{2\lambda_0} \int_x \sigma^2
\end{align}
\end{subequations}
where we are treating $M_{xy} = \sigma(x) C(x,y)$ as a matrix (this computation is reviewed in full in \cref{sec:ONsetup}).
This is a second, explicitly non-local, QFT that encodes the correlators of $\sigma \sim\,  \normord{\phi_i \phi_i}$ in the original theory, and by construction has the same partition function $Z_{\gO(N)}$.

We here have an infinite number of couplings, but to any fixed order in $1/N$ we need only consider a finite number of them; since we want to calculate $F$ to order $1/N$, we go only to $q=4$.
Since the $\phi^2$ mass term is tuned to vanish by assumption, the $\sigma$ tadpole must vanish: $\tr M=0$.
Thus, the $\sigma$ propagator that is generated is the inverse of $-\half C(x,y)^2$, treated as a matrix, which is
\begin{equation}\label{eq:generatedSigmaProp}
\expval{\sigma(x)\sigma(y)} = \frac{C_{\sigma,0}}{s(x,y)^{2\Delta_\sigma^\UV}} + O(1/N), \quad \Delta_\sigma^\UV\equiv d- 2\Delta_\phi^\UV,
\end{equation}
for some normalization $C_{\sigma,0}$.
This is a conformal propagator if we take $s(x,y)$ to be either the chordal distance on the sphere, defined in \cref{app:sphereKineticTerms}, or $\abs{x-y}$ in flat space.
Of course, for the usual short-range model $\Delta_\phi^\UV = \tfrac{d-2}{2}$, so $\sigma^2$ has the naive irrelevant scaling dimension $4>d$; hence we are justified in dropping it when approaching the IR.
We drop it for all the theories that we consider, as we want to work at the conformal fixed point.

If it were easy to regularize this action with a hard cutoff, the computation of $F$ would now be straightforward.
It is not.
Instead, we opt for an analytic regularization scheme DREG$+\delta$, where we analytically continue in two parameters:
\begin{enumerate}
  \item DREG: a standard technique in QFT, we essentially evaluate integrals in dimension $d$ low enough that they converge, and analytically continue back to the region $d$ of interest. 
  As usual, this allows us to ignore power-law divergences.
  \item A shift of the UV scaling dimension of $\sigma$: $\Delta_\sigma^\UV \mapsto \Delta_\sigma^\UV - \delta$.
\end{enumerate}
The reason for the $\Delta_\sigma^\UV$ shift is that DREG no longer suffices to regulate this action once we go beyond the leading order of the critical vector model; this is because DREG works by making a marginal interaction slightly relevant, which turns logarithmic divergences $\log \Lambda$ into poles $1/\epsilon$.
Unfortunately, $\sigma \phi_i \phi_i$, the bare interaction, is marginal and hence has logarithmic divergences for all $d$.

When computing the free energy, the $\Delta_\sigma^\UV$ shift must be implemented consistently in order to find the correct value of $\Ft$; this can only be done by modifying the kinetic term in the action such that the field has scaling dimension $\Delta_\sigma^\UV-\delta$. 
First, we drop the irrelevant $\sigma^2$ term from \eqref{eq:sigmaEffectiveAction}.
We are trying to subtract off a dynamically generated kinetic term; hence we need to add an $O(N^0)$ term to the action, of schematic form
\begin{equation}\label{eq:addedKinetic}
S_{\sigma,\delta} = \frac{1}{g_0^2}\int_{x,y} \half \sigma(x) \left(\frac{c_\delta}{s(x,y)^{2(d-\Delta_{\sigma}^\UV +\delta)}} - \frac{c \mu^{2\delta}}{s(x,y)^{2(d-\Delta_\sigma^\UV)}}\right) \sigma(y)
\end{equation}
for some constants $c,c_\delta$.
The first term is the new kinetic term, and the second should almost entirely subtract off the old one.
If this seems strange, recall that the renormalization procedure generally involves adding some new UV physics. 
This is usually done by changing the bare coupling: adding a counterterm. 
Here, we are changing the UV physics more explicitly by actually modifying the kinetic term in the UV; this improves the UV behaviour of the diagrams, turning the logarithmic divergences into $1/\delta$ poles, as required. %

Before this, $\sigma$ only appeared linearly in the $\lambda_0 \to \infty$ action, and the normalization of the $\sigma$ field could therefore be changed freely to set the coefficient of $-\tfrac{Z_\phi}{\sqrt{N}} \sigma \phi_i \phi_i$ to $1$. 
With this new quadratic term, the normalization of $\sigma$ is now meaningful.
To canonically normalise the field, we set $\sigma_\text{new} =\sigma_\text{old}/g_0$, making the interaction $\frac{Z_\phi g_0}{\sqrt{N}} \sigma \phi_i \phi_i$.
The critical point of this regulated theory now lies in the limit $g_0 \to \infty$, which is the deep IR. 
\textit{We work directly at the critical point}; to control this limit, we write $g_0 = \mu^\delta g Z_g$, where $Z_g$ takes the form of a power series in $1/N$, with each term containing $1/\delta$ poles: thus as $\delta\to 0$, we implement the aforementioned $g_0\to \infty$.

In the effective action for $\sigma$ the second term of \eqref{eq:addedKinetic} partially cancels with the dynamically generated kinetic term for $\sigma$, leaving a residual non-local quadratic $\sigma$ term with a coefficient $\sim Z_{g,1}/(\delta N)$.
This must be treated as an interaction term; taking care to take the correct order of limits, which is $N\to\infty$ first, followed by $\delta \to 0$, we find that the diagram that it multiplies is of order $\delta$. 
Hence, we find an order
\begin{equation}
  \sim \frac{\delta}{\delta} \frac{Z_{g,1}}{N}
\end{equation}
correction to the naive free energies of these models.

\subsection{Discussion}\label{sec:discussion}

In this paper we computed the free energy of the vector models in the large-$N$ limit to order $1/N$, confirming the conjectures of \cite{Tarnopolsky:2016vvd}, which matched the $\epsilon$-expansion results of \cite{Fei:2015oha,Giombi:2024zrt}.
The mismatch of \cite{Tarnopolsky:2016vvd} was resolved by a careful accounting of the analytic regulator, a shift in the UV scaling dimension; this combined with the counterterms to give a contribution, essentially due to a factor $\lim_{\delta \to 0}\delta/\delta$. 
This demonstrates another danger of regularizing by analytically continuing in an operator dimension, on top of the known danger of breaking supersymmetry \cite{Gerchkovitz:2014gta}.
A similar approach to ours here could also be applied to, for example, the $\gO(N)^r$ models of \cite{Jepsen:2023pzm}, or large-$N$ QED$_{d}$ \cite{Giombi:2015haa} (which could then be checked by comparison to the $d=3$ result of \cite[(5.8)]{DiPietro:2019hqe}).

Our results also were naturally generalizable to the long-range vector models and the $\Box^k$ CFTs.
A notable feature of all of these $F$s at this order is their simplicity when expressed in terms of the scaling dimensions, \cref{eq:FsimpleSummary}.
This fact follows naturally from how vacuum diagrams are obtained from two-point function diagrams by the addition of a propagator.
At the next order in $N$ there are many more diagrams, with some giving non-zeta contributions \cite{Derkachov:1993uw,Broadhurst:1996yc}; however, it would be interesting to confirm whether the contributions to the long-range model continue to be of the form $F\supset F'(s)(\tfrac{1}{m_i} \tfrac{\Gamma_i}{N^2})$ for $\frac{\Gamma_i}{N^2} \subset \gamma_\sigma$, with analogous results for the short-range model coming from the limit $s \to s_\mathrm{SR}$; this could be done relatively straightforwardly by comparison to the $\epsilon$ expansion.
We also observed that the short-range models lie at the extrema of the long-range free energy, giving a minimal encoding of the otherwise cumbersome $\Delta_\phi^\mathrm{SR}$; it would be nice to prove this in general.

It is worth noting that, in agreement with \cite{Giombi:2024zrt}, we found no sign of the additional scalar field $\chi$ noted in \cite{Behan:2017emf,Chai:2021arp} at the point of crossover from the long-range to the short-range theory: we found 
\begin{equation}
\lim_{s \to s_\mathrm{SR}}F_\mathrm{LR}(s) = F_\mathrm{SR},
\end{equation}
where our $F_\mathrm{SR}$ also matched the $\epsilon$ expansion.
The expectation from \cite{Behan:2017emf} was that
\begin{equation}
\lim_{s \to s_\mathrm{SR}}F_\mathrm{LR}(s) \stackrel{?}{=} F_\mathrm{SR}+N F_b(\Delta_\chi = \tfrac{d+s}{2}),
\end{equation}
since $\chi_i$ was speculated to be a vector of $N$ decoupled generalized free fields with scaling dimensions equal to the shadow of $\phi_i$ (and so $F_b(\tfrac{d+s}{2})=-F_b(\tfrac{d-s}{2})=-F_b(\Delta_\phi)$). 
This does not agree with our analysis, and so requires further study.

Now, CFTs are usually considered on flat space. 
However, $F$ must be computed on the sphere, as otherwise the free energy diverges with the volume: this CFT is the same as the flat-space one in the sense that it shares the same conformal \href{https://arxiv.org/pdf/1809.05111#subsubsection.3.3.4}{data} $\{(-\Delta_i,\rho_i), C_{ijk}\}$ as the flat-space CFT. 
To justify this: we expect it to be generically possible to couple a flat-space CFT to a background metric $g_{\mu\nu}$ in such a way that the conformal symmetry descends from Weyl symmetry \cite{Farnsworth:2021ycg,Benedetti:2021wzt}. %
If so, then because the sphere is related to flat space by a Weyl map, $g_{\mu\nu}^{S^d} = \Omega^2(x) \eta_{\mu\nu}$, then the CFTs must be the same.
We can also show by explicit calculation that the CFT data is the same in either space, which is what we have done for this paper.
Thus, the quantity $\Ft$ can be associated with the abstract CFT.
However, in general that coupling to $g$ may be non-trivial (or even impossible): we must know the curvature couplings, which we can ignore for the flat-space CFT.
At our order $1/N$, we need only consider the curvature couplings coming from the kinetic term.
However, at subleading orders we must consider the curvature counterterms, as has been seen already in the $\epsilon$ expansion \cite{Fei:2015oha,Giombi:2024zrt}.

Finally, there is debate at present over whether the critical $\gO(N)$ $\phi^4$ vector model and the critical $\gO(N)$ NLSM are indeed the same. 
However, it was optimistically speculated that the two theories match to all orders in the $1/N$ expansion \cite{DeCesare:2025ukl}, in which case the computations here apply to both.

\section{The sphere free energies}\label{sec:Fresults}

We now summarise our results for all the vector models of interest.
In the following, we will make heavy use of
\begin{equation}
A(\Delta) \equiv \frac{\Gamma(\tfrac{d}{2}-\Delta)}{\Gamma(\Delta)}, \quad A_f(\Delta)\equiv \frac{\Gamma \left(\frac{d}{2}-\Delta +\frac{1}{2}\right)}{\Gamma \left(\Delta +\frac{1}{2}\right)},
\end{equation}
which come from the Fourier transforms of bosonic and fermionic propagators respectively (see \cref{app:propsAndAs}). 
We shall see that it is more compact to first present the long-range case, and then give the short-range version. 
We note that the (universal part of the) free energy density with respect to the regularized volume of hyperbolic space\footnote{The computation of the regularized $\vol \mathbb{H}^{d+1}$ is discussed in the appendices of \cite{Fraser-Taliente:2024hzv}, but amounts to computing the volume of hyperbolic space in DREG such that it is finite for generic $d$. 
The factor of $\half \vol S^{d+1}$ in \eqref{eq:Ftdef} is not strictly necessary, being smooth and positive for all $d \ge 0$, but is there by convention from \cite{Giombi:2014xxa}.}, 
\begin{equation}\begin{aligned}\label{eq:Ftdef}
\Ft \equiv \half \frac{\vol S^{d+1}}{\vol \mathbb{H}^{d+1}} [F]_\text{universal} = - \sin(\tfrac{\pi d}{2}) F\rvert_{S^d,\mathrm{DREG}} = \sin(\tfrac{\pi d}{2}) \log Z\rvert_{S^d,\mathrm{DREG}}
\end{aligned}\end{equation}
is the more natural quantity to consider.
$\Ft$ is believed to be finite, independent of the sphere radius, with a smooth analytic continuation in $d$, and satisfy for a general RG flow
\begin{equation}
\Ft_\mathrm{UV} > \Ft_\mathrm{IR};
\end{equation}
it is therefore thought to serve as a measure of the number of degrees of freedom of a (unitary) CFT \cite{Giombi:2014xxa,Fei:2015oha}.
We shall move freely between the two, writing $F$ in general but $\Ft$ when we want to consider its sign.
Happily, we write the CFT $F$s below in such a way that we need only map $F_x(\Delta) \mapsto \Ft_x(\Delta)$ to find the $\Ft$s.
Note also that the $+O(1/N^2)$s are left implicit for all the $F$s below.

\subsection{\ONpdf{} vector model}

The free energy of a free scalar field of scaling dimension $\Delta$ is known to be
\begin{equation}\label{eq:bosonF}
F_b(\Delta) \equiv \int_{\frac{d}{2}}^{\Delta} \odif{\Delta} \, F_b'(\Delta), \quad F_b'(\Delta) \equiv -\frac{\pi}{\sin(\tfrac{\pi d}{2}) \Gamma(d+1)}\frac{1}{A(\Delta)A(d-\Delta)}. %
\end{equation}
We give a compact derivation of $F$ for an arbitrary conformal field in \cref{sec:Fcomputation}. %
There are two features of note. 
\begin{itemize}
\item $\Ft_b'(\tfrac{d}{2}-k)=0$. When modified with the usual factor, $\Ft_b''(\tfrac{d}{2}-k)(-1)^k > 0$ for all $d>2k$: hence the standard free field is a maximum of $\Ft_b(\Delta)$. 
This fact is of great importance for the short-range vector models.
\item We have chosen the standard $F_b(\tfrac{d}{2})=0$, agreeing with the identity $\tr \delta(x-y)=0$ that is forced upon us by the lack of a regulating scale in DREG. 
\end{itemize}

\subsubsection{Long-range \ONpdf}

In the IR of the long-range $\gO(N)$ vector model for generic $s$ we find
\begin{equation}\label{eq:LRSDs}
  \Delta_\phi = \frac{d-s}{2}, \quad \Delta_\sigma = s + \frac{\Gamma_{\mathrm{tet}} + \Gamma_{\mathrm{pr}}}{N} +O(1/N^2),
\end{equation}
where $\Delta_\phi$ is exactly $\tfrac{d-s}{2}$ to all orders in $1/N$, and the $\gamma_{\sigma}$ contributions come from the tetrahedral and prism diagrams respectively\footnote{These results, as well as the later counterterms and normalizations, correct various typos in \cite{Chai:2021arp}.},
\begin{equation}\label{eq:ONtetpr}
\Gamma_{\mathrm{tet}}(s) \equiv -\frac{4}{\Gamma \left(\frac{d}{2}\right)} \frac{A\left(\frac{d-s}{2}\right)^2}{A(d-s)}, \quad \Gamma_{\mathrm{pr}}(s) \equiv \frac{8}{\Gamma \left(\frac{d}{2}\right)}  \frac{A\left(d-\tfrac{3 s}{2}\right) A(s) A\left(\frac{d-s}{2}\right)^3}{A(d-s)^2}.
\end{equation}
The free energy is
\begin{equation}\label{eq:FONfuncS}
F_{\gO(N)}^{\mathrm{LR}} = NF_b(\tfrac{d-s}{2}) + F_b(s) + \frac{F_b'(s)}{N}\left(\half \Gamma_\mathrm{tet} + \frac{1}{3} \Gamma_\mathrm{pr}\right) ,
\end{equation}
which we can rewrite using \eqref{eq:LRSDs}
\begin{equation}\begin{aligned}\label{eq:FONfuncDims}
F_{\gO(N)}^{\mathrm{LR}} = N F_b\left(\Delta_\phi\right) + F_b\left(\Delta_\sigma\right) -\frac{F_b'(s)}{N} \left(\half \Gamma_{\mathrm{tet}} +\frac{2}{3} \Gamma_{\mathrm{pr}} \right)+ O(1/N^2)
\end{aligned}\end{equation}
to make it clear that the correction term is not just absorbed by the $F_b$s.
\subsubsection{Short-range \ONpdf{}}

There are two different ways of finding the free energy of the short-range model: 
\begin{enumerate}
\item Compute directly for the short-range $\gO(N)$ CFT. 
\item Take the limit of the long-range results:
\begin{equation}\label{eq:SRphiDim}
\begin{split}
\Delta_\phi &= \frac{d-s}{2} \to \Delta_\phi^{\mathrm{SR}} = \frac{d-2}{2} + \frac{\hat{\gamma}_{\phi,1}}{N} + \hyperref[eq:gammahatphi2]{O(1/N^2)},\\
\hat{\gamma}_{\phi,1} &\equiv \frac{2F_b'(2)}{F_b''\left(\frac{d-2}{2}\right)} = \frac{-2 \Gamma (d-2)}{\Gamma \left(2-\dotwo\right) \Gamma \left(\dotwo-2\right) \Gamma \left(\dotwo-1\right) \Gamma \left(\dotwo+1\right)}.
\end{split}
\end{equation}
\end{enumerate}
As we show in \cref{sec:assemblingF}, the results found in the two cases are identical; but, crucially, the counterterms required in each case differ. 
Either way, we find
\begin{equation}
  \begin{split}
  F_{\gO(N)}^{\mathrm{SR}}& =N F_b\left(\Delta_\phi^\mathrm{SR}\right) + F_b\left(\Delta_\sigma^{\mathrm{SR}}\right) -\frac{F_b'(2)}{N} \left(\half \Gamma_{\mathrm{tet}} +\frac{2}{3} \Gamma_{\mathrm{pr}} \right) \Big\rvert_{s=2},\\
\Delta_\sigma^\mathrm{SR}&= 2 +\frac{\Gamma_{\mathrm{tet}} + \Gamma_{\mathrm{pr}} - 2 \hat{\gamma}_{\phi,1} }{N}\Big\rvert_{s=2} +O(1/N^2),
  \end{split}
\end{equation}
which also follows from $s\to s_\mathrm{SR} \equiv 2-2\frac{\hat{\gamma}_{\phi,1}}{N} + O(1/N^2)$ in \eqref{eq:FONfuncDims}.
Because $F_b'(\tfrac{d-2}{2})=0$, we only need $\hat{\gamma}_{\phi}$ to leading order in $N$, so this expands
\begin{equation}
  \begin{split}
    F_{\gO(N)}^{\mathrm{SR}}&\, =\,  N F_b\left(\tfrac{d-2}{2}\right)+F_b(2)+\frac{F_b'(2)}{N} \left(\frac{F_b''\left(\tfrac{d-2}{2}\right)}{F_b'(2)} \frac{\hat{\gamma}_{\phi,1}^2}{2} -2\hat{\gamma}_{\phi,1} + \half \Gamma _{\text{tet}}+ \frac{1}{3}\Gamma _{\text{pr}}\right)\\
  &\,\stackrel{\mathmakebox[\widthof{=}]{\eqref{eq:SRphiDim}}}{=}\, N F_b\left(\tfrac{d-2}{2}\right)+F_b(2)+\frac{F_b'(2)}{N} \left(\half \Gamma _{\text{tet}}+ \frac{1}{3}\Gamma _{\text{pr}}-\hat{\gamma}_{\phi,1}\right)\Big\rvert_{s=2},
\end{split}
\end{equation}
where at this order we can evaluate the $\Gamma$s for $s=2$.
This result perfectly matches the conjecture of \cite{Tarnopolsky:2016vvd},
\begin{equation}
  F_{\gO(N)}^{\mathrm{SR}} = N F_b\left(\tfrac{d-2}{2}\right)+F_b(2)+ \left(\frac{3-d}{3 (d-2) (d-1)}+\frac{1}{d}\right) \frac{\hat{\gamma}_{\phi,1}}{N}.
\end{equation}

\subsection{Gross-Neveu model}\label{eq:resultsGN}

Turning now to the fermionic vector model, we recall that the free energy of a single complex component of a free Dirac fermion of dimension $\Delta$ is
\begin{equation}\begin{aligned}\label{eq:fermionF}
F_f(\Delta) &\equiv \int_{\frac{d}{2}}^{\Delta} \odif{\Delta} \, F_f'(\Delta), \quad &F_f'(\Delta) \equiv +\frac{\pi}{\sin(\tfrac{\pi d}{2}) \Gamma(d+1)}\frac{2}{A_f(\Delta)A_f(d-\Delta)}, %
\end{aligned}\end{equation}
where the $2$ in the numerator comes from the complex nature of the field. 
This also satisfies $\Ft_f'(\dotwo -k)=0$ and $\Ft''(\dotwo -k) (-1)^{k-\half} < 0$ for all $d>2k-1$ for half-integer $k$.
Taking the GN model to have a $U(n)$ symmetry, our definition of $N\equiv n\tr \mathbb{I}_s$ \cite{Zinn-Justin:1991ksq} means that $N$ free Dirac fermions contribute $NF_f(\tfrac{d-1}{2})$ to the free energy.
\subsubsection{Long-range GN}
In the IR of the complex Gross-Neveu vector model, for generic $s$
\begin{equation}
\Delta_\psi \equiv \frac{d-s}{2}, \quad \Delta_\sigma = s + \frac{\Gamma_\mathrm{tet}^{\mathrm{GN}}}{N} + O(1/N^2),
\end{equation}
where
\begin{equation}\label{eq:GNtet}
\Gamma_\mathrm{tet}^\mathrm{GN} \equiv \frac{2}{\Gamma(d/2)} \frac{A_f(\tfrac{d-s}{2})^2}{A(d-s)},
\end{equation}
and the equivalent of $\Gamma_\mathrm{pr}^\mathrm{GN}=0$ because there is no cubic vertex for $\sigma$ in the $\sigma$-only QFT.
Then the free energy is
\begin{subequations}
\begin{align}
F_{\mathrm{GN}}^\mathrm{LR} &= N F_f(\tfrac{d-s}{2}) + F_b(s) + \frac{F_b'(s)}{N}\left(\half \Gamma_\mathrm{tet}^\mathrm{GN}\right) \label{eq:FGNfuncS}\\
&= N F_f(\Delta_\psi) + F_b(\Delta_\sigma) - \frac{F_b'(s)}{N}\left(\half \Gamma_\mathrm{tet}^\mathrm{GN}\right) . \label{eq:FGNfuncDims}
\end{align}
\end{subequations}
\subsubsection{Short-range GN}

As before, we could also compute the free energy of the short-range GN CFT directly, but the results are identical to extremizing the LR $F$.
The extremum lies at $s \to s_\mathrm{SR} \equiv 1- 2\frac{\hat{\gamma}_{\psi,1}}{N} + O(1/N^2)$, such that
\begin{equation}\begin{aligned}
\Delta_\psi \to \Delta_\psi^{\mathrm{SR}} &= \frac{d-1}{2} + \frac{\hat{\gamma}_{\psi,1}}{N} +\hyperref[eq:GNboxKpsi2]{O(1/N^2)}, \quad \hat{\gamma}_{\psi,1} \equiv \frac{2F_b'(1)}{F_f''(\tfrac{d-1}{2})}=\frac{\Gamma (d-1)}{\Gamma(2-\dotwo) \Gamma(\dotwo-1)^2 \Gamma(\dotwo+1)} \label{eq:SRpsiDim},\\
\Delta_\sigma^{\mathrm{SR}} &=1 +\frac{\Gamma_{\mathrm{tet}}^\mathrm{GN} -2 \hat{\gamma}_{\psi,1}}{N}.
\end{aligned}\end{equation}
Plugging this in to \eqref{eq:FGNfuncDims} the short-range free energy is
\begin{subequations}
\begin{align}
  F_{\mathrm{GN}}^\mathrm{SR} \,&=\,  N F_f(\Delta_\psi^\mathrm{SR}) + F_b(\Delta_\sigma^\mathrm{SR}) - \frac{F_b'(s)}{N}\left(\half \Gamma_\mathrm{tet}^\mathrm{GN}\right)\\
  & \,\stackrel{\mathmakebox[\widthof{=}]{\eqref{eq:SRpsiDim}}}{=}\, N F_f(\tfrac{d-1}{2}) + F_b(1) + \frac{F_b'(1)}{N}\left(\half \Gamma_\mathrm{tet}^\mathrm{GN}-\hat{\gamma}_{\psi,1}\right)\Big\rvert_{s=1},
\end{align}
\end{subequations}
where again we need only evaluate $\Gamma_\mathrm{tet}^\mathrm{GN}$ for $s=1+O(1/N)$. 
Again, we match \cite{Tarnopolsky:2016vvd},
\begin{equation}
  F_\mathrm{GN}^\mathrm{SR} = N F_f\left(\tfrac{d-1}{2}\right)+F_b(1)+ \frac{1}{d}\frac{\hat{\gamma}_{\psi,1}}{N}.
\end{equation}

\subsection{Supersymmetric model}

For completeness, we now present the short-range\footnote{The supersymmetric long-range models do not appear to be known. 
However, by analogy with the $\gO(N)$ and GN models, we can speculate that the long-range vector SCFT is a SCFT of superfields with scaling dimensions $\Delta_\Phi = \tfrac{d-s}{2}$ and $\Delta_\Sigma = s-1 +O(1/N^2)$.
It should be reachable by RG flow from a UV free theory for suitable ranges of $s$ and $d$: the scaling dimension of $\Phi$ should be unchanged from the UV, whereas that of $\Sigma$ should be modified from a value less than $s-1$ to $\Delta_\Sigma=s-1$.} %
supersymmetric version of the $\gO(N)$ model -- the theory of $N+1$ chiral superfields with an $\gO(N)$-symmetric superpotential $W= \frac{\lambda_0}{2} \Sigma \Phi_i \Phi_i$ presented in \cite{Tarnopolsky:2016vvd}, which has action \cite{Ferreira:1997he}
\begin{equation}
S_\mathrm{SUSY} = \int \odif[order=d]{x}\left[\int \odif[order=4]{\theta}\, (\bar\Phi_i \Phi_i + \bar{\Sigma} \Sigma) -\frac{\lambda_0}{2} \int \odif[order=2]{\theta} \,\Sigma \Phi_i \Phi_i -\frac{\lambda_0}{2} \int \odif[order=2]{\bar{\theta}} \,\bar{\Sigma} \bar\Phi_i \bar\Phi_i \right].
\end{equation}
For the IR CFT, we know that the superpotential must have $R$-charge $2$, so $\Delta_\Sigma + 2\Delta_\Phi = d-1$. 
Localization then permits an exact calculation of the free energy in terms of the unknown $\Delta$s,
\begin{equation}\begin{aligned}\label{eq:FSUSYfull}
F_\mathrm{SUSY} = N F_{S}(\Delta_\Phi) + F_{S} (\Delta_\Sigma),
\end{aligned}\end{equation}
defined in terms of the free energy of a free chiral superfield
\begin{subequations}
\begin{align}
  F_S(\Delta) &= 2 F_b(\Delta) + 2 F_f(\Delta+\thalf) + 2 F_b(\Delta+1),\\
  F_S'(\Delta) &= \frac{1}{\sin(\tfrac{\pi d}{2})}\frac{\pi}{\Gamma(d-1)} \frac{1}{A(\Delta)} \frac{1}{A(d-1-\Delta )},
\end{align}
\end{subequations}
containing a complex scalar, a fermion with 4 real components, and an auxiliary complex scalar. 
As before, $F_f$ is the free energy for a single complex fermion component, and we have analytically continued in $d$ while keeping the dimension of our fermions fixed \cite{Fraser-Taliente:2024rql,Giombi:2014xxa}.
$\Ft$-extremization then tells us that we can determine $\Delta_\Phi$ by extremizing this exact function $F_\mathrm{SUSY}(\Delta_\Phi, \Delta_\Sigma)$, subject to the constraint \cite{Giombi:2014xxa}.

Foreshadowing the next section, we slightly generalize the kinetic term for $\Phi_i$ to be $\sim \Phi_i D^{2(k-1)} \Phi_i$ (and add in suitable counterterms).
We can then expand around $\Delta_\Phi = \tfrac{d}{2}-k + \hat{\gamma}_{\Phi}$ and $\Delta_\Sigma = 2k-1-2\hat{\gamma}_\Phi$ in the large-$N$ limit.
Using $F_S'(\tfrac{d}{2}-k)=0$, extremization finds that
\begin{equation}\label{eq:SUSYanomDims}
\hat{\gamma}_{\Phi,1} = \frac{2F_S'(2k-1)}{F_S''(\tfrac{d}{2}-k)}, \quad \hat{\gamma}_{\Phi,2}= -\hat{\gamma}_{\Phi,1}^2 \left(\frac{F_S'''(\tfrac{d}{2}-k)}{2 F_S''(\tfrac{d}{2}-k)}+\frac{2 F_S''(2k-1)}{F_S'(2k-1)}\right), \quad \Delta_{\Sigma} = 2k-1-2\hat{\gamma}_\Phi,
\end{equation}
which agrees with \cite{Gracey:1990aw,Gracey:1991yz}.
Hence, compactly, the expansion of \eqref{eq:FSUSYfull} begins
\begin{subequations}
\begin{equation}\label{eq:FforSUSY} 
F_\mathrm{SUSY} =N F_S(\tfrac{d}{2}-k) + F_S(2k-1) +\frac{ F_S'(2k-1)}{N}\left(-\hat{\gamma}_{\Phi ,1}\right)
\end{equation}
which is almost identical to \eqref{eq:generalFsforBoxks}, albeit with both $\Gamma$s set to zero.
This continues with
\begin{equation}\label{eq:SUSYFnextOrder}
  +  \frac{1}{N^2} \left(2 F_S''(2k-1) + \frac{1}{6} F_S'''(\tfrac{d}{2}-k) \hat{\gamma}_{\Phi ,1}\right)\hat{\gamma}_{\Phi ,1}^2.
\end{equation}
\end{subequations}
\subsection{\texorpdfstring{$\Box^k$}{Box\textasciicircum k} CFTs}

The results for the free energy of the Wilson-Fisher-like (bosonic) $\Box^k$ CFTs and (fermionic) $\slashed{\partial}\Box^{k-1/2}$ CFT are obtained in the same way as the other short-range CFTs.
In the former, $k$ is integer; in the latter, $k-1/2$ is an integer.

As before, the quickest way to find the results is to plug $s=2k-2 \hat{\gamma}_{\phi,1}/N$ into \eqref{eq:FONfuncS} and $s=2k-2 \hat{\gamma}_{\psi,1}/N$ into \eqref{eq:FGNfuncS}.
Adding in the supersymmetric results for comparison, the leading anomalous dimension of the vector is
\begin{equation}\label{eq:anomDimsBoxK}
  \hat{\gamma}_{\phi,1} \equiv \frac{2F_b'(2k)}{F_b''(\dotwo-k)}, \quad 
  \hat{\gamma}_{\psi,1} \equiv \frac{2F_b'(2k)}{F_f''(\tfrac{d}{2}-k)}, \quad
  \hat{\gamma}_{\Phi,1} \equiv \frac{2F_S'(2k-1)}{F_S''(\tfrac{d}{2}-k)}.
\end{equation}
As was described in \cite{Fraser-Taliente:2024hzv}, this leading correction to the vector anomalous dimension manifestly follows from an analogous $\Ft$-extremization, despite the lack of supersymmetry in the first two cases.
The $\sigma$ field has scaling dimensions that cannot be determined by $\Ft$-extremization,
\begin{equation}\begin{aligned}\label{eq:sigmaDimBoxK}
\Delta_\sigma^{\Box^k} &= 2k+ \frac{\Gamma_\mathrm{tet} + \Gamma_\mathrm{pr}-2\hat{\gamma}_{\phi,1}}{N}\\
\Delta_\sigma^{\partial\Box^{k-1/2}} &= 2k+ \frac{\Gamma_\mathrm{tet}^\mathrm{GN}+0-2\hat{\gamma}_{\psi,1}}{N}\\
\Delta_\Sigma^{D^{2(k-1)}} &= 2k-1+ \frac{0+0-2\hat{\gamma}_{\Phi,1}}{N},
\end{aligned}\end{equation}
which leads to free energies
\begin{equation}\begin{aligned}\label{eq:generalFsforBoxks}
F^{\Box^k}_{\gO(N)} &= N F_b(\dotwo-k) + F_b(2k) + \frac{F_b'(2k)}{N} \left(\half \Gamma_\mathrm{tet} + \frac{1}{3} \Gamma_\mathrm{pr} - \hat{\gamma}_{\phi,1}\right),\\
F^{\slashed{\partial}\Box^{k-1/2}}_\mathrm{GN} &= N F_f(\tfrac{d}{2}-k) + F_b(2k) + \frac{F_b'(2k)}{N} \left(\half \Gamma_\mathrm{tet}^\mathrm{GN} + 0 - \hat{\gamma}_{\psi,1}\right),\\
F_{\mathrm{SUSY}}^{D^{2(k-1)}} &= N F_S(\tfrac{d}{2}-k) + F_S(2k-1) +\frac{ F_S'(2k-1)}{N}\left(0+ 0 -\hat{\gamma}_{\Phi ,1}\right),
\end{aligned}\end{equation}
where again at this order we need only evaluate the $\Gamma$s to leading order in $1/N$ (i.e. for integer $s$).
Note also that these results follow from evaluating \cref{eq:FONfuncDims,eq:FGNfuncDims} for the short-range scaling dimensions.

\subsection{Curious similarities}\label{sec:synchronicity}

It was observed by Gracey \cite{Gracey:1991yz} that the bosonic, fermionic, and supersymmetric computations of anomalous dimensions are all extremely similar, with two, one, and zero of the same diagrams contributing in each case. 
We highlight this here for the free energy also.
The only slight complication is that we are considering a real bosonic model and a complex fermionic model. 
We must therefore take:
\begin{itemize}
\item  $N_b=N$ and $N_f = 2N$, which counts the number of real degrees of freedom of the complex fermionic field. However, since $\psi$ is complex, we also map $F_b \mapsto \thalf F_f$. Thus, $N F_b \mapsto N F_f$.
\item $\tilde{A}_b=A$ and $\tilde{A}_f \equiv  - i A_f$ (accounting for the $-i$ coming from the fermion's Fourier transform in \eqref{eq:FlamsDef}).
\end{itemize} 
Then the expressions for $\hat{\gamma}_{\phi/\psi,1}$ and $\Gamma_{\mathrm{tet}}/N$ become identical in the bosonic and fermionic cases, with $\gamma^\mathrm{LR}_{\phi/\psi}=0$ and
\begin{equation}\begin{aligned}
\gamma_{\phi/\psi}^\mathrm{SR} \supset \frac{\hat{\gamma}_{\phi/\psi,1}}{N} = \frac{2F_b'(2k)}{NF_x''(\dotwo -k)}, \quad \gamma_\sigma \supset \frac{\Gamma_{\mathrm{tet}}}{N} = -\frac{4}{N_x \Gamma(d/2)} \frac{\tilde{A}_x(\tfrac{d-s}{2})^2}{A(d-s)}
\end{aligned}\end{equation}
where, as above, $x=b$ for the $\gO(N)$ model and $x=f$ for the GN model. 
Indeed, we shall later explicitly see this correspondence for the $O(1/N^2)$ correction to $\gamma^\mathrm{SR}_{\phi/\psi}$, in \eqref{eq:GNboxKpsi2}.
Of course, $\Gamma_\mathrm{pr}$ has no analogue in the GN model, as the identity $\tr[\gamma^\mu \gamma^\nu \gamma^\rho] =0$ (which holds for general $d\neq 3$ \cite{Fraser-Taliente:2024rql}) means that there is no three-point interaction term for $\sigma$ (by regularity in dimension this holds also in $d=3$)\footnote{As an aside: for the short-range versions in $d=3$ the parallel is even closer, because in the $\gO(N)$ model $\Gamma_\mathrm{pr}\rvert_{s=2,d=3}=0$.}.
Thanks to our compact forms for $F$, we see that the free energies of these models are also similar.

 Turning to the supersymmetric model, \eqref{eq:generalFsforBoxks} shows that we can obtain both the $\gamma_i$s and $F$ simply by promoting the $2F_b'(\Delta_\sigma^\UV)/F_i''(\Delta_\phi^\UV)$ in \eqref{eq:anomDimsBoxK} to $2F_S'(\Delta_\Sigma^\UV)/F_S''(\Delta_\Phi^\UV)$ and setting both $\Gamma_\text{pr,tet}=0$.
We could now treat, for example, the Popović model \cite{Popovic:1977cq,Fraser-Taliente:2024rql}; the computation there will be analogous, save for a now fermionic auxiliary field, and so we stop here.

\subsection{The short-range theories are the extrema of the long-range theories}

Finally, we note that the short-range theories lie at the extrema of the long-range theories.
That is, if we parametrise the long-range models by $s$, then the short-range models satisfy
\begin{equation}
  \odv{\Ft^\mathrm{LR}(s)}{s}\Big\rvert_{s=s_\mathrm{SR}} = 0,
\end{equation}
which determines
\begin{equation}
  s_\mathrm{SR} = 2k - 2\hat{\gamma}
\end{equation}
to order $1/N^2$ -- i.e. including the unwieldy expressions $\hat{\gamma}_{\phi/\psi,2}$ (which are given in \eqref{eq:gammahatphi2} and \eqref{eq:GNboxKpsi2}).
This provides a principle for picking out the short-range model; $\Ft^\mathrm{LR}(s)$ also appears to be a minimal encoding of the $\hat{\gamma}$s to this order, in the sense that the extrema of a relatively simple function encode the complex $\hat{\gamma}$s.
Calculating the second derivative in each case, we find
\begin{align}
  \odv[order=2]{\Ft^\mathrm{LR}_{\gO(N)}(s)}{s}\Big\rvert_{s=s_\mathrm{SR}} &= \frac{N}{4} \Ft_b''(\dotwo -k) + O(N^0)\\
  \odv[order=2]{\Ft^\mathrm{LR}_{\mathrm{GN}}(s)}{s}\Big\rvert_{s=s_\mathrm{SR}} &= \frac{N}{4} \Ft_f''(\dotwo -k) + O(N^0).
\end{align}
Therefore: in the bosonic theory in $d>2k$, we find a maximum for odd $k$ and a minimum for even $k$; in the fermionic theory, we find a maximum for $k=1/2,5/2,\ldots$, and a minimum for $k=3/2,7/2,\ldots$.

\section{Model setup and conformal data} \label{sec:ONsetup}
\newcommand{\phiphi}{\rho}

In this section, we construct the regulated action of the $\gO(N)$ CFT.
A regularization is required for the theory to be well-defined; we choose to use an analytic regularization, and give the values that the necessary counterterms take at the critical point (the CFT). 
We also provide the two-point function normalizations, which are useful for identifying the correspondence between the long-range and short-range theories.

\subsection{Conformal propagators and their inverses} \label{app:inverseProps}

Consider the unit-normalized conformal propagators of fermions and bosons in flat space,
\begin{equation}
    G_{\phi}(x)\equiv \frac{1}{\abs{x}^{2\Delta_\phi}}, \quad G_{\psi}(x) \equiv \frac{\slashed{x}}{\abs{x}^{2\Delta_\psi+1}}.
\end{equation}
Their inverses are straightforward: they are proportional to the shadow propagators, which are the conformal propagators for operators of scaling dimension $d-\Delta$. 
This is clear from inverting the matrix $G(x,y)\equiv G(x-y)$, which can be done in flat space by diagonalizing (via a Fourier transform), taking the reciprocal of that diagonal momentum-space correlator, and Fourier transforming back:
\begin{subequations}
\begin{align}\label{eq:inverseOfG}
G_\phi^{-1}(x) = \frac{1}{\cN_b(\Delta_\phi)}\frac{1}{x^{2(d-\Delta_\phi)}}, \quad G_\psi^{-1}(x) = \frac{1}{\cN_f(\Delta_\psi)}\frac{\slashed{x}}{x^{2(d-\Delta_\psi)+1}},
\end{align}
where for bosons and fermions
\begin{align}
\cN_\phi& \equiv \cF_{2\Delta_\phi,0} \cF_{2(d-\Delta_\phi),0} =\pi^{d}A(\Delta_\phi)A(d-\Delta_\phi) \label{eq:Nphi}\\
\cN_\psi& \equiv \cF_{2\Delta_\psi,1} \cF_{2(d-\Delta_\psi),1} = -\pi^{d}A_f(\Delta_\psi)A_f(d-\Delta_\psi), \label{eq:Npsi}
\end{align}
\end{subequations}
for the usual $\cF_{2\Delta,s}$ defined in \eqref{eq:FlamsDef}.
Comparing to \cref{eq:bosonF,eq:fermionF}, it is evident that $F_{b/f}'(\Delta) =f(d)/\cN_{b/f}(\Delta)$ -- we shall prove this in general in \cref{sec:Fcomputation}.
On replacing $\abs{x-y} \to s(x,y)$, this inversion relation \eqref{eq:inverseOfG} also holds on the sphere \cite{Benedetti:2021wzt}.

\subsection{Constructing the \texorpdfstring{$\phi^4$}{phi\textasciicircum 4} action}

Consider the $\gO(N)$ $\phi^4$ action in $d$ Euclidean dimensions on the sphere $S^d$. 
Leaving all fields and parameters bare for the moment, we write the partition function and action as
\begin{equation}\begin{aligned}\label{eq:phi4action}
S_{\phi^4}= \int_x \half \phi_i C^{-1} \phi_i + \frac{\lambda_0}{8 N} (\phi_i \phi_i)^2, \quad Z_{\gO(N)} = \int \Dd{\phi} \, e^{-S_{\phi^4}},
\end{aligned}\end{equation}
where $C^{-1}$ is the bare kinetic term and the index $i=1,\ldots, N$.
Taking, for example, the short-range vector model with $k=1$, the kinetic term of the conformally coupled scalar is
\begin{equation}\label{eq:keq1Cval}
    C^{-1} =-\nabla^2+ \frac{d-2}{4(d-1)} \cR.
\end{equation}
Here $\int_x$ indicates $\int\odif[order=d]{x} \sqrt{g}$, and the Ricci scalar curvature of the sphere is $\cR = d(d-1)/R^2$.
We use the sphere's stereographic coordinates, so $\odif{s}^2=\frac{4R^2}{(1+x^2)^2} dx^\mu dx^\mu$. %
Our conventions are reviewed in \cref{app:conventions}.
For arbitrary $s$, $C^{-1}$ is the conformal Laplacian of biweight $(\Delta_\phi^\UV,d-\Delta_\phi^\UV)$ on $S^d$, which we discuss in \cref{app:sphereKineticTerms}.
This ensures that for $\lambda_0=0$, \eqref{eq:phi4action} describes $N$ generalized free fields of conformal scaling dimension $\Delta_\phi^\UV$.

There is a quadratic mass term that should be here, which is always relevant, and which must be tuned away to reach the CFT. 
However, we work in a renormalization scheme (DREG+$\delta$) where we can set the bare value of these parameters to zero.
Indeed, every relevant interaction should be added: hence for the short-range $\Box^k$ CFTs every term $\phi\Box^{m< k}\phi$ must also be tuned -- but in DREG their coefficients just must be tuned to zero.
Also, in some discrete set of dimensions, some monomials become just renormalizable; if we work in generic dimension $d$ this can be avoided, since the quantities that we calculate are regular in dimension \cite{zinn-justin_quantum_2002}.
At this order in $1/N$ it is also not necessary to consider any of the curvature couplings, unlike in the $\epsilon$-expansion \cite{Fei:2015oha,Giombi:2024zrt}.
The same comments apply in the case of the long-range models. %

\subsection{The action with the auxiliary field}

As usual, we introduce a pair of real auxiliary fields. 
The first is $\sigma$, which is integrated along an imaginary contour; the second is $\phiphi(x)$, which we set to equal $ \tfrac{1}{2\sqrt{N}}\phi_i(x)\phi_i(x)$ using a delta function implemented by $\sigma$,
\begin{equation}\begin{aligned} \label{eq:GsigDeltaFunction}
1 = \int \Dd{\phiphi} \,\delta(\phiphi- \tfrac{1}{2\sqrt{N}}\phi_i \phi_i) =  \int \Dd{\phiphi} \DdImag{\sigma} \, e^{-\int_x \sigma(\phiphi-\frac{1}{2\sqrt{N}} \phi_i \phi_i)}.
\end{aligned}\end{equation}
This particular normalization of this bare $\phiphi$ is chosen in order to ensure that eventually $\sigma$ has an order $N^0$ two-point function.
As reviewed in \cref{app:Gaussians}, it is conventional to implement the delta function by defining $\sigma$ with an integral running along the imaginary axis, rather than explicitly adding $i$s to the action: hence if the functional integral were discretized, it would be $\int\DdImag{\sigma}=\prod_x \int_{-i\infty}^{i\infty} \frac{\odif{\sigma(x)}}{2\pi i}$.
Placing this inside $Z$, 
\begin{equation}\begin{aligned}
Z_{\gO(N)} =  \int \Dd{\phi} \DdImag{\sigma}  \Dd{\phiphi}\, \exp - \left(\int_x \half \phi_i C^{-1} \phi_i -\frac{1}{2\sqrt{N}} \sigma \phi_i \phi_i + \half \lambda_0 \phiphi^2 + \sigma \phiphi \right).
\end{aligned}\end{equation}
Factoring $\thalf \lambda_0 (\phiphi+ \sigma/\lambda_0)^2$, we can exactly integrate out the quadratic $\phiphi$, and so
\begin{equation}\label{eq:ZONwithlam0}
Z_{\gO(N)} = Z_\phiphi\int \Dd{\phi} \DdImag{\sigma}\, \exp - \left(\int_x \half \phi_i C^{-1} \phi_i -\frac{1}{2\sqrt{N}} \sigma \phi_i \phi_i -\frac{\sigma^2}{2 \lambda_0}\right),
\end{equation}
where we can drop  $\log Z_\phiphi \equiv -\half\tr \log \left( \lambda_0 \delta(x-y)\right) = 0$ in DREG.
Since $\lambda_0 >0$, the potential for $\sigma$ might appear to be upside-down; however, $\sigma$ runs along an imaginary contour here, and so the $\sigma$ potential is indeed bounded below.

Because $\lambda_0$ is a relevant coupling, in the $k=1$ theory for $2<d<4$ in the IR we expect the last term to drop out. 
That is, to reach the conformal IR we send the fixed (bare) $\lambda_0 \to \infty$, obtaining an IR action (which of course must be regulated)
\begin{equation}\begin{aligned}\label{eq:ONcriticalActionNoReg}
Z_{\gO(N)} = \int \Dd{\phi} \DdImag{\sigma}\, \exp - \left(\int_x \half \phi_i \left(C^{-1} -\frac{1}{\sqrt{N}} \sigma \right)\phi_i\right).
\end{aligned}\end{equation}
Analogous comments apply for higher-$k$ theories. 

\subsection{Analytically regulating the auxiliary action}

Without a regulator, this partition function \eqref{eq:ONcriticalActionNoReg} is divergent.
If we were willing to use hard-cutoff renormalization, we could use it to work exactly at the critical point (see \cite{Gubser:2017vgc} for comments on this approach).
However, doing the required loop integrals is not feasible with a hard cutoff, and so we seek an alternative analytic regulator.
This procedure is briefly described in \cite{zinn-justin_quantum_2002}, but was originally motivated in \cite{Vasiliev:1975mq}.

Because $\sigma$ appears in only one place here, to leading order in $N$ its scaling dimension is $\Delta_\sigma = d-2\Delta_\phi^\UV$. 
Hence, the interaction is marginal in any $d$, and therefore, as described in the introduction, analytically continuing in the dimension does not work. 
However, what if we could analytically shift the UV scaling dimension of $\sigma$ to $s-\delta$, with $\delta \to 0$, while keeping that of $\phi$ at $\tfrac{d-s}{2}$.
The bare scaling dimension of $\sigma \phi_i\phi_i$ would then be $d-\delta$ to leading order, so we could renormalize as usual, cancelling the divergences at each order in $1/N$.
In analogy to the textbook perturbative quantum field theory: the controlling parameter of the formal expansion, which is usually the coupling constant $g$ is here $1/N$; the divergences encoded by $1/\epsilon \sim \log \Lambda$ poles are the $1/\delta$s. 
The analytic continuation in dimension ensures that power law divergences can be ignored (i.e. implicitly cancelled by some Lorentz-invariant local counterterm).
We use a minimal subtraction (MS) scheme where we minimally subtract the $1/\delta$ poles that appear at each order in $N$.

Recall from \eqref{eq:generatedSigmaProp} that the propagator for $\sigma$ that is generated is $K(x,y)$, where
\begin{equation}
K^{-1}(x,y) \equiv -\half \left(\frac{C_{\phi,0}}{s(x,y)^{2\Delta_\phi^\UV}}\right)^2.
\end{equation}
The shift is then done by using a different propagator for $\sigma$ in the Feynman rules, identical to $K$ except for a small shift in $\Delta_\sigma^\UV$:
\begin{equation}
    K_\delta(x,y) = \frac{C_{\sigma,0}}{s(x,y)^{2(s-\delta)}}.
\end{equation}
This can only be implemented in practice by modifying the action: we add $\frac{1}{g_0^2}\half \sigma K^{-1}_\delta \sigma$, and almost entirely subtract off the old kinetic term with a $-\frac{\mu^{2\delta}}{g_0^2}\half \sigma K^{-1} \sigma$. 
However, the cancellation cannot be made perfect.
If the cancellation were perfect, it would be possible to absorb the $Z_g$ coupling by a redefinition of $\sigma$ -- but we need  $Z_g$ to cancel the divergences.
This yields a residual additional two-point counterterm for $\sigma$ that is $\propto \frac{1}{N\delta}K^{-1}$.
Using DeWitt notation for compactness, the new action is
\begin{equation}\begin{aligned}
  S_\mathrm{temp} = \int_{x,y} \frac{Z_\phi}{2} \phi^i_x \left(C^{-1}_{xy} -\frac{\sigma_x \delta_{xy}}{\sqrt{N}}\right) \phi^i_y + \frac{1}{g_0^2}\half \int_{x,y}\sigma_x (K^{-1}_{\delta,xy} - \mu^{2\delta} K^{-1}_{xy})\sigma_y.
\end{aligned}\end{equation}
As described in the introduction, in standard QFT counterterms are a modification of physics at the microscopic scale.
Evidently, in the limit $g_0 \to \infty$ we recover \eqref{eq:ONcriticalActionNoReg}, the interacting CFT; thus the addition of these terms, an order $K_\delta^{-1}-\mu^{2\delta} K^{-1}=O(\delta)$ modification of the original Lagrangian, is also a modification of UV physics.

Rescaling $\sigma_\text{new} = \sigma_\text{old}/g_0$ to canonically normalize $\sigma$, the DREG+$\delta$ regulated action which henceforth will be used for all computations is\footnote{More conventionally, we would write $S_{\gO(N),\delta} = S_{\phi} + S_{\mathrm{int}}+ S_{\sigma,\delta}$ for
\begin{equation}\begin{aligned}
S_\phi &= \int_{x,y} \tfrac{Z_\phi}{2} \phi_i(x) C^{-1}(x,y) \phi_i(y),\quad S_\mathrm{int} = - \int_x \tfrac{Z_\phi Z_g g \mu^\delta}{2\sqrt{N}} \sigma(x) \phi_i(x) \phi_i(x),\\
S_{\sigma,\delta} &= \int_{x,y}\thalf\sigma(x) (K_\delta^{-1}(x,y) - \mu^{2\delta} K^{-1}(x,y)) \sigma(y).
\end{aligned}\end{equation}}
\begin{equation}\begin{aligned}\label{eq:completeAction}
\boxed{ S_{\gO(N),\delta} \equiv \int_{x,y} \frac{Z_\phi}{2} \phi^i_x \left(C^{-1}_{xy} -\frac{Z_g g \mu^{\delta}}{\sqrt{N}} \sigma_x \delta_{xy}\right) \phi^i_y + \half \int_{x,y}\sigma_x (K^{-1}_{\delta,xy} - \mu^{2\delta} K^{-1}_{xy})\sigma_y. }
\end{aligned}\end{equation}
We make the following comments:
\begin{itemize}
\item We treat $C^{-1}$ and $K_\delta^{-1}$ as the bare kinetic terms; the other terms are a three-point and two-point interaction respectively. 
The latter is slightly subtle, since insertions of $K^{-1}$ are not suppressed by factors of $1/N$, but instead cancel with $\phi$ loops as we describe in \cref{app:non2PI}.
We avoid this subtlety: in our calculation of the counterterms, by using the 2PI formalism; when calculating $F$, by integrating out $\phi$ entirely and working with the $\sigma$ theory.
\item $K_\delta$ here implements the scaling dimension shift by $\delta$, and $K=K_{\delta=0}$.
\item $C^{-1}_{xy}$ and $K^{-1}_{xy}$ are the inverses of $C_{xy}$ and $K_{xy}$ treated as matrices, and are given explicitly in \eqref{eq:Sigma0def}.
\item We have written $g_0 = Z_g g \mu^\delta$, but set the dimensionless $g=1$ henceforth. 
The counterterms $Z_\phi$ and $Z_g$ have a perturbative expansion in $1/N$ around $1$, given in \eqref{eq:SRcts} and \eqref{eq:LRcts} for the short- and long-range cases. 
We choose to use MS scheme, so each order in $N$ contains only poles in $\delta$.
\item We could instead non-minimally subtract; however, we are happy for the moment to deal with conformal fields $\phi$ and $\sigma$ that do not have unit-normalized propagators, and therefore only comment on this in \cref{app:nonminimal}.
\item $Z_\phi \neq 1$ is only required in the short-range models, as non-local terms are not renormalized. 
This is because Wilsonian renormalization generates only correction terms that are polynomial in momenta -- that is, local derivative couplings in position space, and hence a non-local term (specifically, bilocal) is unrenormalized \cite{Gubser:2017vgc}. 
\item For the same reason, $K_\delta^{-1}$ and $K^{-1}$ do not need counterterms in generic $d$.
\item Taking the limit $\delta \to 0$ corresponds to removing the regulator. 
\end{itemize}
Our desired partition function is then
\begin{equation}\begin{aligned}
Z_{\gO(N)} \equiv e^{-F_{\gO(N)}} \equiv \lim_{\delta\to 0}\int \Dd{\phi}\DdImag{\sigma} \, e^{-S_{\gO(N),\delta}}.
\end{aligned}\end{equation}
Since we want to work exactly at the critical point, we must tune $Z_\phi$ and $Z_g$ as functions of $\delta$.  
If we wanted to work away from the critical point, we would need to introduce coefficients in front of both $K$ terms \cite{Ciuchini:1999wy} -- but we do not want to.
The critical values of our counterterms must be determined by a standard RG analysis; though standard, we perform it in \cref{app:flatSpace2pt}, as inconsistent regulators have been used in the literature.
We take care to use the same shift by $\delta$ in all diagrams; we also choose to use the 2PI formalism, meaning that we need to consider fewer diagrams.

\subsection{General \texorpdfstring{$s$}{s} conformal data}\label{sec:genSdata}
The counterterms at the infrared critical point for generic $s$ are
\begin{equation}\begin{aligned}\label{eq:LRcts}
Z_\phi = 1, \quad Z_g = 1 + \frac{Z_{g,1}}{N \delta} + \gO(1/N^2), \quad Z_{g,1} \equiv - \frac{1}{2}\left(\Gamma_{\mathrm{tet}} + \half\Gamma_{\mathrm{pr}}\right).
\end{aligned}\end{equation}
With these, we can safely take the limit $\delta \to 0$.
Then, to the $N$-orders shown, the two-point functions are
\begin{equation}\begin{aligned}\label{eq:ONLRSDs}
\expval{\sigma(x) \sigma(y)} &= \frac{C_\sigma}{s(x,y)^{2\Delta_\sigma}}, \quad \Delta_\sigma= s+\gamma_{\sigma}, \quad \gamma_{\sigma} = \frac{\Gamma_{\mathrm{tet}} + \Gamma_{\mathrm{pr}}}{N}, \quad C_\sigma = C_{\sigma,0} \left(1+\frac{\normCor_{\sigma,1}}{N}\right),\\
\expval{\phi_i(x) \phi_j(y)} &= \frac{C_\phi \delta_{ij}}{s(x,y)^{2\Delta_\phi}}, \quad \Delta_\phi = \frac{d-s}{2}, \quad C_\phi = C_{\phi,0} \left(1+\frac{\normCor_{\phi,1}}{N} +\frac{\normCor_{\phi,2}}{N^2}\right),
\end{aligned}\end{equation}
where in the long-range model the scaling dimension of $\Delta_\phi$ is exact to all orders in $N$. 
The lack of anomalous dimension here follows from the non-renormalization of the non-local kinetic term mentioned above, and was proven rigorously by \cite{Lohmann:2017qyq}.

With $A(x) \equiv \Gamma(\tfrac{d}{2}-x)/\Gamma(x)$ as above, the leading contributions to $\gamma_\sigma$ are
\begin{equation}\begin{aligned}
\Gamma_{\mathrm{tet}} \equiv  -\frac{4}{\Gamma \left(\frac{d}{2}\right)} \frac{A\left(\frac{d-s}{2}\right)^2}{A(d-s)}, \quad \Gamma_{\mathrm{pr}} \equiv \frac{8}{\Gamma \left(\frac{d}{2}\right)} A(s) A\left(d-\tfrac{3 s}{2}\right)  \frac{A\left(\frac{d-s}{2}\right)^3}{A(d-s)^2}.
\end{aligned}\end{equation}
The leading normalizations are
\begin{equation}\begin{aligned}\label{eq:BareONtwoPointNorms}
C_{\phi,0} = \frac{\pi ^{-\frac{d}{2}} 2^{-s}}{A\left(\frac{d-s}{2}\right)}, \quad C_{\sigma,0} = \frac{1}{(C_{\phi,0})^2}\frac{-2}{\cN_b(s)} = -2 \frac{4^s A\left(\frac{d-s}{2}\right)^2}{A(d-s)A(s)}.
\end{aligned}\end{equation}
The subleading normalization for $\sigma$ is 
\begin{equation}\begin{aligned}
\normCor_{\sigma,1} &= \normCor_{\sigma,1,\mathrm{tet}} +\normCor_{\sigma,1,\mathrm{pr}} +\normCor_{\sigma,1,\mathrm{pil}}\\
\normCor_{\sigma,1,\mathrm{tet}} &= \Gamma_{\mathrm{tet}} \left(B\left(\tfrac{d-s}{2}\right)-B(d-s)\right)\\
\normCor_{\sigma,1,\mathrm{pr}} &= \Gamma_{\mathrm{pr}} \left(\tfrac{3}{2} B\left(\tfrac{d-s}{2}\right)+\tfrac{1}{2} B\left(d-\tfrac{3 s}{2}\right)-B(d-s)-B(s)\right)\\
\normCor_{\sigma,1,\mathrm{pil}} &= \frac{4F_b'(s)}{F_b'(\tfrac{d-s}{2})} =\frac{4\cN_b(\tfrac{d-s}{2})}{\cN_b(s)} =\frac{4A\left(\frac{d-s}{2}\right) A\left(d-\frac{d-s}{2}\right)}{A(s) A(d-s)},
\end{aligned}\end{equation}
where we have defined $B(x) \equiv -\odv{}{x}\log A(x)$.
For $\phi$ we can go one order further,
\begin{equation}\begin{aligned}
\normCor_{\phi,1} &= -\half \normCor_{\sigma,1,\mathrm{pil}}\\
\normCor_{\phi,2} &= %
\frac{-F_b'(s)}{F_b'(\tfrac{d-s}{2})} \left[\left(\Gamma _{\text{pr}}+\Gamma _{\text{tet}}\right)\left(B\left(\tfrac{d-s}{2}\right)-2 B(d-s)\right) + \Gamma _{\text{tet}}B(s)+\Gamma _{\text{pr}} B\left(d-\tfrac{3 s}{2}\right)\right].
\end{aligned}\end{equation}

\subsection{The short-range models}\label{sec:SRmodels}

It is clear that if we attempt to take the bare scaling dimension $s \to 2k$, all of $\normCor_{\sigma,1,\mathrm{pil}}$, $\normCor_{\phi,1}$, and $\normCor_{\phi,2}$ diverge because 
\begin{equation}
\lim_{s\to 2k} \frac{1}{F_b'(\tfrac{d-s}{2})} \sim \Gamma(-k)\Gamma(k)= \infty
\end{equation}
for all integer $k$.
This comes from the fact that the limits $\delta \to 0$ and $s \to 2k$ do not commute; we cannot take $\delta \to 0$ for arbitrary $s$ and then take $s\to 2k$. 
Thus to directly reach the short-range models, $s=2k$ must be fixed first, and only then may we take the regulator to zero.

To keep the short-range model's two-point functions finite as $\delta \to 0$, we are forced to modify $Z_\phi$,
\begin{equation}\begin{aligned}\label{eq:SRcts}
Z_\phi &=1-\frac{\hat{\gamma}_{\phi,1}}{N \delta} + O(1/N^2)\\
Z_g &= 1-\frac{1}{2N \delta}\left(\Gamma_\mathrm{tet} + \half \Gamma_\mathrm{pr}-2 \hat{\gamma}_{\phi,1}\right)_{s=2k} + O(1/N^2).
\end{aligned}\end{equation}
Note that the product $Z_\phi Z_g = 1-\frac{1}{2N \delta}\left(\Gamma_\mathrm{tet} + \half \Gamma_\mathrm{pr}\right)_{s=2k}$ is identical to the general-$s$ case; that is, we only need to modify the coefficient of the kinetic term of $\phi$ in the action.

Then the IR scaling dimensions are
\begin{equation}\begin{aligned}
\Delta_\phi &= \frac{d}{2}-k + \frac{\hat{\gamma}_{\phi,1}}{N} + \frac{\hat{\gamma}_{\phi,2}}{N^2} + O(1/N^3),\\
 \Delta_\sigma &= 2k + \frac{1}{N}\left(\Gamma_{\mathrm{tet}} + \Gamma_{\mathrm{pr}}-2\hat{\gamma}_{\phi,1}\right)_{s=2k} + O(1/N^2),
\end{aligned}\end{equation}
where, as usual, $\hat{\gamma}_{\phi,1} = \frac{2F_b'(2k)}{F_b''(\dotwo -k)}$, and the $O(1/N^2)$ term is
\begin{equation}\begin{aligned}\label{eq:gammahatphi2}
    \hat{\gamma}_{\phi,2} = & \frac{\hat{\gamma}_{\phi,1}}{2}  \left[ (\Gamma _{\text{tet}} + \Gamma_\mathrm{pr})\left(B(\dotwo -k)-2B(d -2 k)\right) + \Gamma _{\text{tet}} B(2 k) + \Gamma _{\text{pr}} B(d -3 k)\right]\\
    &-\hat{\gamma}_{\phi,1}^2 \underbrace{\Bigg(\frac{F_b'''(\dotwo -k)}{2 F_b''(\dotwo -k)}+\frac{2 F_b''(2 k)}{F_b'(2 k)}\Bigg)}_{\mathclap{\left(\psi ^{(0)}\left(\dotwo -k\right)-\psi ^{(0)}\left(\dotwo + k\right)-\tfrac{1}{k}\right)+2 (B(2 k)-B(d-2 k))}},\end{aligned}\end{equation}
The remaining normalization data is given in \cref{sec:SRdiags}.
Clearly the first line here is $ \lim_{\delta \to 0} (\delta\,\normCor_{\phi,2}\rvert_{s=2k-\delta})$, and the second line is identical to that which follows from $\Ft$-extremization (compare the supersymmetric result \eqref{eq:SUSYanomDims})\footnote{\cite{Vasiliev:1981dg} described this term as the contribution of the Hartree-Fock diagrams.}. 
As discussed in the introduction, $\Delta_\sigma$ here indeed coincides at this order with the scaling dimensions of the long-range model \eqref{eq:ONLRSDs} evaluated for $s_\mathrm{SR} = 2-2\hat{\gamma}_\phi$.
Further, this value of $\hat{\gamma}_{\phi}$, including the $O(1/N^2)$ correction, does indeed extremize $\Ft^{\mathrm{LR}}_{\gO(N)}$.

For the Gross-Neveu model, following the recipe described in \cref{sec:synchronicity}, we map $\Gamma_\mathrm{pr} \to 0$, $\Gamma_\mathrm{tet} \to \Gamma_{\mathrm{tet}}^\mathrm{GN}$, and $F_b(\dotwo-k) \mapsto \thalf F_f(\dotwo-k)$. 
Then, we map $B(\dotwo-k)\mapsto B_f(\dotwo -k)$, where $B_f(x) \equiv -\odv{}{x}\log A_f(x)$, and all other $B$s stay the same.
This yields the correct result \cite{Gracey:1993kc}
\begin{equation}\begin{aligned}\label{eq:GNboxKpsi2}
    \hat{\gamma}_{\psi,2} &= \frac{\hat{\gamma}_{\psi,1}}{2}  \Gamma _{\text{tet}}^\mathrm{GN} \left(B(2k) + B_f(\dotwo -k)-2B(d -2 k)\right) 
    \\
    &\quad -\hat{\gamma}_{\psi,1}^2 \Bigg(\frac{F_f'''(\dotwo -k)}{2 F_f''(\dotwo -k)}+\frac{2 F_b''(2 k)}{F_b'(2 k)}\Bigg).
\end{aligned} \end{equation}

\section{Free energy calculation}\label{sec:calculatingF}

We have correctly regulated this partition function; let us now integrate it.
The computation for the Gross-Neveu model exactly parallels this computation, so we only provide some comments in \cref{app:GNFdetails}.

\subsection{Integrating out \texorpdfstring{$\phi$}{phi}}

Performing the Gaussian integral \eqref{eq:GaussianIntegral} of $\phi_i$ in \eqref{eq:completeAction}, we find a 
\begin{equation}
    \half \Tr_{x,i} \log\left(\delta_{ij} C^{-1}(x,y) - \frac{g_0}{\sqrt{N}} \delta_{ij} \sigma(x) \delta(x-y)\right) = \frac{N}{2} \tr \log\left(C^{-1} - \frac{g_0}{\sqrt{N}} \sigma \mathbb{I} \right)
\end{equation}
appearing in the effective action -- the $\Tr_{x,i}$ here is over space and the $\gO(N)$ index, whereas all other $\tr$s are only over space. 
If we now define
\begin{equation}
Z_{\phi,\text{free}}^N \equiv \exp\left(-\frac{N}{2} \tr \log C^{-1}\right)= \exp\left(- N F_b (\tfrac{d-s}{2})\right),
\end{equation}
which we discuss in the next section,
the $\tr \log C^{-1}$ can be factored out
\begin{subequations}
\begin{equation}\begin{aligned}
\frac{Z_{\gO(N),\delta}}{Z_{\phi,\text{free}}^N} &= \int \DdImag{\sigma} \exp-\left(\frac{N}{2} \tr \log\left[\mathbb{I} - \frac{g_0}{\sqrt{N}} M\right] + \half \int_{x,y}\sigma_x (K^{-1}_{\delta,xy} - \mu^{2\delta}K^{-1}_{xy})\sigma_y\right)\\
&= \int \DdImag{\sigma} \exp-\left(\sum_{q=1}^\infty -\frac{g_0^q}{2q} \frac{\tr M^q}{N^{q/2-1}} + \half \int_{x,y}\sigma_x (K^{-1}_{\delta,xy} - \mu^{2\delta}K^{-1}_{xy})\sigma_y\right).
\end{aligned}\end{equation}
We again use the matrix $M(x,y) = \sigma(x) C(x,y)$ to simplify notation.
Note that the value of the counterterm $Z_\phi$ does not impact the universal part of the free energy, as it disappeared when $\phi$ was integrated out.

The $q=1$ term $\int_x \sigma(x) C(x,x)$ ruins conformality if it is non-zero \cite{zinn-justin_quantum_2002}. 
However, by assumption, we are working at the critical point, and therefore must tune the relevant interaction $+\thalf m_0^2 \phi_i \phi_i$. 
Adding such an interaction to the action \eqref{eq:completeAction} produces a $\sigma$ tadpole term that we can use to cancel the $q=1$ term.
Happily, in DREG this is unnecessary as $C(x,x)=0$ \cite{Benedetti:2021wzt}, and so the $\sigma$ tadpole automatically vanishes: $\tr M=0$, and so we can just set $m_0=0$.

Recalling that $K^{-1}(x,y) = -\half C(x,y)^2$, we see that $\exp-(\tfrac{g_0^2}{2} \int \sigma K^{-1} \sigma)$ is the first non-zero term coming from the sum.
This only partially cancels the additional $K^{-1}$ from the regulator, since the critical $g_0^2 = \mu^{2\delta}(1 + \frac{2Z_{g,1}}{N\delta})$. 
Thus
\begin{equation}
    = \int \DdImag{\sigma} \exp-\left(\half \int_{x,y}\sigma_x (K^{-1}_{\delta,xy} -(\mu^{2\delta}-g_0^2)K^{-1}_{xy})\sigma_y  -\frac{g_0^3}{6\sqrt{N}}\tr M^3 -\frac{g_0^4}{8N} \tr M^4 + \cdots \right)
\end{equation}
\end{subequations}
Dropping the subleading terms in $N$, we find a second quadratic integral, and so
\begin{equation}\label{eq:leadingFterms}
F_{\gO(N)} = NF_b(\tfrac{d-s}{2})  + F_b(s- \delta) + O(1/N),
\end{equation}
where we have used $\int \DdImag{\sigma} \, e^{-\half \int \sigma K_\delta^{-1} \sigma} = \half \tr \log (-K_\delta^{-1}) = F_b(s-\delta)$ which, as discussed in \cref{app:imagGaussian}, holds when evaluated in DREG.
We now calculate $\half \tr \log C^{-1}$ in full generality.

\subsection{The free energy of a free conformal field of dimension \texorpdfstring{$\Delta$}{delta}}\label{sec:Fcomputation}

We now give a straightforward computation of (the universal part of) the sphere free energy of a free bosonic/fermionic conformal field $\Phi$ of dimension $\Delta_\Phi$ in an arbitrary representation. 
This is a standard result from the AdS/CFT literature for any Lorentz representation of $\Phi$ \cite{Sun:2020ame,Benedetti:2021wzt,Harribey:2022esw,Fraser-Taliente:2024hzv,Fraser-Taliente:2024lea}; %
we provide here a very simple alternative derivation, which demonstrates the relationship of $F$ to the inversion $\cN$s defined in \eqref{eq:inverseOfG}.
We rely on the trivial result that for any matrix $M(\Delta)$
\begin{equation}\label{eq:trlogMdiff}
\odv{}{\Delta}\tr \log M(\Delta) = \tr M^{-1}(\Delta) \odv{}{\Delta} M(\Delta) =\odv{}{\alpha} \tr M^{-1}(\Delta) M(\alpha) \Big\rvert_{\alpha=\Delta}
\end{equation}
which we can use if working in the range of $d$ where everything converges.

Without loss of generality, we can consider a real conformal generalized free field $\Phi$ of arbitrary dimension $\Delta_\Phi$ and representation $\rho$, which has a propagator $G_\Phi$ (suppressing all indices, and ignoring the presence of ghosts, which may be required for some representations). 
Then it has a kinetic term $G_\Phi^{-1}$, which is written down properly, including contact terms, for scalars in \cref{app:sphereKineticTerms}.
Performing the functional integral,
\begin{equation}\begin{aligned}
F_\Phi &\equiv - \log \int \DdImag{\Phi}\, e^{(-1)^{\mathrm{F}^\Phi}\half \int_{x,y} \Phi_x G_{\Phi,xy}^{-1} \Phi_y} \\
&= \half \str \log G_\Phi^{-1}  = (-1)^{\mathrm{F}^\Phi}\half \tr \log G_\Phi^{-1},
\end{aligned}\end{equation}
where $(-1)^{\mathrm{F}^\Phi}$ just gives a minus sign for fermionic reps. 
The trace here is over space, Lorentz indices, and symmetry group indices. 
The normalization of this propagator does not contribute to the universal part, since $\tr \delta(x-y) =0$ in DREG -- so we can take $G_\Phi$ to be unit-normalized.

First, let us elaborate on the propagator inverse relation \eqref{eq:inverseOfG}. 
At separated points, it is generally true that the inverse of a conformal propagator is proportional to the \textit{shadow propagator}, which is the propagator of the field $\tilde{\Phi}$ with scaling dimension $\tilde{\Delta}_\Phi \equiv d-\Delta_\Phi$, transforming in the \textit{reflected representation} $\rho^{\mathrm{ref}}$. 
This is because the inverse propagator and the shadow field's propagator transform identically under conformal symmetry. 
Since both are unique, we must have%
\begin{equation}\begin{aligned}
G_\Phi^{-1}(x,y) &=\frac{1}{\cN_{\Phi}} G_{\tilde\Phi}(x,y), %
\label{eq:inverseIsShadow}
\end{aligned}\end{equation} 
for some $\cN_\Phi(\Delta_\Phi,\rho)$, up to contact terms that vanish for $x\neq y$ (given in, e.g., \eqref{eq:DContactTerms}).
Because the $G$s are unit-normalized, $\cN_\Phi$ must be a purely representation-theoretical quantity\footnote{For a general representation, $\cN_\Phi(\Delta_\Phi, \rho)$ can also be found using \cite{Karateev:2018oml} %
\begin{equation}\begin{aligned}\label{eq:plancherelDef}
\frac{\dim_{\mathbb{R}}(\rho)}{\cN_\Phi} = (-1)^{\mathrm{F}^\Phi} \Omega_d \, \mu(\Phi),
\end{aligned}\end{equation} 
where $\Omega_d=2^d \vol\SO(d)$ is a constant that drops out of all computations,
and $\mu(\Phi)$ is the Plancherel measure of $\SO(d+1,1)$ for $\Phi$'s conformal representation $(-\Delta_\Phi,\rho)$.}, identical for flat space and the sphere, which can be calculated explicitly by taking the inverse in momentum space (we gave the values for bosons and fermions in \eqref{eq:Nphi} and \eqref{eq:Npsi}).

\subsubsection{Computing the trace}
Differentiating and using \eqref{eq:trlogMdiff} we find %
\begin{align}
	F_\Phi' = \odv{F_\Phi}{\Delta_\Phi} &= -(-1)^{\mathrm{F}^\Phi}\half \tr G_\Phi^{-1} \odv{G_\Phi}{\Delta_\Phi} = -(-1)^{\mathrm{F}^\Phi}\half\odv{}{\alpha}\left(\tr G_\Phi^{-1} G_{\Phi,\Delta_\Phi=\alpha}\right)\Big\rvert_{\alpha=\Delta_\Phi},
\end{align}
where we clearly have repeatedly swapped the order of derivatives and integrals, assuming everything to converge.
As mentioned above, we therefore must implicitly work in the range of $d$ where this trace converges as the UV cutoff is taken to infinity, and then analytically continue back up: this is just DREG.
Explicitly writing out the position space part of the trace,
\begin{align}
	F_\Phi' &= -\half \frac{1}{\cN_\Phi} \odv{}{\alpha} \int_{x,y} \tr_\rho \Big[G_\Phi^{-1}(x,y) \underbrace{(-1)^{\mathrm{F}^\Phi}G_{\Phi,\alpha}(y,x)}_{G_{\Phi,\alpha}(x,y)}\Big]\Big\rvert_{\alpha=\Delta_\Phi}\\
  &= -\half \frac{\tr_\rho(\mathbb{I}_\rho)}{\cN_\Phi} \odv{}{\alpha} \int_{x,y} \frac{1}{s(x,y)^{2(d-\Delta_\Phi+\alpha)}}\Big\rvert_{\alpha=\Delta_\Phi},
\end{align}
where we have used the known behaviour of conformal propagators under $x \leftrightarrow y$, and we can drop the contributions of contact terms due to $G(x,x)=0$ and $\tr \delta(x-y)=0$.
By definition of the inverse, we do not need to consider the details of the suppressed Lorentz/symmetry indices, being guaranteed to end up with only the identity $\mathbb{I}_\rho$ in the representation $\rho$. 
The trace of $\mathbb{I}_\rho$ is simply the (real, by assumption) dimension of $\rho$, so we are left with the standard sphere integral
\begin{equation}
    I_2(\Delta) \stackrel{\eqref{eq:CardyI2}}{\equiv} \int_{x,y} \frac{1}{s(x,y)^{2\Delta}} = \frac{\pi^d}{(2R)^{2 (\Delta-d)}} \frac{\Gamma(\dotwo)}{\Gamma(d)} \frac{\Gamma \left(\frac{d}{2}-\Delta \right)}{\Gamma (d-\Delta )},
\end{equation}
i.e.
\begin{equation}\begin{aligned}
F_\Phi' &= -\half \frac{\dim_\mathbb{R}(\rho)}{\cN_\Phi} \odv{}{\alpha} I_2(d-\Delta_\Phi+\alpha)\Big\rvert_{\alpha=\Delta_\Phi}\\
&= -\half \frac{\dim_\mathbb{R}(\rho)}{\cN_\Phi} I_2'(d).
\end{aligned}\end{equation}
We now assume that $F_\Phi(\Delta_\Phi=\tfrac{d}{2})=0$, which follows immediately from the identity in DREG $\tr \log G_\Phi^{-1} = - \tr \log G_\Phi$.
This yields
 \begin{equation}
F_\Phi(\Delta_\Phi,\rho) = \frac{-1}{\sin(\tfrac{\pi d}{2})}\frac{\pi^{d+1}}{\Gamma(d+1)} \int_{\frac{d}{2}}^{\Delta_\Phi} \odif{\Delta'} \frac{\dim_\mathbb{R}(\rho)}{\cN_{\Phi}(\Delta', \rho)}
 \end{equation}
for any conformal GFF.

\subsection{The free energy corrections}

The corrections here are, as usual, minus the sum of the connected vacuum diagrams.
The $N$-scaling that we have chosen for $\sigma$ above makes it easy to see what diagrams contribute.
Ignoring self-loops, which vanish, they are
\begin{equation}\label{eq:ONFcorrections}
-\delta F_{\gO(N),1/N} \equiv
\begin{tikzpicture}[scale=1.0, line cap=round, line join=round,baseline={(0,-\MathAxis pt)}]
\begin{scope}[xshift=-2.5cm]
    \coordinate (C) at (0,0); %
    \draw[sigma] (C) circle (0.5);
    
    \coordinate (X) at (0,0.5);
    \draw[thick] ($(X)+(-0.15,-0.15)$) -- ($(X)+(0.15,0.15)$);
    \draw[thick] ($(X)+(-0.15,0.15)$) -- ($(X)+(0.15,-0.15)$);
    
    \node[below] at (0,-0.75) {$I_{\mathrm{ct}}: \half$};
    \node[right] at (0.85,0) {$+$};
\end{scope}
\begin{scope}
\coordinate (TL) at (-0.5, 0.5); %
\coordinate (BL) at (-0.5,-0.5); %
\coordinate (TR) at ( 0.5, 0.5); %
\coordinate (BR) at ( 0.5,-0.5); %
\draw[thick] (TL) -- (TR) -- (BR) -- (BL) -- cycle;
\draw[sigma]
    (TL) .. controls ($(TL)+(-0.3,0)$) and ($(BL)+(-0.3,0)$) .. (BL);
\draw[sigma] (TR) .. controls ($(TR)+(0.3,0)$) and ($(BR)+(0.3,0)$) .. (BR);
\node[below] at (0,-0.75) {$I_{\mathrm{pil}}: \frac{1}{4} $};
\node[right] at (1.05,0) {$+$};
\end{scope}

\begin{scope}[xshift=2.5cm]
    \coordinate (TL) at (-0.5, 0.5); %
    \coordinate (BL) at (-0.5,-0.5); %
    \coordinate (TR) at ( 0.5, 0.5); %
    \coordinate (BR) at ( 0.5,-0.5); %
  \draw[thick] (TL) -- (TR) -- (BR) -- (BL) -- cycle;
  \draw[sigma] (BL) -- (TR);
  \draw[sigma] (TL) -- (BR);
  \node[below] at (0,-0.75) {$I_{\mathrm{tet}}: \frac{1}{8}$};
  \node[right] at (0.7,0) {$+$};
\end{scope}

\begin{scope}[xshift=5cm]
  \coordinate (A1) at (-0.5,0);
  \coordinate (A2) at (-1,0.5);
  \coordinate (A3) at (-1,-0.5);

  \coordinate (B1) at (0.5,0);
  \coordinate (B2) at (1,0.5);
  \coordinate (B3) at (1,-0.5);

  \draw[thick] (A1) -- (A2) -- (A3) -- (A1);
  \draw[thick] (B1) -- (B2) -- (B3) -- (B1);
  \draw[sigma] (A1) -- (B1);
  \draw[sigma] (A2) -- (B2);
  \draw[sigma] (A3) -- (B3);

  \node[below] at (0,-0.75) {$I_{\mathrm{pr}}: \frac{1}{12}$};
\end{scope}
\end{tikzpicture}\; ,
\end{equation}
which we term the counterterm diagram, the pillow diagram, the tetrahedron diagram, and the prism diagram respectively.
The cross indicates an insertion of $-\mu^{2\delta} \frac{2Z_{g,1}}{N\delta}K^{-1}$, thus making the diagram order $1/N$. 
We would have also considered the diagram \begin{tikzpicture}[scale=0.5,baseline={(0,-0.1)}]
  \coordinate (A1) at (-0.5,0);
  \coordinate (A2) at (-1,0.5);
  \coordinate (A3) at (-1,-0.5);
  \coordinate (B1) at (0.5,0);
  \coordinate (B2) at (1,0.5);
  \coordinate (B3) at (1,-0.5);
  \draw[phi] (A1) -- (A2) -- (A3) -- (A1);
  \draw[phi] (B1) -- (B2) -- (B3) -- (B1);
  \draw[sigma] (A1) -- (B1);
    \draw[sigma] (A2) .. controls ($(A2)+(-0.3,0)$) and ($(A3)+(-0.3,0)$) .. (A3);
    \draw[sigma] (B2) .. controls ($(B2)+(0.3,0)$) and ($(B3)+(0.3,0)$) .. (B3);
\end{tikzpicture}, with symmetry factor $1/8$, but it is automatically zero even for finite $\delta$ (and even with additional infinitesimal shifts of the propagators), something which can be checked using its Mellin-Barnes representation.

\subsubsection{The non-counterterm contributions}

The diagrams, evaluated with general conformal propagators, are
\begin{subequations}\label{eq:Idefs}
\begin{align}
I_{\mathrm{pil}} &= \frac{1}{N} \frac{1}{4}\int_{x_{1,2,3,4}} \frac{C_\sigma^2}{(s_{12} s_{34})^{2\Delta_\sigma}} \frac{C_\phi^4}{(s_{12} s_{23} s_{34} s_{41})^{2\Delta_\phi}}\\
I_{\mathrm{tet}} &= \frac{1}{N} \frac{1}{8} \int_{x_{1,2,3,4}} \frac{C_\sigma^2}{(s_{13} s_{24})^{2\Delta_\sigma}} \frac{C_\phi^4}{(s_{12} s_{23} s_{34} s_{41})^{2\Delta_\phi}}\\
I_{\mathrm{pr}} &= \frac{1}{N}\frac{1}{12} \int_{x_{1 \cdots 6}} \frac{C_\phi^3}{(s_{12} s_{23} s_{31})^{2\Delta_\phi}} \frac{C_\sigma^3}{(s_{14} s_{25} s_{36})^{2\Delta_\sigma}} \frac{C_\phi^3}{(s_{45} s_{56} s_{64})^{2\Delta_\phi}},
\end{align}
\end{subequations}
where $s_{ij} = s(x_i, x_j)$ is the sphere chordal distance; we leave the coefficients and scaling dimensions $C_i$ and $\Delta_i$ general for the moment.
These diagrams are computed with \texttt{MBresolve}, following exactly the procedure described in \cite{Fei:2015oha}, which we comment on in \cref{app:sphereInts}.
In particular, we take advantage of the sphere's symmetries to fix one of the integrated points. This simplifies the computation, and also, somewhat surprisingly, means that no further regulators (analogous to the $\eta$ shifts in \cref{app:flatSpace2pt}) are required.

The shift of $\Delta_\sigma^\UV$ to $s-\delta$ ensures that $\Delta_\sigma + 2 \Delta_\phi = d-\delta \neq d$, and also means that the Mellin-Barnes representations of the diagrams are finite.
Defining a small parameter $\varkappa$,
\begin{equation}
    \varkappa \equiv d -\Delta_\sigma - 2 \Delta_\phi \ll 1,
\end{equation}
that then allows a compact presentation of the results for arbitrary $\Delta_\phi$.
Performing the Mellin-Barnes integrals, and expanding for small $\varkappa$ to linear order, we find expressions of the form $\prod_i \Gamma(\alpha_i)^{p_i} \times (1+ \sum_i c_i p_i \, x \, \psi^{(0)}(\alpha_i))$ for some $p_i$s and $c_i$s. 
Since in general
\begin{equation}\label{eq:polygamma}
    \prod_i \Gamma(\alpha_i)^{p_i}\left(1+ \sum_i c_i p_i \, \varkappa \, \psi^{(0)}(\alpha_i)  +O(\varkappa^2) \right) = \prod_i \Gamma(\alpha_i+ c_i \varkappa)^{p_i} +O(\varkappa^2),
\end{equation}
we can write the evaluations of \eqref{eq:Idefs} compactly (albeit non-uniquely):
\begin{subequations}
\begin{align}
    \frac{I_{\mathrm{tet}}}{C_\sigma^2 C_\phi^4} &= \frac{3}{8N} \frac{2\pi ^{2 d}}{\Gamma(d)} e^{2 \gamma  \varkappa } (2R)^{4 \varkappa } \Gamma (2 \varkappa -\tfrac{d}{2})  A(\Delta_\sigma +\tfrac{\varkappa }{3}) A(\Delta _{\phi }+\tfrac{\varkappa}{3})^2 + O(\varkappa^2);\\
    \frac{I_{\mathrm{pr}}}{C_\sigma^3 C_\phi^6} &= \frac{\frac{5}{2}}{12 N}\frac{2\pi^{3d}}{\Gamma(d)} e^{3\gamma \varkappa} (2R)^{6\varkappa} \Gamma(3 \varkappa -\tfrac{d}{2})A(\Delta_\sigma +\tfrac{\varkappa }{5})^3 A(\Delta_\phi +\tfrac{3 \varkappa }{5})^3   A(-\tfrac{d}{2} +3\Delta_\phi +\tfrac{3 \varkappa }{5}) + O(\varkappa^2).
\end{align}
\end{subequations}
We need to evaluate these diagrams at $\varkappa = \delta$ and for $C_i=C_{i,0}$; however, these expressions are valid for any small shift of these scaling dimensions.

We can solve the pillow exactly in terms of hypergeometric functions, and a complete expression is presented in \eqref{eq:IpillowTotallyAnalytic}.
The expression for the pillow is $O(\delta)$ for generic $s$, but $O(\delta^0)$ for $s=2k$. 
This reflects the fact that the pillow only contributes to the anomalous dimensions in the short-range models.
\begin{equation}
\frac{I_{\mathrm{pil}}}{C_\sigma^2 C_\phi^4} = \begin{cases}
    \frac{1}{4 N} \frac{2 \pi ^{2 d}\Gamma \left(\frac{d}{2}\right) \Gamma \left(-\frac{d}{2}\right)}{\Gamma (d)} \frac{\Gamma \left(\frac{s}{2}\right) \Gamma \left(-\frac{s}{2}\right)}{\Gamma \left(\frac{d-s}{2}\right) \Gamma \left(\frac{d+s}{2}\right)} \varkappa + O(\varkappa^2) & s \neq 2k \\
    \frac{12}{4 N} \frac{\pi ^{2 d}}{\Gamma(d)} \frac{(-1)^k}{4k} \frac{\Gamma\left(-\dotwo\right) \Gamma\left(\dotwo\right)}{\Gamma \left(\dotwo-k\right) \Gamma \left(\dotwo+k\right)} +O(\varkappa) & s = 2k
\end{cases}. \label{eq:ItetEvals}
\end{equation}

We now fix to
\begin{equation}
\Delta_\phi \to \frac{d-s}{2}, \quad \Delta_\sigma \to s-\delta,
\end{equation}
and therefore set
\begin{equation}
\varkappa=\delta, \quad C_\sigma C_\phi^2 \to C_{\sigma,0} (C_{\phi,0})^2 \stackrel{\eqref{eq:BareONtwoPointNorms}}{=} -2/\cN_b(s),
\end{equation}
which makes
\begin{subequations}
\begin{equation}
 I_{\mathrm{tet}} = -3\times \frac{1}{2N} F_b'(s) \Gamma_\mathrm{tet} + O(\delta), \quad I_{\mathrm{pr}} = -\frac{5}{2} \times \frac{1}{3N} F_b'(s) \Gamma_\mathrm{pr}  + O(\delta),
\end{equation}
and
\begin{equation}
    I_{\mathrm{pil}} = \begin{cases}
        O(\delta) & s \neq 2k\\
        3\times \frac{1}{N} F_b'(2k) \hat{\gamma}_{\phi,1} +O(\delta) & s = 2k
    \end{cases}.
\end{equation} 
\end{subequations}

\subsubsection{The counterterm contribution}

The counterterm diagram is
\begin{equation}\label{eq:IctDef}
    I_{\mathrm{ct}} = \half \left(-\mu^{2\delta}\frac{2 Z_{g,1}}{N\delta}\right) \tr K^{-1} K_\delta
\end{equation}
which can be computed using $I_2(\Delta)$,
\begin{equation}
    \tr K^{-1} K_\delta = \frac{1}{\cN_{s}} \int_{x,y} \frac{1}{s(x,y)^{2\left((d-s)+ s-\delta\right)}} = \frac{1}{\cN_{s}}I_2\left(d-\delta\right).
\end{equation}
Taking the limit $\delta \to 0$, we find a finite correction from the counterterm for any $s$\footnote{
   This same correction also follows from doing the Gaussian $\sigma$ integral and expanding
    \begin{equation}
    F_{\gO(N)} \supset \half \tr \log \left(K^{-1}_{\delta} + \frac{2Z_{g,1} \mu^{2\delta}}{N \delta}K^{-1}\right) = \half \tr \log K^{-1}_{\delta} +\frac{Z_{g,1} \mu^{2\delta}}{N \delta} \tr K^{-1} K_\delta + O(1/N^2).
    \end{equation}},
\begin{equation}
    F_{\gO(N)} \supset -I_{\mathrm{ct}} =  \frac{Z_{g,1}}{N} \frac{1}{\cN_{s}} \frac{2}{-\sin(\tfrac{\pi d}{2})}\frac{\pi^{d+1}}{\Gamma(d+1)}+O(\delta) = \frac{2 Z_{g,1}}{N} F_b'(s) + O(\delta).
\end{equation}

\subsubsection{Assembling \FtextOrPDF} \label{sec:assemblingF}

Recalling that $Z_g$ differs in each of the short-range and long-range cases, we see that the extra factors of $3$ and $5/2$ that appeared in the integral evaluations cancel with the contributions from $Z_{g,1}$ in the $1/N$ correction to $F$.
Writing the counterterm contributions on a separate row, we find
\begin{equation}\begin{aligned}
F_{\gO(N),1/N} &= \begin{matrix}
    & -I_\mathrm{tet} - I_\mathrm{pr} - I_\mathrm{pil}\\
    & -I_\mathrm{ct}
    \end{matrix}\\
\frac{F_{\gO(N),1/N}}{F_b'(s)/N} &=\begin{matrix}
&+3\times\half \Gamma_\mathrm{tet} &+ \frac{5}{2} \times \frac{1}{3} \Gamma_\mathrm{pr} \\
&-2\times\half \Gamma_\mathrm{tet} &- \frac{3}{2} \times \frac{1}{3} \Gamma_\mathrm{pr}
\end{matrix}
  + \begin{cases}
   0 & s \neq 2k\\
    \begin{matrix} 
        &+3 \times \half(-2\hat{\gamma}_{\phi,1})\\ 
        &-2 \times  \half(-2\hat{\gamma}_{\phi,1})
    \end{matrix} & s=2k \end{cases},
\end{aligned}
\end{equation}
where we have evaluated $Z_{g,1}$ using \eqref{eq:SRcts} for SR and \eqref{eq:LRcts} for LR.
Thus, we obtain
\begin{subequations}\label{eq:FinalFs}
\begin{align} 
    F_{\gO(N)}^\mathrm{SR,\Box^k} &= N F_b(\tfrac{d}{2}-k) + F_b(2k) + \frac{F_b'(2k)}{N}\left(\half(-2 \hat{\gamma}_{\phi,1}) + \half \Gamma_\mathrm{tet} +\frac{1}{3} \Gamma_\mathrm{pr}\right)_{s=2k},\\
      F_{\gO(N)}^\mathrm{LR} &= N F_b(\tfrac{d-s}{2}) + F_b(s) + \frac{F_b'(s)}{N}\left(\half \Gamma_\mathrm{tet} +\frac{1}{3} \Gamma_\mathrm{pr}\right).
\end{align}
\end{subequations}
Using $F_b'(\dotwo-k)=0$ and $\hat{\gamma}_\phi = 2F_b'(2k)/F_b''(\dotwo -k)$, the identity
\begin{equation}
F^\mathrm{LR}_{\gO(N)} (s= 2k -2 \hat{\gamma}_\phi) = F^{\mathrm{SR},\Box^k}_{\gO(N)}
\end{equation}
becomes obvious, as does the compact form \eqref{eq:FsimpleSummary}. 
We conclude with two comments about why $F$ takes this form.

\subsubsection{Why do anomalous dimensions appear in \FtextOrPDF?} \label{sec:whyGammas}

The schematic form for the free energy corrections $\sim \frac{F_b'(s)}{N} \Gamma_i$ can be understood by studying the two-point function equations. 
Manifestly, any $\expval{\sigma\sigma}$ diagram can be converted into a vacuum diagram by closing the loop. 
If $\Sigma_{xy}=\expval{\sigma(x)\sigma(y)}$ is the $\sigma$ propagator, then any vacuum diagram satisfies
\begin{equation}\label{eq:IproptoDiff}
I_i \propto \int_{xy} \fdv{I_i}{\Sigma_{xy}} \Sigma_{xy};
\end{equation}
but this functional derivative is just an amputated self-energy diagram, as described in \cref{app:counterterms}.
The self-energy diagrams that we find are, as usual for a self-energy diagram, proportional to the inverse propagator, with a pole that gives the anomalous dimension contribution of that diagram:
\begin{equation}
\fdv{I_i}{\Sigma_{xy}} \propto \Sigma_{xy}^{-1} s(x,y)^{2m \delta} \times \left(\frac{\Gamma_i}{N}  \frac{1}{\delta} + O(\delta^0)\right)
\end{equation}
for some integer $m$ -- this is clear from inspection of
 \cref{eq:dItetdSigma,eq:dIprdSigma}. 
 Performing the integral in \eqref{eq:IproptoDiff} then yields the finite result
\begin{equation}
I_i \propto \frac{1}{N} \frac{\Gamma_i}{\cN_b(s)} \lim_{\delta \to 0} \frac{1}{\delta} I_2(d-m \delta) \sim \frac{F_b'(s)}{N}\Gamma_i.
\end{equation}
This heuristic argument explains how the vacuum diagrams pick out the poles of the two-point function diagrams. 
However, it does not explain the particular numerical factors of $\thalf$ and $\tfrac{1}{3}$ that we find in the final $F$s \eqref{eq:FinalFs} -- those require a full calculation that also accounts for the counterterms.

\subsection{\FtextOrPDF for the UV of our alternative field theory}

Recall that the way that the $4-\epsilon$ expansion works is that the QFT \eqref{eq:ONaction} describes both the free theory of $N$ fields, with $g=0$, and the interacting Wilson-Fisher CFT with $g \sim \epsilon$ in $d=4-\epsilon$.
Because these two theories collide for $\epsilon=0$, taking $\epsilon$ to be small, we solve for the WF CFT by perturbing around the solvable free CFT. 
The WF free energy is \cite{Fei:2015oha}
\begin{equation}
F_{\gO(N), d=4-\epsilon} = F_\text{free} + O(\epsilon^2), \quad F_\text{free} = N F_b(\tfrac{d}{2}-1).
\end{equation}
The corrections are proportional to a positive power of the small parameter that controls the deviation of the CFT from some solvable theory (in this case, a free field theory).

The same occurs in the large-$N$ case.
Evidently, the DREG$+\delta$-regulated field theory's free energy can be calculated by sending $g \to 0$ and $\mu \to 0$ in \eqref{eq:completeAction}, giving
\begin{equation}\label{eq:UVfreeenergy}
F_{\gO(N),g=0} = NF_b(\tfrac{d-s}{2}) + F_b(s-\delta).
\end{equation}
Recall that we want to improve the UV convergence of this theory, and therefore we choose $\delta>0$; hence sending $\mu\to 0$ has led to the $K_\delta^{-1}$ dominating over $-K^{-1}$.
Strictly speaking, this involves working away from the critical point, and so really we should be considering coupling constants in front of the $\sigma$ kinetic terms, and finding their beta functions, but this would have the same effect.
The interacting free energies that we found are clearly perturbations around \eqref{eq:UVfreeenergy} proportional to the small parameter $1/N$, just as in the $\epsilon$-expansion case.

\eqref{eq:UVfreeenergy} tells us that in the UV this regulated theory has an extra scalar of dimension $s$, compared to the $\phi^4$ theory. 
This makes sense, as in the large-$N$ limit we access the $\gO(N)$ CFT from an alternative field theory that is a distance $1/N$ away, as was shown in \cref{fig:duality}.
This is what we mean when we say that the analytic regulator makes $\sigma$ dynamical in the UV. 
However, this extra scalar does not have the $\Delta=\tfrac{d+s}{2}$ required to be the extra scalar $\chi$ proposed in \cite{Behan:2017emf} to resolve problems with the crossover from short- to long-range. %

\acknowledgments

We are grateful to John Wheater for discussions and comments on the draft.
The author is supported by the Dalitz Scholarship from the University of Oxford and Wadham College.

For the purpose of open access, the author has applied a CC BY public copyright licence to any Author Accepted Manuscript (AAM) version arising from this submission. %

\appendix
\section{Conventions}\label{app:conventions}

\tikzset{
    dot/.style={circle, fill, inner sep=1.2pt}, %
    bigdot/.style={circle, fill, inner sep=2pt}, %
    line label/.style={font=\scriptsize, inner sep=2.5pt, anchor=south}, %
    linelabel2/.style={font=\scriptsize}, %
    point name/.style={font=\tiny, inner sep=1pt}, %
    arrow style/.style={->, >=Stealth}, %
    fermion/.style={
	decoration={markings, mark=at position 0.55 with {\arrow[->, >=Stealth]{>}}},
	postaction={decorate}
	},
	antifermion/.style={
	decoration={markings, mark=at position 0.55 with {\arrow[->, >=Stealth]{<}}},
	postaction={decorate}
	},
}
For convenience, we summarise here our conventions.

\subsection{Gaussian integrals}\label{app:Gaussians}
The Gaussian integrals are
\begin{subequations}\label{eq:GaussianIntegral}
	\begin{equation}
	    \begin{aligned}
	\int \Dd{\phi}\,  e^{-\half \phi_i [C^{-1}]_{ij} \phi_j + J_i \phi_i} &=  \prod_i \int_{-\infty}^\infty \frac{\dd \phi_i}{\sqrt{2\pi}} e^{-\half \phi^T C^{-1} \phi + J^T \phi} =  (\det C^{-1})^{-\half}  e^{+\half J^T C J}\\
	&= \exp\left(-\half \log \det C^{-1} + \half J^T C J\right)
	    \end{aligned}
	\end{equation}
	and its real Grassmann cousin
	\begin{equation}\begin{aligned}
	\int \Dd{\psi}\, e^{+\half \psi_i [C^{-1}]_{ij} \psi_j + \eta_i \psi_i}&= \int \odif{\psi_{n}} \cdots \odif{\psi_1} e^{+\half \psi^T C^{-1} \psi_j + \eta_i \psi_i} = (\det C^{-1})^{\half} e^{+\half \eta^T C \eta}\\
	& = \exp\left(+\half\log \det C^{-1} + \half \eta^T C \eta\right),
	\end{aligned} \end{equation}
	in the conventions of \cite{Zinn-Justin:1991ksq}.  %
	The complex version is	%
	\begin{equation}\begin{aligned}
		\int \Dd{\psi} \Dd{\psibar}\, e^{\psibar_i [C^{-1}]\indices{^i_j} \psi^j + \bar{\eta}_i \psi^i + \psibar_i \eta^i}&= \int \prod_i \odif{\psi^i} \odif{\psibar_i} \, e^{\psibar\, C^{-1} \psi + \bar{\eta} \psi + \psibar \eta} = (\det C^{-1}) e^{-\bar{\eta}\, C \eta}\\
		& = \exp\left(+\log \det C^{-1}- \bar{\eta}\, C \eta\right).
	\end{aligned}
	\end{equation}
	\end{subequations} %
	We stress the different signs in the action, $S_b=\half \phi \, C^{-1} \phi$ and $S_f = - \psibar \, C^{-1} \psi$; the latter sign follows purely from our chosen conventions for Grassmann integrals.

	If we consider these integrals as written in DeWitt notation, these are exactly infinite-dimensional functional integrals.
That is, if we upgrade $\phi_i \mapsto \phi(x)$, $[C^{-1}]_{ij} \mapsto C^{-1}(x,y)$, $\mathbb{I} =\delta_{ij} \mapsto \delta(x-y)$ and $\sum_i \mapsto \int_x$, these expressions describe the functional Gaussian integrals -- of course, we have been careful to define the bosonic measure $\Dd{\phi}$ with the factors of $(\sqrt{2\pi})^\infty$. 
We evaluated the universal part of these determinants for conformal sphere propagators $C$ in \cref{sec:Fcomputation}. %

\subsubsection{Imaginary Gaussian integrals}\label{app:imagGaussian}

Recall the definition of the Dirac delta function,
\begin{equation}
	\delta(H) \equiv \int_{-\infty}^{\infty} \frac{\odif{\rho}}{2\pi} e^{i\rho H} \stackrel{\sigma\equiv -i\rho}{=} \int_{-i\infty}^{i\infty} \frac{\odif{\sigma}}{2\pi i} e^{-\sigma H}.
\end{equation}
Upgrading to the functional version, it is conventional to define a field $\sigma(x)$ with an imaginary contour \cite{Moshe:2003xn}, such that
\begin{equation}
\delta(H(x))=\int \DdImag{\sigma} \, e^{-\int_x \sigma(x) H(x)} \equiv \left(\prod_x \int_{-i\infty}^{+i\infty} \frac{\odif{\sigma(x)}}{2\pi i}\right)\, e^{-\int_x\sigma(x) H(x)}.
\end{equation}
Alternatively, we could keep $\sigma$ on a real contour at the cost of $i$s in the action; by convention, we do not do so.

The $\sigma$ action takes the form $\half \sigma K^{-1} \sigma$, for $-K^{-1}$ positive definite.
Due to the imaginary contour and the normalization of $\DdImag{\sigma}$,
\begin{equation}\begin{aligned}
&\int \DdImag{\sigma}\, e^{-\half \int_{x,y} \sigma_x K^{-1}_{xy} \sigma_y + \int_x J_x \sigma_x} \\
\stackrel{\rho = i \sigma}{=} & \left(\prod_x \int_{-\infty}^{\infty} \frac{\odif{\rho_x}}{\sqrt{2\pi}}\right) \, e^{-\int_{x,y}\half \rho_x (-K_{xy}^{-1}) \rho_y -i \int_x J_x \rho_x - \int_x \half \log 2\pi}\\
\stackrel{\eqref{eq:GaussianIntegral}}{=} & \, e^{-\half \tr \log (-K^{-1})- \half  \log (2\pi) \tr \id} e^{-\half J^T (-K) J}.
\end{aligned} \end{equation}
$\sum_x 1= \tr \id = \int_{x,y} \delta(x-y)$ manifestly gives a UV divergent contribution to the free energy on the sphere. 
However, this divergence vanishes in DREG;  
equivalently, it can be cancelled by a local counterterm that modifies the constant part of the action, $S \mapsto S - \int_x \half \log(2\pi)$.
Therefore, the universal part of this integral is simply its computation in DREG, which we write schematically as
\begin{equation}
	\int \DdImag{\sigma}\, e^{-\half \int \sigma K^{-1} \sigma + \int J \sigma } = e^{-\half \tr \log (-K^{{-1}}) + \half J^T K J}.
\end{equation}
In the main text, we calculated this trace for $-K^{-1} \mapsto  -K_\delta^{-1}= \half (C_{\phi,0})^2/s(x,y)^{2(d-s+\delta)}$. 

\subsection{Kinetic terms and position space}\label{app:space}
In position space, on an arbitrary geometry, we write
\begin{equation}
\int_x = \int \odif[order=d]{x} \sqrt{g(x)}.
\end{equation}
When working in flat space, we can transform to momentum space, where we define
\begin{equation}\begin{aligned}
\quad\int_p= \int \frac{\odif[order=d]{p}}{(2\pi)^d}.
\end{aligned}\end{equation}
For the bosonic theory, we write
\begin{equation}
S_b = + \half \int_x \phi \, C^{-1} \phi,
\end{equation}
where any fields without argument are assumed to be at $x$.
As usual, the propagator is the matrix inverse of the kinetic term $C^{-1}$. 
So, for the flat space $C^{-1} = -\partial^2$,
\begin{align}
 \expval{\phi(p) \phi(q)} &= C(p,q) = C(p)\delta(p+q) = +\frac{1}{p^2} \delta(p+q)\\
 \expval{\phi(x) \phi(y)} &= C(x,y) = C(x-y), \quad C(x) \equiv \frac{C_{\phi,0}}{\abs{x}^{2\Delta_\phi^\UV}}, \quad \Delta_\phi^\UV \equiv \tfrac{d-2}{2}
\end{align}
In the fermionic models, we keep the dimension of spinor space $\tr \spinid$ (i.e. the dimension of the gamma matrices) general, folding it into $N$ as described in the main text.
Our Euclidean gamma matrix conventions and Gaussian integral conventions are those of \cite{zinn-justin_quantum_2002}, such that the free fermion action is
\begin{equation}
	S_f = -\int_x \psibar \, C^{-1} \psi, 
\end{equation}
with $\psibar \equiv -\psi^\dagger \gamma_0$, and of course $\{\gamma_\mu, \gamma_\nu\} = 2g_{\mu\nu}$.
Once again, for $C^{-1} = \slashed{\partial}$,
\begin{align}
 \expval{\psi(p) \psibar(q)} &= C(p,q) = C(p)\delta(p+q) = +\frac{1}{i\slashed{p}} \delta(p+q)\\ %
 \expval{\psi(x) \psibar(y)} &= C(x,y) = C(x-y), \quad C(x) \equiv \frac{C_{\psi,0} \slashed{x}}{\abs{x}^{2\Delta_\psi^\UV+1}}, \quad \Delta_\psi^\UV \equiv \tfrac{d-1}{2} \label{eq:FermionPropPos}
\end{align}
where position and momentum space propagators are distinguished implicitly by argument. 

We heavily overload our notation, such that we can freely move between the local derivative implementations $C^{-1} =-\partial^2$ and $\slashed{\partial}$, the real space matrix $C(x,y)=C(x-y)$, and the momentum space $C(p,q)=C(p)\delta(p+q)$.
Thus, the most general kinetic term is
\begin{equation}
\half\int_x \phi\, C^{-1}\phi = \half\int_{x,y} \phi(x) C^{-1}(x,y) \phi(y)
\end{equation}
which in flat space would be $\thalf \int_x \phi(x) (-\partial^2)^{\frac{d}{2}- \Delta_\phi^{\mathrm{free}}} \phi(x) = \thalf \int_{p} \tilde{\phi}(p) (p^2)^{\frac{d}{2}- \Delta_\phi^{\mathrm{free}}} \tilde{\phi}(-p)$.

For a fermion in flat space, we have $C^{-1} = \slashed{\partial}$ for the canonical free fermion, with
\begin{equation}
C^{-1} = \slashed{\partial} (-\partial^2)^{\tfrac{d-1}{2}-\Delta_\psi^\text{free}}= \frac{\slashed{\partial}}{\sqrt{-\partial^2}} (-\partial^2)^{\tfrac{d}{2}-\Delta_\psi^\text{free}}
\end{equation}
for general $\Delta_\psi^{\text{free}}$.

\subsection{Sphere kinetic terms for scalars}\label{app:sphereKineticTerms}

We work in stereographic coordinates on the Euclidean sphere $S^d$.
These coordinates are defined by
\begin{align}
	\odif{s^2} &= \Omega(x)^2  \odif{x^\mu} \odif{x^\mu}, \quad \Omega(x) \equiv \frac{2R}{1+x^2}, \quad \cR_{\mu\nu} = \frac{d-1}{R^2} g_{\mu\nu}, \quad \cR = \frac{d(d-1)}{R^2},
\end{align}
The volume of the unit sphere then follows from
\begin{align}
	R^d \vol S^d &\equiv \int_x 1 = \int\odif[order=d]{x}\, \sqrt{g} = R^d \frac{2 \pi^{\frac{d+1}{2}}}{\Gamma(\tfrac{d+1}{2})}
\end{align}
We define the sphere chordal distance to be
\begin{equation}
s(x,y) \equiv \sqrt{\Omega(x)\Omega(y)} \abs{x-y}.
\end{equation}

If we want a free propagator of the form
\begin{equation}\label{eq:desiredSphereProp}
C(x,y) \equiv \frac{C_\phi}{s(x,y)^{2\Delta}}
\end{equation}
on the sphere, we can schematically write
\begin{equation}\label{eq:schematicC}
S = \half \int_{x,y} \phi(x) C^{-1}(x,y) \phi(y),\quad C^{-1}(x,y) \approx \frac{1}{C_\phi} \frac{1}{\cN_b(\Delta)} \frac{1}{s(x,y)^{2(d-\Delta)}}.
\end{equation}
However, this is not quite right, as this integral is (a) divergent, requiring contact terms to make it converge; and (b) clearly zero for $\Delta = \dotwo -k$.
The correct way to write the inverse $C^{-1}$ that satisfies
\begin{equation}
\int_y C(x,y)C^{-1}(y,z)= \frac{1}{\sqrt{g(x)}}\delta(x-z)
\end{equation}
is with the conformal Laplacian of biweight $(\Delta,d-\Delta)$ on $S^d$, which we will now explain, following \cite{Benedetti:2021wzt}.
 Defining the standard $\zeta = \dotwo - \Delta = \tfrac{s}{2}$, we can obtain \eqref{eq:desiredSphereProp} from
\begin{equation}
	S= \half \int_x \phi(x) C^{-1} \phi(x),\quad  C^{-1}=\frac{\cD_{\zeta}}{C_\phi \cF_{2\Delta,0} }, \quad \cD_{\zeta}= \frac{1}{R^{2\zeta}} \frac{\Gamma(R \cD_\half + \thalf + \zeta)}{\Gamma(R \cD_\half + \thalf -\zeta)},
\end{equation}
where $\cD_{\half}$ is defined with the Laplace-Beltrami operator $\nabla^2_{S^d}$,
\begin{align}
\cD_{\half} \equiv \sqrt{-\nabla^2_{S^d} + \left(\tfrac{d-1}{2R}\right)^2}.
\end{align}
On the sphere, $-\nabla^2_{S^d}$ has eigenvalues $\omega_n$ with degeneracy $g_n$ for $n \ge 0$, 
\begin{equation}
    \omega_{n\ge 0} = \frac{1}{R^2} n(n+d-1), \quad g_n \equiv \frac{(n+d-2)!(2n+d-1)}{n!(d-1)!} = (1+\tfrac{2n}{d-1}) \binom{d+n-2}{n}.
\end{equation}
Thus, $\cD_{\half}$ and $\cD_\zeta$ have eigenvalues with the same degeneracy
\begin{align}
     \omega_n^{(\half)} &\equiv \tfrac{1}{R}\left(\tfrac{d-1}{2} + n\right), \quad 
\omega_n^{(\zeta)} \equiv \frac{1}{R^{2\zeta}} \frac{\Gamma(n+\dotwo + \zeta)}{\Gamma(n+\dotwo-\zeta)}.
\end{align}
To calculate the universal part of the free energy, we compute
\begin{equation}
    F_b(\Delta\equiv \dotwo-\zeta) \equiv \half \tr \log \cD_\zeta = \half \sum_{n=0}^\infty g_n \log\omega_n^{(\zeta)}.
\end{equation}
This matches \eqref{eq:bosonF} as computed using the methods of \cref{sec:Fcomputation} after differentiating with respect to $\zeta$; the resulting sum contains terms of the form $0^{-d}$, and so is finite for $d<0$ -- it can then be analytically continued up to the $d$ of interest.

We make three observations:
\begin{itemize}
\item $\cD_{k}$ is manifestly a finite polynomial of order $k$ in $(\cD_\half)^2$ for integer $k$, beginning
\begin{equation}
\cD_{k\in \mathbb{Z}} = (\cD_{\half}^2)^{k} + \frac{k}{12}\frac{(1-4k^2)}{R^2} (\cD_{\half}^2)^{k-1} +  \cdots,
\end{equation}
which makes explicit the intuition that for the $\Box^k$ models $C^{-1} \sim (-\Box)^k + \cdots$.
It also matches \eqref{eq:keq1Cval} for $\zeta = k=1$ (the constant term serves to remove the zero mode).
\item $\cD_\zeta \to (\cD_\half)^{2\zeta}$ in the flat-space limit of $R\to \infty$ (equivalently, in the large-momentum limit), making explicit the intuition that the flat-space $C^{-1} \sim (-\Box)^{s/2}$.
\item For all $\Delta$, a calculation of $C(x,x)$ in DREG gives \cite[\S D]{Benedetti:2021wzt}
\begin{equation}
C(x,x) = 0.
\end{equation}
\end{itemize}
For $0<s<2$, this conformal Laplacian can also be written as an explicitly non-local operator using a subtracted hypersingular integral (the $r$ limit is essentially a principal value -- with $R$ factors corrected from \cite{Benedetti:2021wzt}):
\begin{align}
\cD_{\dotwo -\Delta} \phi(x) & = \lim_{r\to 0} \frac{1}{\cF_{2(d-\Delta),0}}\int_{y, \, s(x,y)>r} \frac{\phi(y) -\phi(x)}{s(x,y)^{2(d-\Delta)}} + \frac{1}{R^{d-2\Delta}}\frac{\Gamma(d-\Delta)}{\Gamma(\Delta)} \phi(x). \label{eq:DContactTerms}
\end{align}
This is a rigorous implementation of the schematic \eqref{eq:schematicC} (since $\cF_{2\Delta,0} \cF_{2(d-\Delta),0} = \cN_b(\Delta)$).

For generic $0<s\neq 2k$, we must add further derivative terms to ensure that the Green's function is exactly the conformal propagator.
In flat space $g=\delta$, we find \cite[(2.13)]{Gubser:2019uyf} (note that their $D^s=\cD_{s/2}/(2\pi)^s$ in our notation),
\begin{equation}
\begin{aligned}
\cD_{\dotwo -\Delta} \phi(x) \rvert_{g=\delta} & = \lim_{r\to 0} \frac{1}{\cF_{2(d-\Delta),0}}\int_{y, \, \abs{x-y}>r} \frac{1}{\abs{x-y}^{2(d-\Delta)}}  \left(\phi(y) - \sum_{n=0}^{\floor{s/2}} \abs{x-y}^{2n} b_n \Box^n \phi(x)\right)\\
b_n &= \frac{\Gamma(\dotwo)}{2^{2n} \Gamma(n+\dotwo) \Gamma(n+1)},\label{eq:sGreaterThan2}
\end{aligned}
\end{equation}
which indeed satisfies
\begin{equation}
    \int_k \phi(k) \abs{k}^{2s} \phi(-k)  = \int_x \phi(x) \cD_{s/2} \phi(x)|_{g=\delta}
\end{equation}
(recalling our Fourier transform conventions).
Thus, simple $\abs{k}^s$ terms in momentum space correspond to carefully tuned non-local position-space actions that contain both terms $\sim\phi(x) \phi(y)$ and $\sim\phi(x) \phi(x)$.

These integral representations are not strictly valid for $s=2k$, because, using \eqref{eq:FlamsDef}, $1/\cF_{2(d-\Delta),0} = 0$  for $\Delta = \dotwo -k$; however, they can still be used to define the local operator $(-\partial^2)^k$ as a \enquote{resonant} limit \cite[\S A]{Diaz:2008hy}.

\subsection{Sphere kinetic terms for fermions}\label{app:sphereKineticTermsFermions}

For fermions, the fractional Dirac operator was defined in \cite[Prop. 2.5]{maalaouiConformalFractionalDirac2025} (generalizing the integer-power work of \cite{fischmannConformalPowersDirac2014}). 
It leads to a fermion sphere propagator which is a Weyl transformation of \eqref{eq:FermionPropPos} from flat space to the sphere.
In embedding coordinates $\eta^{a=1,\cdots,d+1}$, $\sum_{ab} \delta_{ab} \eta^a \eta^b=R^2$, with $(d+1)$-dimensional gamma matrices $\{\alpha_a, \alpha_b\} = 2\delta_{ab}$, this is simply \cite{Giombi:2015haa}
\begin{equation}\label{eq:fermionPropSphereEmbedding}
   C(x,y) =  C_\psi \frac{\alpha_a (\eta_1 -\eta_2)^a}{\abs{\eta_1-\eta_2}^{2\Delta_\psi+1}}
\end{equation}

To get this from an action of the form $S=-\int_x \bar{\psi} C^{-1} \psi$, we define $\lambda \equiv \frac{d}{2}-\Delta_\psi = \frac{s}{2}$; then, taking the sphere Dirac operator to be $D= \slashed{\nabla}_{S^d}$, we write\footnote{The $i$ here follows from $C(p,q) =\expval{\psi(p) \bar\psi(q)} =\cF_{2\Delta,1} C_\psi \slashed{p}/p^{d-2\Delta_\psi+1} \delta(p+q)$ after applying \eqref{eq:Gfouriers} to the flat-space version of \eqref{eq:fermionPropSphereEmbedding}; hence clearly $S = -\half \int \bar{\psi}(q) C^{-1} \psi(p) = -\half \int \bar{\psi}(q) (\cF_{2\Delta,1} C_\psi)^{-1} \slashed{p} p^{d-1-2\Delta_\psi} \delta(q+p) \psi(p)$; then $p \to -i\slashed{\nabla}$ as usual.}
\begin{equation}
   C^{-1} = \frac{-i\cD_\lambda}{C_\psi \cF_{2\Delta_\psi,1}}, \quad \cD_{\lambda} = \frac{1}{R^{2\lambda}} \frac{\Gamma(R\abs{D} + \half + \lambda)}{\Gamma(R\abs{D} + \half - \lambda)} \frac{D}{\abs{D}}
\end{equation}
We do not venture a subtracted hypersingular integral representation here, though it should exist.

Since $D$ has two sets of eigenspinors $\hat{\phi}_{k,\pm}$ with eigenvalues and degeneracies\footnote{Recall that we have fixed $\tr \spinid =T$, ignoring the dependence on $d$ of the dimension of the gamma matrices, which is usually $\tr \spinid =2^{\floor{\dotwo}}$.}\cite[(3.44,3.58)]{Camporesi:1995fb}
\begin{equation}
    \omega_{n \ge 0,\pm} = \pm(\dotwo + n)/R, \quad g_{n,\pm} = T \frac{(d+n-1)!}{n! (d-1)!} = T \binom{n+d-1}{n}, %
\end{equation}
The same $\hat{\phi}_{k,\pm}$ have eigenvalue $\abs{\omega_n^{(\pm)}}$ for $|D|$ (in the notation of \cite[\S A.2]{maalaouiConformalFractionalDirac2025}, $D= \nu \cdot D_{h_0}$), so
\begin{equation}
    \omega_{n,\pm}^{(\lambda)} = \pm \frac{1}{R^{2\lambda}} \frac{\Gamma(\frac{d+1}{2}+n+ \lambda)}{\Gamma(\frac{d+1}{2}+n- \lambda)}.
\end{equation}
We define $F_f$ as the universal part of the free energy of a single complex component of a Dirac fermion of dimension $\Delta$,
\begin{align}
    F_f(\Delta=\dotwo-\lambda) \equiv -\frac{\tr \log \cD_\lambda}{T} = -\sum_{\sigma=\pm}\sum_{n \ge 0} \frac{g_{n,\sigma}}{T} \log \omega_{n,\sigma}^{(\lambda)}=-2\sum_{n \ge 0} \frac{g_{n,+}}{T} \log \omega_{n,+}^{(\lambda)}.
\end{align}
The last equality follows from $\sum_{n\ge 0}g_{n,\pm}\propto (1-1)^{-d}=0$.
Again, after differentiating and analytically continuing from $d<0$, this matches \eqref{eq:fermionF} as computed using the methods of \cref{sec:Fcomputation}.

We note finally that for $\lambda = k+ \half$, $k\in \mathbb{N}_0$,
\begin{equation}\begin{aligned}
   \cD_{k+\half}|_{R=1} &= (\abs{D}+k) \cdots (\abs{D}-1) D (\abs{D}+1) \cdots (\abs{D}+k)\\
   &= (D+k) \cdots (D-1) D (D+1) \cdots (D+k),
\end{aligned} \end{equation}
which holds because $\cD_{\frac{2k+1}{2}}/D$ depends only on $\abs{D}^2=D^2$ (which itself is clear from $\abs{\omega_{n \ge 0,\pm}}^2=(\omega_{n \ge 0,\pm})^2$); thus we match \cite[(5.13)]{fischmannConformalPowersDirac2014}.

\subsection{Propagators and associated functions}\label{app:propsAndAs}
In flat space, we define generic boson/fermion propagators (now unit-normalized) to be
\begin{align}
	G_{\phi}(x_{12})\equiv \frac{1}{|x_{12}|^{2\Delta_\phi}} = \begin{tikzpicture}[transform shape, baseline={([yshift=-1ex]current bounding box.center)}]
	    \node[dot] (l) at (0,0) {}; \node[below=1pt of l, point name] {$x_1$};
	    \node[dot] (r) at (1,0) {}; \node[below=1pt of r, point name] {$x_2$};
	    \draw (l) -- (r) node[midway, line label] {$\Delta_\phi$};
	\end{tikzpicture}
	,\quad
	G_{\psi}(x_{12}) \equiv \frac{\slashed{x}_{12}}{|x_{12}|^{2\Delta_\psi+1}} = \begin{tikzpicture}[transform shape, baseline={([yshift=-1ex]current bounding box.center)}]
	    \node[dot] (l) at (0,0) {}; \node[below=1pt of l, point name] {$x_1$}; 
	    \node[dot] (r) at (1,0) {}; \node[below=1pt of r, point name] {$x_2$};
	    \draw[fermion] (r) -- (l) node[midway, line label] {$\Delta_\psi$};
	\end{tikzpicture},
\end{align}
and they are implicitly differentiated by their propagators.
We can Fourier transform these:
\begin{equation}\begin{aligned}\label{eq:Gfouriers}
G_{\phi}(p) = \int_x e^{-i p \cdot x} G_{\phi}(x) = \cF_{2\Delta_\phi,0}\frac{1}{p^{d-2\Delta_\phi}},\quad
G_{\psi}(p) = \int_x e^{-i p \cdot x} G_{\psi}(x) = \cF_{2\Delta_\psi,1}\frac{\slashed{p}}{p^{d-2\Delta_\psi +1}},
\end{aligned}\end{equation}
where we define the general Fourier transform of a traceless propagator \cite{Karateev:2018oml}
\begin{equation}\begin{aligned}\label{eq:FlamsDef}
&\cF_{\lambda, s} \equiv i^{-s}2^{d-\lambda}\pi^{d/2} \frac{\Gamma(\frac{d+s-\lambda}{2})}{\Gamma(\frac{\lambda+s}{2})},\\ %
\text{where } & \cF_{\lambda,s} (p^{\mu_1} \cdots p^{\mu_s} -\text{traces})\, p^{-(d-\lambda)-s}=\int_x e^{-i p\cdot x} (x^{\mu_1}\cdots x^{\mu_s}-\text{traces})\, x^{-\lambda-s}.
\end{aligned}\end{equation}
Then the inverse transforms follow
\begin{equation}\begin{aligned} 
\int_p e^{i p \cdot x} \frac{1}{p^{d-2\Delta_\phi}} = \frac{\cF_{d-2\Delta_\phi,0}}{(2\pi)^d} \frac{1}{x^{2\Delta_\phi}},\quad \int_p e^{i p \cdot x} \frac{\slashed{p}}{p^{d-2\Delta_\psi+1}} = -\frac{\cF_{d-2\Delta_\psi,1}}{(2\pi)^d} \frac{\slashed{x}}{x^{2\Delta_\psi+1}}
\end{aligned}\end{equation} %
Since $\int_p e^{-ip\cdot x} = \delta(x)$, consistency demands that indeed $\cF_{2\Delta,s} \cF_{d-2\Delta,s} = i^{-2s}(2\pi)^d$; the $i^{-2s}$ then cancels with the necessity of taking $-p$ in \eqref{eq:FlamsDef} when inverse transforming.
Hence,
\begin{equation}
	\frac{1}{\cF_{\lambda,s}} (x^{\mu_1} \cdots x^{\mu_s} -\text{traces})\, x^{-\lambda-s} =\int_p e^{i p\cdot x} (p^{\mu_1}\cdots p^{\mu_s}-\text{traces})\, p^{-(d-\lambda)-s}.
\end{equation}

The main functions that we find in this paper are particular combinations of Gamma functions that arise from Fourier transforms \eqref{eq:Gfouriers} We drop the $i$s, $2$s, and $\pi$ that almost always cancel, and so define $\cF_{2\Delta,s}/(-i^s 2^{d-2\Delta}\pi^{d/2})$ for $s=0,1$:
\begin{equation}\begin{aligned}
A(\Delta) \equiv \frac{\Gamma(\tfrac{d}{2}-\Delta)}{\Gamma(\Delta)}, \quad
A_f(\Delta) \equiv \frac{\Gamma(\tfrac{d}{2}-\Delta + \thalf)}{\Gamma(\Delta + \thalf)},
\end{aligned}\end{equation}
which both satisfy $A_x(\Delta) A_x(\dotwo-\Delta) =1$.
From Taylor expanding these functions, we frequently also see
\begin{equation}\begin{aligned}
	B(\Delta) &\equiv -\odv{}{\Delta}\log A(\Delta) =\psi ^{(0)}(\Delta ) + \psi ^{(0)}(\dotwo -\Delta ),\\
	B_f(\Delta) &\equiv -\odv{}{\Delta}\log A_f(\Delta) =\psi ^{(0)}\left(\Delta +\thalf\right)+ \psi ^{(0)}\left(\dotwo - \Delta + \thalf\right),
\end{aligned}
\end{equation}
which both satisfy 
\begin{equation}
B_x(\Delta)= B_x(\dotwo-\Delta), \quad A_x(\Delta +\epsilon) = A_x(\Delta)\left[1 - \epsilon B_x(\Delta) + O(\epsilon^2)\right].
\end{equation}
For the uniqueness relations, we then define
\begin{equation}\begin{aligned}
U(a,b,c) &= \pi^{d/2} A(a) A(b) A(c)\\
U_\mathrm{bff}(a,b,c) &= \pi^{d/2} A(a) A_f(b) A_f(c).
\end{aligned}\end{equation}

\subsection{Position space loop integrals and the method of uniqueness} \label{app:uniqueness}

To make this paper self-contained, we summarise here the method of uniqueness \cite{Preti:2018vog}, which is used to calculate diagrams in flat space.  Let us begin with the loop integrals
\begin{align} %
	\begin{tikzpicture}[baseline={(0,-\MathAxis pt)}]
	    \node[dot] (l) at (0,0) {};
	    \node[dot] (r) at (1.5,0) {};
	    \draw (l) to[bend left=35] node[midway,line label] {$a$} (r);
	    \draw (l) to[bend right=35] node[midway,line label] {$b$} (r);
	\end{tikzpicture}
	&= \begin{tikzpicture}[baseline={(0,-\MathAxis pt)}]
	    \node[dot] (l) at (0,0) {};
	    \node[dot] (r) at (1.5,0) {}; %
	    \draw (l) -- node[line label, midway] {$a+b$} (r);
	\end{tikzpicture}
	,\quad
	\begin{tikzpicture}[baseline={(0,-\MathAxis pt)}, >=Stealth]
	    \node[dot] (l) at (0,0) {};
	    \node[dot] (r) at (1.5,0) {};
	    \draw[fermion]
		(r) to[bend right=35] node[midway, line label] {$a$} (l);
	    \draw (l) to[bend right=35] node[midway,line label] {$b$} (r);
	\end{tikzpicture}
	= \begin{tikzpicture}[baseline={(0,-\MathAxis pt)}, >=Stealth]
	    \node[dot] (l) at (0,0) {};
	    \node[dot] (r) at (1.5,0) {};
	    \draw[fermion]
		(r) -- node[line label, midway] {$a+b$} (l);
	\end{tikzpicture}
	,\quad
	\begin{tikzpicture}[baseline={(0,-\MathAxis pt)}, >=Stealth]
	    \node[dot] (l) at (0,0) {};
	    \node[dot] (r) at (1.5,0) {};
	    \draw[fermion]
		(l) to[bend left=35] node[midway, line label] {$a$} (r);
	    \draw[antifermion] %
		(l) to[bend right=35] node[midway, line label] {$b$} (r);
	\end{tikzpicture}
	= \begin{tikzpicture}[baseline={(0,-\MathAxis pt)}, >=Stealth]
	    \node[dot] (l) at (0,0) {};
	    \node[dot] (r) at (1.5,0) {};
	    \draw
		(l) -- node[midway, line label] {$a+b$} (r);
	\end{tikzpicture}
	\times (-\spinid).
	\end{align}
Next, the propagator merging relations, which can be derived by Fourier transforming both propagators and then reverse Fourier transforming the combination.
With $x_{ij} \equiv x_i- x_j$,
\begin{equation} \label{eq:propMerge}\begin{split}
\int_{x_m} \frac{1}{|x_{1m}|^{2a}} \frac{1}{|x_{m2}|^{2b}} &=
\begin{tikzpicture}[transform shape,baseline={(0,-\MathAxis pt)}]
    \node[dot] (l) at (0,0) {}; \node[below=1pt of l, point name] {$x_1$};
    \node[bigdot] (c) at (1,0) {}; \node[below=1pt of c, point name] {$x_m$};
    \node[dot] (r) at (2,0) {}; \node[below=1pt of r, point name] {$x_2$};
    \draw (l) -- (c) node[midway, line label] {$a$};
    \draw (c) -- (r) node[midway, line label] {$b$};
\end{tikzpicture}  = \begin{tikzpicture}[transform shape, baseline={(0,-\MathAxis pt)}]
    \node[dot] (l) at (0,0) {}; \node[below=1pt of l, point name] {$x_1$};
    \node[dot] (r) at (2.0,0) {}; \node[below=1pt of r, point name] {$x_2$};
    \draw (l) -- (r) node[midway, line label] {$a+b-\dotwo$};
\end{tikzpicture}
\times U(a,b,d-a-b)
\\
\int_{x_m} \frac{\slashed{x}_{1m}}{|x_{1m}|^{2a+1}} \frac{1}{|x_{m2}|^{2b}}  &=
\begin{tikzpicture}[transform shape, baseline={(0,-\MathAxis pt)}]
    \node[dot] (l) at (0,0) {}; \node[below=1pt of l, point name] {$x_1$}; 
    \node[bigdot] (c) at (1,0) {}; \node[below=1pt of c, point name] {$x_m$};
    \node[dot] (r) at (2,0) {}; \node[below=1pt of r, point name] {$x_2$};
    \draw[antifermion] (l) -- (c) node[midway, line label] {$a$};
    \draw (c) -- (r) node[midway, line label] {$b$};
\end{tikzpicture} = \begin{tikzpicture}[transform shape,baseline={(0,-\MathAxis pt)}]
    \node[dot] (l) at (0,0) {}; \node[below=1pt of l, point name] {$x_1$};
    \node[dot] (r) at (2.0,0) {}; \node[below=1pt of r, point name] {$x_2$};
    \draw[antifermion] (l) -- (r) node[midway, line label] {$a+b-\dotwo$};
\end{tikzpicture} \times U_\mathrm{bff}(b,a,d-a-b)%
\\
\int_{x_m} \frac{1}{|x_{1m}|^{2a}} \frac{\slashed{x}_{m2}}{|x_{m2}|^{2b+1}}  &=
\begin{tikzpicture}[transform shape,baseline={(0,-\MathAxis pt)}]
    \node[dot] (l) at (0,0) {}; \node[below=1pt of l, point name] {$x_1$}; 
    \node[bigdot] (c) at (1,0) {}; \node[below=1pt of c, point name] {$x_m$};
    \node[dot] (r) at (2,0) {}; \node[below=1pt of r, point name] {$x_2$};
    \draw (l) -- (c) node[midway, line label] {$a$};
    \draw[fermion] (r) -- (c) node[midway, line label] {$b$};
\end{tikzpicture} = \begin{tikzpicture}[transform shape,baseline={(0,-\MathAxis pt)}]
    \node[dot] (l) at (0,0) {}; \node[below=1pt of l, point name] {$x_1$};
    \node[dot] (r) at (2.0,0) {}; \node[below=1pt of r, point name] {$x_2$};
    \draw[antifermion] (l) -- (r) node[midway, line label] {$a+b-\dotwo$};
\end{tikzpicture} \times U_\mathrm{bff}(a,b,d-a-b)%
\\
\int_{x_m} \frac{\slashed{x}_{1m}}{|x_{1m}|^{2a+1}} \frac{\slashed{x}_{m2}}{|x_{m2}|^{2b+1}}  &=
\begin{tikzpicture}[transform shape, baseline={(0,-\MathAxis pt)}]
    \node[dot] (l) at (0,0) {}; \node[below=1pt of l, point name] {$x_1$}; 
    \node[bigdot] (c) at (1,0) {}; \node[below=1pt of c, point name] {$x_m$};
    \node[dot] (r) at (2,0) {}; \node[below=1pt of r, point name] {$x_2$};
    \draw[fermion] (r) -- (c) node[midway, line label] {$b$};
    \draw[fermion] (c) -- (l) node[midway, line label] {$a$};
\end{tikzpicture} =
\begin{tikzpicture}[transform shape, baseline={(0,-\MathAxis pt)}]
    \node[dot] (l) at (0,0) {}; \node[below=1pt of l, point name] {$x_1$};
    \node[dot] (r) at (2.0,0) {}; \node[below=1pt of r, point name] {$x_2$};
    \draw (l) -- (r) node[midway, line label] {$a+b-\dotwo$};
\end{tikzpicture} \times (-1) U_\mathrm{bff}(d-a-b,a,b).
\end{split} \end{equation} %
In this final relation we have used the standard convention that gamma matrix expressions are written by starting at the beginning of the arrow ($x_2$) and writing down each component from right to left.
Then the uniqueness relation (derived in appendix A of \cite{Gracey:2018ame}), which is valid for $a+b+c=d$, is
	\begin{align}
	\int_{x_m} \frac{1}{|x_{1m}|^{2a} |x_{2m}|^{2b} |x_{3m}|^{2c}} &= U(a, b, c)\frac{1}{|x_{12}|^{d-2c} |x_{13}|^{d-2b} |x_{23}|^{d-2a}}\\
	\begin{tikzpicture}[baseline={(0,-\MathAxis pt)}]
	    \coordinate (c) at (0,0);
	    \node[bigdot] at (c) {};
	    \coordinate (p2_leg) at (150:0.9cm); 
	    \coordinate (p3_leg) at (30:0.9cm); 
	    \coordinate (p1_leg) at (270:0.9cm);
	    \draw (c) -- (p2_leg) node[midway, below left, inner sep=1pt, xshift=-1pt, linelabel2] {$a$};
	    \draw (c) -- (p3_leg) node[midway, below right, inner sep=1pt, xshift=1pt,linelabel2] {$b$};
	    \draw (c) -- (p1_leg) node[midway, right, inner sep=1pt,linelabel2] {$c$};
	    \node[circle, fill, inner sep=1.5pt] at (p1_leg) {};
	    \node[circle, fill, inner sep=1.5pt] at (p2_leg) {};
	    \node[circle, fill, inner sep=1.5pt] at (p3_leg) {};
	\end{tikzpicture}
	&= U(a,b,c)\times
	\begin{tikzpicture}[baseline={(0,-\MathAxis pt)}]
	    \node[circle, fill, inner sep=1.5pt] (vbeta) at (150:0.9cm) {}; 
	    \node[circle, fill, inner sep=1.5pt] (valpha) at  (30:0.9cm) {};
	    \node[circle, fill, inner sep=1.5pt] (vgamma) at (270:0.9cm) {};
	    \draw (valpha) -- node[midway, above,linelabel2, inner sep=1pt] {$\dotwo-c$}(vbeta) -- node[midway, left, linelabel2, inner sep=3pt] {$\dotwo-b$} (vgamma) -- node[midway, right, linelabel2, inner sep=3pt] {$\dotwo-a$} (valpha);
	\end{tikzpicture} \notag.
	\end{align}
For a Yukawa vertex, again requiring $a+b+c=d$,
	\begin{align}
	\int_{x_m}\frac{1}{|x_{1m}|^{2a}} \frac{\slashed{x}_{2m}}{|x_{2m}|^{2b+1}} \frac{\slashed{x}_{m3}}{|x_{m3}|^{2c+1}} 
	&= U_\mathrm{bff}(a,b,c)
	  \frac{\slashed{x}_{21}}{|x_{21}|^{d-2c+1}} \frac{\slashed{x}_{13}}{|x_{13}|^{d-2b+1}}\frac{1}{|x_{23}|^{d-2a}} 
	  \\
	\begin{tikzpicture}[baseline={(0,-\MathAxis pt)}]
			\coordinate (c) at (0,0);
			\node[bigdot] at (c) {};
			\coordinate (p2_leg) at (150:0.9cm); 
			\coordinate (p3_leg) at (30:0.9cm); 
			\coordinate (p1_leg) at (270:0.9cm);
			\draw (c) -- (p2_leg) node[midway, below left, inner sep=1pt, xshift=-1pt, linelabel2] {$a$};
			\draw[fermion] (c) -- (p3_leg) node[midway, below right, inner sep=1pt, xshift=1pt,linelabel2] {$b$};
			\draw[antifermion] (c) -- (p1_leg) node[midway, right, inner sep=2pt,linelabel2] {$c$};
			\node[circle, fill, inner sep=1.5pt] at (p1_leg) {};
			\node[circle, fill, inner sep=1.5pt] at (p2_leg) {};
			\node[circle, fill, inner sep=1.5pt] at (p3_leg) {};
			\node[left=2pt of p2_leg, point name] {$x_1$};
			\node[right=2pt of p1_leg, point name] {$x_3$};
			\node[right=2pt of p3_leg, point name] {$x_2$};
		    \end{tikzpicture}
	&= U_\mathrm{bff}(a,b,c)\times
		    \begin{tikzpicture}[baseline={(0,-\MathAxis pt)}]
			\node[circle, fill, inner sep=1.5pt] (vbeta) at (150:0.9cm) {}; 
			\node[circle, fill, inner sep=1.5pt] (valpha) at  (30:0.9cm) {};
			\node[circle, fill, inner sep=1.5pt] (vgamma) at (270:0.9cm) {};
			\draw[antifermion] (valpha) -- node[midway, above,linelabel2, inner sep=1pt] {$\dotwo-c$}(vbeta);
			\draw[antifermion] (vbeta) -- node[midway, left, linelabel2, inner sep=3pt] {$\dotwo-b$} (vgamma);
			\draw (vgamma) -- node[midway, right, linelabel2, inner sep=3pt] {$\dotwo-a$} (valpha);
		    \end{tikzpicture}\notag. %
	\end{align}
The order of the gamma matrices remains unchanged, as we simply replace the $m$s by $1$s: $\slashed{x}_{2m}\slashed{x}_{m3} \to\slashed{x}_{21}\slashed{x}_{13}$. %

We note that these equations also hold under a map to the sphere: namely, every appearance of the metric is replaced by the sphere metric, and every appearance of $\abs{x-y}$ is replaced by $s(x,y)$. %

\section{Counterterm calculations: 2PI formalism/skeleton diagrams}\label{app:counterterms}

We use the action \eqref{eq:completeAction} contained in the main text, and provide here a standard computation of the counterterms $Z_{g}$ and $Z_\phi$ given in \eqref{eq:LRcts} and \eqref{eq:SRcts} via the 2PI formalism.
A review of this formalism is available in \cite{Fraser-Taliente:2024hzv}, with a longer one in \cite{Benedetti:2018goh}.
Essentially, it allows us to automatically resum self-energy diagrams, and work only with the full propagators; but at its core it is just a Legendre transform with respect to a two-point source.
This is very well suited to CFTs, where we know exactly the functional form of the propagators.

We now give a simplified sketch of the derivation of the 2PI effective action. 
For simplicity, we can consider a partition function with only a two-point source $k(x,y)$. 
That is, we assume that the relevant couplings are tuned such that the field VEVs are zero, so we don't need a one-point source. 
Thus, defining
\newcommand{\bfPhi}{\Phi}
\begin{equation}\begin{aligned}
	\mathbf{W}[k] \equiv \log \int \cD \varphi \, \exp\left(- S[\varphi] + \int_{x,y} \half \varphi(x) k(x,y) \varphi(y)\right),
\end{aligned}\end{equation} 
the propagator in the presence of the source $k$ is just the functional derivative
\begin{align}
	\mathbf{G}(x, y) &= \expval{\varphi(x)\varphi(y)}_{k} = 2 \fdv{\mathbf{W}}{k(x,y)}.
\end{align}
Then the Legendre transform of $\mathbf{W}[k]$ is the two-particle effective action $\Gamma[\mathbf{G}]$, defined by
\begin{equation}\begin{aligned} \label{eq:GammaAndWLT}
	\Gamma[\mathbf{G}] + \mathbf{W}[k] = \half \int_{x, y} \mathbf{G}(x, y) k(x, y),
\end{aligned}\end{equation}
such that
\begin{equation}
	\fdv{\Gamma[\mathbf{G}]}{\mathbf{G}(x,y)} = \half k(x,y).
\end{equation}
Hence, the true propagator (without sources) is simply whatever solves the two-point Schwinger-Dyson equation
\begin{equation}
	\fdv{\Gamma[\mathbf{G}]}{\mathbf{G}(x,y)} =0.
\end{equation}
The formalism is useful because $\Gamma[\mathbf{G}]$ is extremely easy to construct, as we will show shortly: we just sum up all the diagrams that are two-particle-reducible (i.e. the diagrams that do not contain a self-energy insertion on any of their lines). 
The two-particle effective action can be functionally differentiated to give the full two-point Schwinger-Dyson equations.
This process is trivially generalized to multiple fields by adding indices to $\varphi$ and $k$; then the skeleton diagram technique of \cite{Vasiliev:1981dg,Vasiliev:1981yc,Vasiliev:1982dc}, reviewed in \cite{Gracey:2018ame}, is exactly identical to just self-consistently solving these Schwinger-Dyson equations.

\subsection{Implementing the 2PI formalism}

We can ignore here the fact that $\sigma$ has an imaginary contour, as the extra minus sign that would arise in $\log(-\Sigma^{-1})$ that was noted in \cref{app:imagGaussian} just gives a constant and can be dropped.
Tuning the relevant couplings to set the field VEVs to zero ($\expval{\phi_i} =0$, $\expval{\sigma}=0$), we find
\begin{equation}
\Gamma_{\mathrm{2PI}}[G,\Sigma] = \half \tr \log G^{-1} + \half \tr G_0^{-1} G +\half \tr \log \Sigma^{-1} + \half \tr \Sigma_0^{-1} \Sigma + \Gamma_2[G,\Sigma],
\end{equation}
where we have also assumed that the $\gO(N)$ is unbroken, meaning that $\expval{\phi_i \sigma}=0$.
$\Gamma_2$ is minus the sum of the two-particle irreducible graphs, and $G_0$ and $\Sigma_0$ are the quadratic terms in the original action\footnote{We have ignored here the contact terms in \eqref{eq:DContactTerms} which are required only to make the inversions $C^{-1}$ and $K^{-1}$ rigorous, and could be easily restored. 
In any case, they can be thrown away here, as we are considering the two-point functions for separated points only.},
\begin{align}
 G_0^{-1} &= \frac{Z_\phi}{C_{\phi,0}} \frac{1}{\cN_{\tfrac{d-s}{2}}} \frac{1}{s(x,y)^{d+s}},\\
\Sigma_0^{-1} &= K_{\delta}^{-1} - \mu^{2\delta} K^{-1} = \frac{1}{C_{\sigma,0}} \left(\frac{s(x,y)^{-2(d-s+\delta)}}{\cN_b(s-\delta)}  -\frac{\mu^{2\delta} s(x,y)^{-2(d-s)}}{\cN_b(s)} \right), \label{eq:Sigma0def}
\end{align}
with the bare normalizations given by \eqref{eq:BareONtwoPointNorms}.
As an aside, in the original theory \eqref{eq:ZONwithlam0} regulated with a hard cutoff, we would be using $\Sigma_0^{-1}(x,y) = -\frac{1}{\lambda_0} \delta(x-y)$.
Variation of this effective action yields the equations of motion
\begin{equation}\label{eq:twoPointeoms}
G^{-1} - G_0^{-1} - 2 \fdv{\Gamma_2}{G} = 0, \quad
\Sigma^{-1} - \Sigma_0^{-1} - 2 \fdv{\Gamma_2}{\Sigma} = 0.
\end{equation}

At order $N^0$, only the melon contributes to $\Gamma_2$; at order $1/N$, we have only the tetrahedron and the prism. 
The pillow diagram is two-particle-reducible, being a melon diagram with a self-energy insertion, and so we do not need to consider it.
Thus, using the vertex rule $-Z_\phi g_0/\sqrt{N}$, we find%
\begin{align*}
-\Gamma_\mathrm{2} &= \quad \begin{tikzpicture}[scale=1.0, line cap=round, line join=round,baseline={(0,-\MathAxis pt)}]
	\begin{scope} 
			\coordinate (L) at (-0.7,0); \coordinate (R) at (0.7,0);	
			\draw[sigma] (L) to[bend left=55] (R);
			\draw[phi] (L) to[bend right=55] (R);
			\draw[phi] (R) to[bend right=0] (L);
		\node[below] at (0,-0.75) {$I_\mathrm{mel}: \frac{1}{4} $};
		\node[right] at (1.0,0) {$+$};
	\end{scope}
	\begin{scope}[xshift=2.5cm]
	    \coordinate (TL) at (-0.5, 0.5);
	    \coordinate (BL) at (-0.5,-0.5); 
	    \coordinate (TR) at ( 0.5, 0.5);
	    \coordinate (BR) at ( 0.5,-0.5);
	  \draw[phi] (BL) rectangle (TR); %
	  \draw[sigma] (BL) -- (TR);
	  \draw[sigma] (TL) -- (BR);
	  \node[below] at (0,-0.75) {$I_{\mathrm{tet}}: \frac{1}{8}$};
	\node[right] at (0.75,0) {$+$};
	\end{scope} %
	\begin{scope}[xshift=5.1cm]
	  \coordinate (A1) at (-0.5,0);
	  \coordinate (A2) at (-1,0.5);
	  \coordinate (A3) at (-1,-0.5);
	  \coordinate (B1) at (0.5,0);
	  \coordinate (B2) at (1,0.5);
	  \coordinate (B3) at (1,-0.5);
	  \draw[phi] (A1) -- (A2) -- (A3) -- (A1);
	  \draw[phi] (B1) -- (B2) -- (B3) -- (B1);
	  \draw[sigma] (A1) -- (B1);
	  \draw[sigma] (A2) -- (B2);
	  \draw[sigma] (A3) -- (B3);
	  \node[below] at (0,-0.75) {$I_{\mathrm{pr}}: \frac{1}{12}$};
	\end{scope}
	\end{tikzpicture}\; \, \, \, +\,\,\,  O(1/N^2),\\
& I_{\mathrm{melon}}= \frac{1}{4} \frac{Z_\phi^2 g_0^2}{N} \int_{x,y} G_{ij}(x,y) G_{ji}(y,x) \Sigma(x,y)\\
& I_{\mathrm{tet}}= \frac{1}{8} \frac{Z_\phi^4 g_0^4}{N^2} \int_{x_{1,2,3,4}} G_{ij}(x_1,x_2) G_{jk} (x_2,x_3) G_{kl}(x_3,x_4) G_{lm}(x_4,x_1) \Sigma(x_1,x_3) \Sigma(x_2,x_4)\\
& I_{\mathrm{pr}}=\frac{1}{12} \frac{Z_\phi^6 g_0^6}{N^3} \int_{x_{1,2,3,4,5,6}}(G_{ij}(x_1,x_2) G_{jk}(x_2,x_3) G_{kl}(x_3,x_1))  
\\
& \quad \times\Sigma(x_1,x_4) \Sigma(x_2,x_5) \Sigma(x_3,x_6)
(G_{i'j'}(x_4,x_5) G_{j'k'}(x_5,x_6) G_{k'l'}(x_6,x_4)).
\end{align*}

We note that on shell (i.e. with the solutions for $G$ and $\Sigma$ plugged in), we have that $\Gamma_\mathrm{2PI}\rvert_\star = -\log Z_{\gO(N)} = F_{\gO(N)}$ by construction.
Thus, an alternative way of calculating the free energy is just to directly evaluate this at the critical point (this was done in \cite{Fraser-Taliente:2024hzv}). 
Although slightly simpler in practice, it is conceptually clearer to stick to the standard Feynman-diagrammatic approach to the evaluation of the free energy, rather than Legendre transforming it as we have done here.

\subsection{The two-point Schwinger-Dyson equations}

The two-point Schwinger-Dyson equations follow by plugging $\Gamma_2$ in to \eqref{eq:twoPointeoms}.
Assuming no $\gO(N)$-symmetry-breaking, so $G_{ij}(x,y) = G(x,y) \delta_{ij}$, we ansatz the conformal solution
\begin{equation}
\frac{\expval{\phi_i(x) \phi_j(y)}}{\delta_{ij}} = G(x,y) = \frac{C_\phi}{s(x,y)^{2\Delta_\phi}}, \quad \expval{\sigma(x) \sigma(y)} = \Sigma(x,y) = \frac{C_\sigma}{s(x,y)^{2\Delta_\sigma}}.
\end{equation}
A derivative with respect to $\Sigma$ of $\Gamma_\mathrm{2PI}$ clearly does not modify the $N$-scaling of the terms; a derivative with respect to $G_{ij}$ does, but it breaks open a single  $\phi$ loop in each term, and therefore uniformly reduces the $N$-scaling of each term by $1$; hence no other diagrams will start contributing here. 
Henceforth, we drop all references to the flavour indices, as the $N$-scaling of each term is now explicit.
Writing in diagrammatic form, \eqref{eq:twoPointeoms} becomes
\pgfmathsetmacro{\stublen}{0.3}
\begin{subequations}\label{eq:twoPointEomsDiags}
\begin{align}
	G^{-1} - G_0^{-1} &+ \frac{Z_\phi^2 g_0^2}{N} 
	\begin{tikzpicture}[baseline={(0,-\MathAxis pt)}]
		\coordinate (L) at (0,0); \coordinate (R) at (1.2,0);
		\draw[phi] (-\stublen,0) -- (L);
		\draw[phi] (R) -- (1.2+\stublen,0);
		\draw[sigma] (L) to[bend left=55] (R);
		\draw[phi] (L) to[bend right=55] (R);
	\end{tikzpicture}
	+ \frac{Z_\phi^4 g_0^4}{N^2} 
	\begin{tikzpicture}[baseline={(0,-\MathAxis pt)}]
		\coordinate (L) at (0.3,0); \coordinate (R) at (1.5,0);
		\coordinate (T) at (0.9,0.6); \coordinate (B) at (0.9,-0.6);
		\draw[phi] (0.3-\stublen,0) -- (L);
		\draw[phi] (L) to (T);
		\draw[phi] (T) to (B);
		\draw[phi] (B) to (R);
		\draw[phi] (R) -- (1.5+\stublen,0);
		\draw[sigma] (L) to (B);
		\draw[sigma] (T) to (R);
	\end{tikzpicture}
	+ \frac{Z_\phi^6 g_0^6}{N^2} \begin{tikzpicture}[baseline={(0,-\MathAxis pt)}]
		\coordinate (BL) at (0,0);
		\coordinate (BR) at (1.5,0);
		\coordinate (BM) at ($(BL)!0.5!(BR)$);
		\coordinate (TL) at (0,1);
		\coordinate (TR) at (1.5,1);
		\coordinate (V) at ($(TL)!0.5!(TR) - (0,0.6)$);
		\draw[phi] (-\stublen,0) -- (BL) -- (BR) -- (1.5+\stublen,0);
		\draw[sigma] (BL) -- (TL);
		\draw[sigma] (BR) -- (TR);
		\draw[phi] (TL) -- (TR) -- (V) -- cycle;
		\draw[sigma] (V) -- (BM);
		\end{tikzpicture}
	 = 0, \\
	\Sigma^{-1} - \Sigma_0^{-1} &+ \frac{Z_\phi^2 g_0^2}{2} 
	\begin{tikzpicture}[baseline={(0,-\MathAxis pt)}]
		\coordinate (L) at (0,0); \coordinate (R) at (1.2,0);
		\draw[sigma] (-\stublen,0) -- (L);
		\draw[sigma] (R) -- (1.2+\stublen,0);
		\draw[phi] (L) to[bend left=55] (R);
		\draw[phi] (L) to[bend right=55] (R);
	\end{tikzpicture} 
	+ \frac{Z_\phi^4 g_0^4}{2N} 
	\begin{tikzpicture}[baseline={(0,-\MathAxis pt)}]
		\coordinate (L) at (0.3,0); \coordinate (R) at (1.5,0);
		\coordinate (T) at (0.9,0.6); \coordinate (B) at (0.9,-0.6);
		\draw[sigma] (-\stublen,0) -- (L);
		\draw[sigma] (R) -- (1.5+\stublen,0);
		\draw[phi] (L) -- (T) -- (R) -- (B) -- (L);
		\draw[sigma] (T) -- (B);
	\end{tikzpicture}
	+\frac{Z_\phi^6 g_0^6}{2N} 
	\begin{tikzpicture}[baseline={(0,-\MathAxis pt)}]
		\coordinate (l1) at (-0.8,0.5);  \coordinate (l2) at (-1.3,0); \coordinate (l3) at (-0.8,-0.5);
		\coordinate (r1) at (0.8,0.5);   \coordinate (r2) at (1.3,0);  \coordinate (r3) at (0.8,-0.5);
		\draw[sigma] (-1.3-\stublen,0) -- (l2);
		\draw[sigma] (r2) -- (1.3+\stublen,0);
		\draw[phi] (l1) -- (l2) -- (l3) -- cycle;
		\draw[phi] (r1) -- (r2) -- (r3) -- cycle;
		\draw[sigma] (l1) -- (r1);
		\draw[sigma] (l3) -- (r3);
		\end{tikzpicture} = 0,\label{eq:CsigmaEquation}
	\end{align}
\end{subequations}
which we evaluate using 
\begin{equation}
\Delta_\sigma = s- \delta + \gamma_\sigma,\quad \Delta_\phi =\frac{d-s}{2} + \gamma_\phi,
\end{equation}
and then take $N\to\infty$ followed by $\delta \to 0$. 

We evaluate these diagrams in two different ways.
In \cref{sec:applyingUniqueness}, we work in flat space ($s(x,y) = \abs{x-y}$), where
we can use uniqueness to directly calculate the subleading $\sigma$ diagrams by hand.
In \cref{app:sphereInts} we work on the sphere, where the diagrams can be evaluated using Mellin-Barnes integral representations and \texttt{MBresolve}.
Naturally, the same results are obtained in each case, though the method of uniqueness is less powerful (it no longer works for the calculation of $\hat{\gamma}_{\phi,2}$).
We can then use these results to solve the two-point functions.

\subsection{Solving the two-point functions for the long-range models, \texorpdfstring{$s\neq 2k$}{s≠2k}}\label{app:flatSpace2pt}

\subsubsection{Leading order in \NPDF for \texorpdfstring{$s\neq 2k$}{s≠2k}}
We begin with the first corrections to the propagators. We take $g_0 = \mu^\delta Z_g g$, and freely set $X= \mu s(x,y)$ and $\mu=1$ to reduce clutter.
Stripping off the $\delta_{ij}$s, we find
\begin{align}
G^{-1} - G_0^{-1} &= -2\fdv{I_{\mathrm{mel}}}{G}= - \, \frac{Z_\phi^2 g_0^2}{N}\begin{tikzpicture}[baseline={(0,-\MathAxis pt)}]
	\coordinate (L) at (0,0); \coordinate (R) at (1.2,0);
	\draw[phi] (-\stublen,0) -- (L);
	\draw[phi] (R) -- (1.2+\stublen,0);
	\draw[sigma] (L) to[bend left=55] (R);
	\draw[phi] (L) to[bend right=55] (R);
\end{tikzpicture} = -\frac{1}{N} \frac{Z_\phi^2 g_0^2 C_\sigma C_\phi}{X^{2(s- \delta + \gamma_\sigma + \frac{d-s}{2} + \gamma_\phi)}} + O(1/N^2)
\\
\Sigma^{-1} - \Sigma_0^{-1} &= -2\fdv{I_{\mathrm{mel}}}{\Sigma} = - \frac{Z_\phi^2 g_0^2}{2} \, \begin{tikzpicture}[baseline={(0,-\MathAxis pt)}]
	\coordinate (L) at (0,0); \coordinate (R) at (1.2,0);
	\draw[sigma] (-\stublen,0) -- (L);
	\draw[sigma] (R) -- (1.2+\stublen,0);
	\draw[phi] (L) to[bend left=55] (R);
	\draw[phi] (L) to[bend right=55] (R);
\end{tikzpicture} 
  = -\half \frac{Z_\phi^2 g_0^2 C_\phi^2}{X^{2(d-s + 2\gamma_\phi)}} + O(1/N).
\end{align}
It is convenient to define $z = C_\sigma C_\phi^2 = z_0 + \frac{z_1}{N} + O(1/N^2)$, and self-consistent at this order to take $Z_\phi Z_g = 1+ O(1/N)$ (recall that $g=1$). 
Substituting for $G$ and $\Sigma$, we find
\begin{align}
\frac{1}{\cN_b(\tfrac{d-s}{2} + \gamma_\phi)} - \frac{C_\phi Z_\phi}{C_{\phi,0}} \frac{X^{-2\gamma_\phi}}{\cN_b(\tfrac{d-s}{2})} + \frac{z_0}{N}X^{-2(\gamma_\sigma +2\gamma_\phi-\delta)} + O(1/N^2)&=0 \label{eq:phiEquationLO}
\\
\frac{1}{\cN_b(s-\delta+\gamma_\sigma)} - \frac{C_\sigma}{C_{\sigma,0}} \left(\frac{X^{-2\gamma_\sigma}}{\cN_b(s-\delta)}  -\frac{X^{2(\delta-\gamma_\sigma)}}{\cN_b(s)} \right)+\frac{z_0}{2} X^{-2(\gamma_\sigma + 2 \gamma_\phi - \delta)} + O(1/N) &=0.
\end{align}
Note that $1/\cN(\dotwo -k)=0$, which complicates the analysis; we therefore delay discussion of $s=2k$ to \cref{sec:SRdiags}, and
choose to focus first on the generic and simpler case $s \neq 2k$. 
We then expand around $\gamma_{\sigma/\phi}\to \gamma_{\sigma/\phi,1}/N + \cdots$ and $C_{\sigma/\phi} \to C_{\sigma/\phi,0}(1+\normCor_{\sigma/\phi,1}/N + \cdots)$, and therefore also $z \to z_0 + z_1/N + \cdots$.
First taking $N \to \infty$ and then taking $\delta\to 0$ yields
\begin{equation}\begin{aligned}
0 &= \tfrac{1}{\cN_b(\tfrac{d-s}{2})} \Bigg[1-\tfrac{C_{\phi,0} Z_\phi}{C_{\phi,0}}\left(1 + \tfrac{\normCor_{\phi,1} - 2 \gamma_\phi \log(\mu X)}{N} \right) + \tfrac{z_0 \cN_b(\tfrac{d-s}{2})- \gamma_{\phi,1} \tfrac{\cN_b'(\tfrac{d-s}{2})}{\cN_b(\tfrac{d-s}{2})}}{N} +O(1/N^2)\Bigg]\\
0 &= \frac{2}{\cN_b(s)} +z_0+ O(1/N).
\end{aligned}\end{equation} 

Solving this order by order in $N$, we can take $Z_\phi =1$ finite and
\begin{equation}
	z_0 = \frac{-2}{\cN_b(s)}
\end{equation}
so indeed $\Delta_\phi$ is unmodified,
\begin{equation}
\gamma_{\phi,1} =0, \quad \normCor_{\phi,1} = z_0 \, \cN_b(\tfrac{d-s}{2}).
\end{equation}

\subsubsection{Subleading order in \NPDF for \texorpdfstring{$s\neq 2k$}{s≠2k}}

Let us now consider the next order.
Both on the sphere, $X=\mu s(x,y)$ and in flat space, $X= \mu \abs{x-y}$, the subleading diagrams in \eqref{eq:twoPointEomsDiags} can be calculated to order $1/N$,  %
\begin{subequations}\label{eq:2ptEvaluations}
\begin{align}
2\fdv{I_{\mathrm{tet}}}{G} &= \frac{Z_\phi^4 g_0^4}{N^2} \frac{C_\sigma^2 C_\phi^3}{X^{d+s-4\delta}} \left(\frac{K_1}{\delta} + \Sigma_1' + O(\delta)\right)\\
2\fdv{I_{\mathrm{pr}}}{G} &= \frac{Z_\phi^6 g_0^6}{N^2} \frac{C_\sigma^3 C_\phi^5}{X^{d+s-6\delta}} \left(\frac{K_2}{\delta} + \Sigma_2' + O(\delta)\right)\\
2\fdv{I_{\mathrm{tet}}}{\Sigma} %
& = \frac{Z_\phi^4 g_0^4}{2N} \frac{C_\sigma C_\phi^4}{X^{2 (d-s)-2\delta}} \left(\frac{K_1}{\delta} + \Pi_1' + O(\delta)\right) \label{eq:dItetdSigma}\\ 
2\fdv{I_{\mathrm{pr}}}{\Sigma} %
& = \frac{Z_\phi^6 g_0^6}{2N} \frac{C_\sigma^2 C_\phi^6}{X^{2(d-s)-4 \delta}}  \left(\frac{K_2}{\delta} + \Pi_2' + O(\delta)\right). \label{eq:dIprdSigma}
\end{align}
\end{subequations}
The exponents of $X$ here should contain $\gamma_{\phi,\sigma}$s, but they have been dropped as they contribute only at the next order in $N$.
We use the conventions of \cite{Vasiliev:1981dg}, 
\begin{equation}\begin{aligned}
K_1 = \frac{2 \pi ^d}{\Gamma \left(\tfrac{d}{2}\right)} A(s) A\left(\tfrac{d-s}{2}\right)^2, \quad K_2 = \frac{\pi ^{2 d}}{\Gamma \left(\tfrac{d}{2}\right)} A(s)^3 A\left(\tfrac{d}{2} + \tfrac{d-s}{2} -s\right) A\left(\tfrac{d-s}{2}\right)^3
\end{aligned}\end{equation}
\begin{equation}\begin{aligned}
\Sigma_1' &= \frac{K_1}{2} (B(s) -B(\tfrac{d-s}{2})),\quad &\Sigma_2'& = 2 K_2  (B(s)-B(\tfrac{d-s}{2})) \\
\Pi_1' &= K_1 (B(s)-B(\tfrac{d-s}{2})),\quad &\Pi_2'& = K_2 (4 B(s) - 3 B(\tfrac{d-s}{2}) - B(\tfrac{d}{2} + \tfrac{d-s}{2} -s)),
\end{aligned}\end{equation}
and 
\begin{equation}
p(\Delta) \equiv \frac{1}{\cN_b(\Delta)} = \frac{\Gamma (d+1)}{\pi^{d+1}} \Ft_b'(\Delta)=-\sin(\tfrac{\pi  d}{2}) \frac{\Gamma (d+1)}{\pi^{d+1}} F_b'(\Delta).
\end{equation}
Now, recall that $Z_g$ is divergent at order $1/N$, while $Z_\phi$ we can take to be finite and equal to $1$,
\begin{equation}
    Z_\phi = 1, \quad \text{ for generic } s \neq 2k.
\end{equation}
Thus, in minimal subtraction, we have
\begin{equation}\begin{aligned}
Z_\phi^2 g_0^2 = \mu^{2\delta} \left(1+ \frac{1}{N} \frac{2Z_{g,1}}{\delta} + O(1/N^2)\right)
\end{aligned}\end{equation}
Putting this together for the $\sigma$ equation, %
\begin{equation}\begin{aligned}
0&= p(s-\delta+ \gamma_\sigma) - \frac{C_\sigma}{C_{\sigma,0}} \left(p(s-\delta) X^{-2\gamma_\sigma}  -p(s) X^{2(\delta-\gamma_\sigma)} \right) \\
&+ \frac{z}{2}\left(1 + \frac{2Z_{g,1}}{N \delta}\right) X^{2(\delta-\gamma_\sigma-2\gamma_\phi)} \\
& + \frac{z^2}{2N} X^{4 \delta}  \left(\frac{K_1}{\delta} + \Pi_1'\right)\\
& +  \frac{z^3}{2N} X^{6\delta} \left(\frac{K_2}{\delta} + \Pi_2'\right)+  O(1/N^2),
\end{aligned} \end{equation}
We now expand $z$ and $\gamma_\sigma$ in $N$ and demand that each order in $1/N$ is zero to $O(\delta)$.
In minimal subtraction we can therefore ignore the $\Sigma_0^{-1}$ contribution as it is $O(\delta)$, and so
\begin{equation}\begin{aligned}
z= z_0 + \frac{z_1}{N} + O(1/N^2), \quad \gamma_\sigma = \frac{\gamma_{\sigma,1}}{N} + O(1/N^2),
\end{aligned}\end{equation}
we find at orders $1$ and $1/(N \delta)$
\begin{equation}
p(s) + \frac{z_0}{2} =0, \quad \frac{z_0}{2 N \delta} \left(2Z_{g,1} + z_0 K_1 + z_0^2 K_2\right)=0.
\end{equation}
Looking at the coefficients of $\log X$,
\begin{equation}\begin{aligned}
\frac{z_0}{2N} \left(-2\gamma_\sigma - 4 \gamma_\phi +  z_0 (2 K_1) + z_0^2 (4 K_2)\right)  \log X =0.
\end{aligned}\end{equation}
Since this equation must hold for all values of $X$, we conclude that 
\begin{equation}\begin{aligned}
z_0 &= -2 p(s) = -2/\cN_b(s)\\
\Delta_\sigma + 2\Delta_\phi -d &= \gamma_\sigma + 2 \gamma_\phi = -\varkappa_1 = z_0 K_1 + 2 z_0^2 K_2 = \Gamma_\mathrm{tet} + \Gamma_{\mathrm{pr}},\\
2Z_{g,1} &= -z_0 K_1 -z_0^2 K_2 = -\Gamma_\mathrm{tet} - \half \Gamma_{\mathrm{pr}},
\end{aligned}\end{equation}
Continuing on, we find 
\begin{equation}\begin{aligned}\label{eq:LRzeq1}
z_1 = -z_0^2 \Pi_1' - z_0^3 \Pi_2' +2 \varkappa_1 p'(s).
\end{aligned}\end{equation}
The $\phi$ equation is automatically finite at this order thanks to the $Z_g$ counterterm, so
\begin{align}
	\gamma_{\phi,2} &= 0\\
	\normCor_{\phi,2} &= \frac{1}{p\left(\frac{d-s}{2}\right)}\left(-2 (\Gamma_\mathrm{tet}+\Gamma_{\mathrm{pr}}) p'(s)+z_0^2(\Sigma_1' -\Pi_1') +z_0^3 (\Sigma_2' - \Pi_2')\right),
\end{align}
confirming the lack of anomalous dimension in the long-range model.

The renormalization condition here has been input by requiring that $z$ be finite at each order in $1/N$, and it is easy to confirm by taking $C_\sigma = z/C_\phi^2$ that the above indeed matches the data in \eqref{sec:genSdata}.

\subsection{Solving the two-point functions for the short-range models, \texorpdfstring{$s=2k$}{s=2k}}\label{sec:SRdiags}

If computing directly at the short-range points (i.e. with $C^{-1} = \Box^k$), the difference is that $1/\cN_b(\dotwo -k)=0$. 
This only affects the $\phi$ equation of motion. 
The $\sigma$ equation is the same, and hence to the given orders in $N$, some of the above results still hold, but with $s=2k$ and, crucially, $\gamma_\phi \neq 0,Z_\phi\neq 1$:
\begin{equation}\begin{aligned}\label{eq:generalSStillHold}
z_0 &= -2/\cN_b(s) \rvert_{s=2k}, \quad &z_1& = \eqref{eq:LRzeq1} \rvert_{s=2k}\\
\gamma_\sigma + 2 \gamma_\phi &= %
 \frac{\Gamma_\mathrm{tet} + \Gamma_\mathrm{pr}}{N}\Big\rvert_{s=2k}, \quad & Z_\phi Z_g & =  1- \frac{\Gamma_\mathrm{tet} + \half \Gamma_\mathrm{pr}}{2N} \Big\rvert_{s=2k}.
\end{aligned}\end{equation}
The only differences are that now we also need to renormalize the $\phi$ propagator, since for integer $k$ it is local and therefore \textit{can receive corrections}. %
Our use of DREG here implies that, as usual, we have implicitly put a Lorentz-invariant counterterm for each of the possible relevant short-range propagators $\phi \Box^{0\le m < k}$, to cancel all power law divergences (and all the irrelevant ones too, in fact -- but they don't matter). 

Considering the $\phi$ equation of motion, $1/\cN_b(\dotwo -k) =0$ means that $Z_\phi$ drops out of the equation.
This is because the free conformal propagator has no inverse, because it is in an exceptional irrep of the conformal group \cite{Karateev:2018oml,Dobrev:1977qv}.
Instead, we avoid inverting the conformal propagators, and consider the geometric series version of the $\phi$ SDE \eqref{eq:twoPointeoms},
\begin{equation}
G = G_0 + G_0\star \Pi \star G %
, \quad \Pi_{xy} = -2 \fdv{\Gamma_2}{G_{xy}},
\end{equation}
where the convolution just means matrix multiplication with respect to the indices $x,y$.
We could have used this equation for $\sigma$ as well, but it would have been messier.
Using the propagator merging relations \eqref{eq:propMerge}, we find
\begin{equation}
	G = \frac{1}{Z_\phi}\frac{C_{\phi,0}}{X^{d-2k}} \left(1+ \frac{Z_\phi^2 z_0}{N} X^{2\delta} U\left(\dotwo-k,\dotwo-\delta ,\delta +k\right) U\left(\dotwo-k,\dotwo-\delta +k,\delta \right) +O(1/N^2) \right).
\end{equation}
To minimally subtract the divergence we choose
\begin{equation}
	Z_\phi = 1 - \frac{\hat{\gamma}_{\phi,1}}{N \delta} + O(1/N^2),
\end{equation}
where indeed the new anomalous dimension for $\phi$ is equal to $\hat{\gamma}_{\phi,1}/N +O(1/N^2)$, with
\begin{equation}
\hat{\gamma}_{\phi,1} = \frac{2F_b'(2k)}{F_b''(\tfrac{d}{2}-k)}.
\end{equation}
Through \eqref{eq:generalSStillHold}, it is of course necessary to also modify $Z_g$.
Putting everything together, we find the results of \cref{sec:SRmodels}, including $\hat{\gamma}_{\phi,2}$ which agrees with the $k=1$ vector model results in \cite{Vasiliev:1981dg, Goykhman:2019kcj}\footnote{For explicit matching with the results of Vasil'ev et al.: $\mu= d/2$, $\alpha =\Delta_\phi=\mu -1+\eta/2$, $\beta= \Delta_\sigma=2-\eta-\varkappa$, $\gamma_\phi = \eta/2$, and $\gamma_\sigma = -\eta - \varkappa$; thus $ \Delta_\sigma+ 2\Delta_\phi -d = -\varkappa$. %
Compare also \cite{Henriksson:2022rnm}, which contains a summary of $\gO(N)$ model data.
}.
The remaining short-range normalization data can then be obtained from
\begin{align}
	\hat{\normCor}_{\phi,1} &= \hat{\gamma}_{\phi,1} \Bigg(\frac{F_b'''(\dotwo -k)}{2 F_b''(\dotwo -k)}\Bigg) = \hat{\gamma}_{\phi,1}  \left(\psi ^{(0)}\left(\dotwo -k\right)-\psi ^{(0)}\left(\dotwo + k\right)-\tfrac{1}{k} \right) %
	\end{align}
and \eqref{eq:generalSStillHold}. %

\subsection{A different scheme: non-minimal subtraction} \label{app:nonminimal}

We could instead unit-normalize the propagators\footnote{Unit-normalization is not always the natural choice, as it can lead to spurious divergences of OPE coefficients \cite{Douglas:2010ic}.}; if we do so, then the long-range model counterterms at finite $\delta$ would be identical to those of the short-range model. 
This would be necessary if we wanted the canonically normalized OPE coefficients, as in \cite{Goykhman:2019kcj, Chai:2021wac}.

In that case, we would need to define the following while still at finite $\delta$:
\begin{equation}
Z_\phi' = Z_\phi^\text{MS} C_\phi, \quad Z_g' = Z_g ^\text{MS} C_\sigma^\half,
\end{equation}
with $C_{\phi/\sigma}$ as defined by \eqref{eq:ONLRSDs}.
However, the short-range and long-range theories would be distinguished by the difference that the limit 
\begin{equation}
	\lim_{\delta \to 0} \lim_{s \to 2k}
\end{equation}
is not the same as 
\begin{equation}
	\lim_{s \to 2k} \lim_{\delta \to 0},
\end{equation}
which is clear from consideration of $\normCor_\mathrm{\phi/\psi,1,\mathrm{pil}}$.
This shows that the difference in counterterms for the SR \eqref{eq:SRcts} and LR \eqref{eq:LRcts} cases is only due to our choice of minimal subtraction.
Defining the unit-normalized arbitrary-$s$ theory at finite $\delta$ to be $T_\delta(s)$, the SR/LR crossover is then expressed by
\begin{equation}
	\lim_{s \to 2k-2\hat{\gamma}_\phi} \lim_{\delta \to 0} T_\delta(s) = \lim_{\delta \to 0} \lim_{s \to 2k} T_\delta(s).
\end{equation}
\subsection{Non-2PI formalism}\label{app:non2PI}

If we did not work in the 2PI formalism, the subtlety mentioned below \eqref{eq:completeAction} comes into play. 
If we approach \eqref{eq:completeAction} as a QFT, using $1/N$ as the expansion parameter, and start writing down all possible diagrams, the two possible $O(N^0)$ self-energy insertions on any $\sigma$ line are
\begin{equation}
\begin{tikzpicture}[scale=1.0, line cap=round, line join=round,baseline={(0,-\MathAxis pt)}]
	\begin{scope}[xshift=-4cm]
		\coordinate (L) at (0,0); \coordinate (R) at (1.2,0);
	\draw[sigma] (L) -- (R);

	    \coordinate (X) at (0.6,0);
	    \draw[thick] ($(X)+(-0.15,-0.15)$) -- ($(X)+(0.15,0.15)$);
	    \draw[thick] ($(X)+(-0.15,0.15)$) -- ($(X)+(0.15,-0.15)$);
	    
	    \node[below] at (0.5,-0.75) {$\mu^{2\delta} K^{-1}(x,y)$};
	\end{scope}
	\coordinate (L) at (0,0); \coordinate (R) at (1.2,0);
	\draw[sigma] (-0.7,0) -- (L);
	\draw[sigma] (R) -- (1.2+0.7,0);
	\draw[phi] (L) to[bend left=55] (R);
	\draw[phi] (L) to[bend right=55] (R);
	\begin{scope}
		\coordinate (L) at (0,0); \coordinate (R) at (1.2,0);
	\draw[sigma] (-0.7,0) -- (L);
	\draw[sigma] (R) -- (1.2+0.7,0);
	\draw[phi] (L) to[bend left=55] (R);
	\draw[phi] (L) to[bend right=55] (R);
	    \node[below] at (0.5,-0.65) {$\half \times \frac{N Z_\phi^2 g_0^2}{N} C(x,y)^2$};
	\end{scope}
\end{tikzpicture}.
\end{equation}
Since each contribute at order $N^0$ (we have included the symmetry factor $\half$ on the $\phi$ loop), we naively would need to resum them to all orders. 

However, take any $\sigma$ line in any diagram $D$, i.e.,
\begin{equation}
D=\int_{w,z} f(w) K_\delta(w,z) g(z)
\end{equation}
and insert each of these separately on that line. 
The sum of the two diagrams that arise is
\begin{equation}
	\int_{w,x,y,z} f(w) K_\delta(w,x) \Bigg(\underbrace{\mu^{2\delta} K^{-1}(x,y)+ \half \times \frac{N Z_\phi^2 g_0^2}{N} C(x,y)^2}_{=\mu^{2\delta}(1- Z_g^2) K^{-1}(x,y)}\Bigg) K_\delta(y,z)
\end{equation}
which is order $1/N$.
So, we only need to consider a finite number of insertions of $\mu^{2\delta}(1- Z_g^2) K^{-1}(x,y)\propto 1/N$ -- as many as are required at the given order in $N$ \cite{Vasiliev:1975mq}.
This is all quite fiddly; this is why the 2PI formalism is the best approach here, as any diagram including such an insertion as this is manifestly 2PI, and so need not be considered.

\section{Regulated action construction for Gross-Neveu} \label{app:GNsetup}

We now perform the analysis of \cref{sec:ONsetup} for the GN model.
As before, we introduce a pair of real auxiliary fields:
\begin{enumerate}
	\item $\sigma$, which has an imaginary contour of integration;
	\item $\phiphi(x)$, which we set to equal $ \tfrac{1}{\sqrt{N}}\psibar_i(x)\psi_i(x)$.
\end{enumerate}
Then we insert
\begin{equation} \label{eq:GsigDeltaFunctionGN}
1 = \int \Dd{\phiphi} \,\delta(\phiphi- \tfrac{1}{\sqrt{N}}\psibar_i \psi_i) =  \int \Dd{\phiphi} \DdImag{\sigma} \, e^{+\int_x \sigma(\phiphi-\frac{1}{\sqrt{N}} \psibar_i \psi_i)}
\end{equation}
into the partition function \eqref{eq:GNaction}. 
For convenience, we have switched the sign of $\sigma$ in the delta function implementation compared to the same identity in the $\gO(N)$ case; this is to ensure that we end up with $\log(C^{-1} - \sigma \mathbb{I}/\sqrt{N})$ in the $\sigma$ effective action, which eliminates extra minus signs in the Feynman rules in both the $\gO(N)$ and GN models.
We find
\begin{align}
	Z_{\mathrm{GN}} &= \int  \Dd{\psi} \Dd{\psibar} \DdImag{\sigma}  \Dd{\phiphi}\, \exp \left(\int_x \psibar_i \left(C^{-1} -\frac{1}{\sqrt{N}} \sigma \right)\psi_i - \half \lambda_0 \phiphi^2 + \sigma \phiphi \right)\\
	&= \int \DdImag{\sigma} \, \exp \left( + \tr \log \left(C^{-1} -\frac{1}{\sqrt{N}} \sigma \mathbb{I} \right) +\int_x \frac{1}{2\lambda_0} \sigma^2\right) \label{eq:criticalGNnoreg},
\end{align}
where we again used $Z_\phiphi =1$.
This action now matches \eqref{eq:GNactionwithSig}.
In the IR, we drop the UV kinetic term for $\sigma$\footnote{Here this is simply a delta function. However, for the general Gross-Neveu-Yukawa theory that is marginal in $d=2k-\epsilon$ ($k\in \Z$), there are other terms that must be dropped, because $S_\text{GNY} = \int_x -\bar\psi_i \left[ C^{-1}- g_0 \sigma\right] \psi^i + \half  \sigma(-\partial^2)^{k-1}\sigma + \text{counterterms}$ \cite{Tarnopolsky:2016vvd}; in all cases, we find in the IR an action identical to $\lim_{\lambda_0 \to \infty}\eqref{eq:criticalGNnoreg}$.}. 
As before, if we used a hard cutoff, we could choose to work with this action at the critical point ($\lambda_0 \to \infty$). This is too hard, so instead we introduce the analytic regulator $\delta$, and define
\begin{align}
	Z_{\mathrm{GN},\delta} &= \int  \Dd{\psi} \Dd{\psibar} \DdImag{\sigma}  e^{-S_{\mathrm{GN},\delta}},
\end{align}
with action 
\begin{align}
\boxed{ S_{\mathrm{GN},\delta} \equiv \int_{x,y} - Z_\psi \psibar_{i,x} \left(C^{-1}_{xy} -\frac{Z_g g \mu^{\delta}}{\sqrt{N}} \sigma_x \delta_{xy}\right) \psi^i_y + \half \int_{x,y}\sigma_x (K^{-1}_{\delta,xy} - \mu^{2\delta} K^{-1}_{xy})\sigma_y.}
\end{align}
The regulator terms are identical other than a new value for $C_{\sigma,0}$,
\begin{equation}
K_\delta = \frac{C_{\sigma,0}}{s(x,y)^{2(s-\delta)}}, \quad K = K_{\delta=0}.
\end{equation}
The minimal subtraction counterterms are
\begin{equation}
	Z_\psi = \begin{cases} 1 & s \notin 2 \mathbb{Z} + 1\\
		1 - \frac{1}{N\delta} \hat{\gamma}_{\psi,1} + O(1/N^2) & s =2k \in 2 \mathbb{Z} + 1
	\end{cases}, \quad Z_\psi Z_g = 1- \frac{1}{2N \delta} \Gamma^{\mathrm{GN}}_\mathrm{tet} + O(1/N^2),
\end{equation}
with the usual $\hat{\gamma}_{\psi,1} = 2F_b'(2k)/F_f''(\dotwo-k)$ in the short-range case.
The calculation of these counterterms is identical to that of \cref{app:flatSpace2pt}, save for the differing Lorentz representation, and is therefore not reproduced here.

\subsection{Conformal data}
We find the conformal data given in \cref{eq:resultsGN} for the long-range model, which agrees with \cite{Chai:2021wac}, and the short-range data agrees with \cite{Derkachov:1993uw}.
The normalizations are
\begin{equation}
C_{\sigma/\psi}= C_{\sigma/\psi,0} \left(1+ \frac{\normCor_{\sigma/\psi,1}}{N} + \frac{\normCor_{\sigma/\psi,2}}{N^2}\right),
\end{equation}
for
\begin{equation}
\begin{aligned}
	C_{\sigma,0} &= \frac{1}{(C_{\psi,0})^2} \frac{-1}{\cN_b(s)} = -\frac{4^s A_f\left(\frac{d-s}{2}\right)^2}{A(d-s)A(s)}, \\
	\normCor_{\sigma,1} &= \normCor_{\sigma,1,\mathrm{tet}} +\normCor_{\sigma,1,\mathrm{pil}}\\
	\normCor_{\sigma,1,\mathrm{tet}} &= \Gamma_{\mathrm{tet}}^\mathrm{GN} \left(B_f(\tfrac{d-s}{2}) - B(d-s)\right), \quad \normCor_{\sigma,1,\mathrm{pil}}= \frac{4 F_b'(s)}{F_f'(\tfrac{d-s}{2})},
\end{aligned}
\end{equation}
and
\begin{align}
	C_{\psi,0}&= \frac{-i}{\cF_{2(\tfrac{d-s}{2}),1}} =\frac{\pi^{-\dotwo} 2^{-s}}{A_f(\tfrac{d-s}{2})}\\
	\normCor_{\psi,1} &= -\half \normCor_{\sigma,1,\mathrm{pil}},\quad  \normCor_{\psi,2} = -\frac{F_b'(s)}{F_f'(\tfrac{d-s}{2})}  \Gamma _{\text{tet}}^\mathrm{GN} \left(B(s) + B_f(\tfrac{d-s}{2})-2B(d -s)\right). %
\end{align}
The short-range results follow through identically, including the short-range $O(1/N^2)$ correction \eqref{eq:GNboxKpsi2} to $\hat{\gamma}_\psi$.
Following the steps in \cref{sec:calculatingF}, we find the first correction
	\begin{equation}
		\frac{F_{\mathrm{GN},1/N}}{F_b'(s)/N} = 
		\begin{matrix}
		&+3\times\half \Gamma_\mathrm{tet}^\mathrm{GN} \\
		&-2\times\half \Gamma_\mathrm{tet}^\mathrm{GN}
		\end{matrix}
		  + \begin{cases}
		    \begin{matrix} 
			&+3 \times \half(-2\hat{\gamma}_{\psi,1})\\ 
			&-2 \times  \half(-2\hat{\gamma}_{\psi,1})
		    \end{matrix} & s=2k\\ 0 & s \neq 2k \end{cases},
		\end{equation}
and hence the free energy agrees with \cite{Tarnopolsky:2016vvd}:
\begin{equation}
F_{\mathrm{GN}} = -\lim_{\delta \to 0} \log Z_{\mathrm{GN},\delta} = \eqref{eq:FGNfuncS}.
\end{equation}
We provide some notes on these vacuum diagram computations in \cref{app:GNFdetails}.
As expected, we also find that
\begin{equation}
	F^\mathrm{LR}_\mathrm{GN} (s= 2k -2 \hat{\gamma}_\psi) = F^{\mathrm{SR},\slashed{\partial}\Box^{k-1/2}}_\mathrm{GN}
\end{equation}
extremizes $F^\mathrm{LR}_\mathrm{GN}(s)$.

We conclude with a note that the natural expression for the $\cN=1$ (real) supersymmetric model is then for a real scalar superfield $\Sigma$,
\begin{align}
	Z_{\mathrm{SUSY}} &= \int \DdImag{\Sigma} \, \exp \left( -\half \Str \log \left(C^{-1} -\frac{1}{\sqrt{N}} \Sigma \, \mathbb{I} \right) +\int_{S} \frac{1}{2\lambda_0} \Sigma^2\right)
\end{align}
for the suitable $C^{-1} \sim D^2$ in superspace $S$; of course, this action would then need to be modified by a regulator. %

\section{Evaluating diagrams}

In this paper we have had to calculate both two-point function diagrams and vacuum diagrams -- though of course, as we commented in \cref{sec:whyGammas}, these calculations are manifestly related.
The former we calculated both on the sphere and in flat space, while the latter is finite (and so calculable) only on the sphere.
While on the sphere we can explicitly calculate both kinds of diagrams via Mellin-Barnes integrals, for flat space the easiest approach to finding the leading anomalous dimensions is to use the uniqueness relations given above -- which can be done manually. 
To make this paper self-contained, in this appendix we explain the calculation of flat-space diagrams via the method of uniqueness, and then discuss diagram evaluation by means of Mellin-Barnes integrals.

\subsection{Using uniqueness to calculate flat-space diagrams} \label{sec:applyingUniqueness}

The method of uniqueness only works for the $\sigma$ equation, \eqref{eq:CsigmaEquation}; our goal will be to derive in flat space the results \cref{eq:dItetdSigma,eq:dIprdSigma}.
Now, we cannot apply uniqueness directly to these diagrams, as none of their vertices satisfy the sum requirement $a+b+c=d$; however, we can consider diagrams which differ from the diagram we want only up to terms that are $O(\delta^2)$, but that are unique.

In flat space, let us take both of our original diagrams, throw away the anomalous dimensions, and define new diagrams \eqref{eq:tetShifted} and \eqref{eq:prShifted}, where each of the exponents $\{\tfrac{d-s}{2},s\}$ have been shifted by small amounts: by some parameter $\eta$ for $2\fdv{I_{\mathrm{tet}}}{\Sigma}$ and $\eta$ and $\eta'$ for $2\fdv{I_{\mathrm{pr}}}{\Sigma}$. 
This is carefully done such that (1) the diagrams become manifestly invariant under, separately,
\begin{equation}
   \eta \to -\eta, \quad \text{ or } \eta' \to -\eta'.
\end{equation} 
This means that the diagrams evaluated as a function of these shifts differ only at $O(\eta^2)$ or $O(\eta'^2)$ (with no cross term) from the originals:
\begin{align}\label{eq:requiredTaylorForm}
   2\fdv{I_{\mathrm{tet}}}{\Sigma}(\delta,\eta) &= 2\fdv{I_{\mathrm{tet}}}{\Sigma}(\delta,0)\times (1+ O(\eta^2))\\
   2\fdv{I_{\mathrm{pr}}}{\Sigma}(\delta,\eta,\eta') &= 2\fdv{I_{\mathrm{pr}}}{\Sigma}(\delta,0,0) \times (1+ O(\eta^2)+O(\eta'^2)).
\end{align}
We also ensure that the shifts are (2) such that for some particular values $\eta, \eta' \propto \delta$ we can apply uniqueness. 
Since the original diagrams can only have at worst a $1/\delta$ pole, they are identical to the shifted diagrams at this order.

The required shifts must be determined by inspection, and are illustrated here. 
The solid lines are propagators of scaling dimension $\tfrac{d-s}{2}$, and the dashed lines are propagators of scaling dimension $s$; the numbers on the lines then indicate the shifts away from that value.
We define the shifted diagram
\begin{align}\label{eq:tetShifted}
	2\fdv{I_{\mathrm{tet}}}{\Sigma}(\delta,\eta)& \equiv \frac{g_0^4}{2N} \, \begin{tikzpicture}[scale=1.0, line cap=round, line join=round,font=\small,baseline={(0,-\MathAxis pt)}]
	\coordinate (TL) at (-1.5, 0); %
	\coordinate (BL) at (0, -1.5); %
	\coordinate (BR) at (1.5,0); %
	\coordinate (TR) at (0, 1.5); %
    \draw[thick] (TL) -- node[above left] {$+\eta$}  (TR) --node[above right] {$+\eta$} (BR) --node[below right] {$-\eta$} (BL) --node[below left] {$-\eta$} cycle;
      \draw[sigma] (BL) -- (TR) node [midway,right] {$-\delta$};
	\draw[sigma] (-2,0) -- (TL);
	\draw[sigma] (BR) -- (2,0);
    \end{tikzpicture}\,,
   \end{align}
which is manifestly invariant under $\eta \to -\eta$. 
The prism diagram requires two shifts
   \begin{align}\label{eq:prShifted}
    2\fdv{I_{\mathrm{pr}}}{\Sigma}(\delta,\eta,\eta') &= \frac{g_0^6}{2N} \, \begin{tikzpicture}[scale=1.0, line cap=round, line join=round,font=\small,baseline={(0,-\MathAxis pt)}]
	\coordinate (TL) at (-1.5, 0); %
	\coordinate (BL) at (0, -1.5); %
	\coordinate (TR) at (0, 1.5); %
    \draw[thick] (TL) -- node[above left] {$\eta+\eta'$}  (TR) --node[right] {$+\eta'$} (BL) --node[below left] {$-\eta-\eta'$} cycle;
	\coordinate (TLp) at (3.5, 0); %
	\coordinate (BLp) at (2, -1.5); %
	\coordinate (TRp) at (2, 1.5); %
    \draw[thick] (TLp) -- node[above right] {$\eta'-\eta$}  (TRp) --node[left] {$-\eta'$} (BLp) --node[below right] {$\eta-\eta'$} cycle;
      \draw[sigma] (BL) -- (BLp) node[midway,above] {$-\delta$};
      \draw[sigma] (TR) -- (TRp) node[midway,below] {$-\delta$};
      \draw[sigma] (-2,0) -- (TL);
      \draw[sigma] (TLp) -- (4,0);
    \end{tikzpicture}\,,
\end{align}
and so is a little more complicated.
Because Lorentz invariance means that these diagrams are invariant under $x \leftrightarrow y$, the identities
\begin{equation}
   2\fdv{I_{\mathrm{pr}}}{\Sigma}(\delta,\eta,0) \equiv 2\fdv{I_{\mathrm{pr}}}{\Sigma}(\delta,-\eta,0),\quad 2\fdv{I_{\mathrm{pr}}}{\Sigma}(\delta,\eta,\eta')=2\fdv{I_{\mathrm{pr}}}{\Sigma}(\delta,\eta,-\eta')
\end{equation}
are true by inspection\footnote{Specifically, as diagrams $\mathrm{pr}_{-\eta,0}=\reflectbox{$\mathrm{pr}_{\eta,0}$}\stackrel{\star}{=}\mathrm{pr}_{\eta,0}$ and $\mathrm{pr}_{\eta,-\eta'}= \scalebox{-1}[-1]{$\mathrm{pr}_{\eta,\eta'}$}\stackrel{\star}{=}\mathrm{pr}_{\eta,\eta'}$, where the equalities $\stackrel{\star}{=}$ follow from the symmetry under the exchange of $x$ and $y$.}.
Hence, the Taylor expansions in the $\eta$s must take the form \eqref{eq:requiredTaylorForm}.
We then notice that for $\eta=-\delta/2$ and $\eta=\eta'=-\delta$ we have unique triangles.
Hence, we can calculate the desired
\begin{align}
   2\fdv{I_{\mathrm{tet}}}{\Sigma}(\delta,-\delta/2) &= \frac{g_0^4}{2N} \frac{C_\sigma C_\phi^4}{X^{2 (d-s)-2\delta}} U(2 \delta ,\mu -\delta ,\mu -\delta)  U(s-\delta ,\tfrac{d-s+\delta}{2},\tfrac{d-s+\delta}{2}),\\
   2\fdv{I_{\mathrm{pr}}}{\Sigma}(\delta,-\delta, -\delta) &= \frac{g_0^6}{2N} \frac{C_\sigma^2 C_\phi^6}{X^{2(d-s)-4\delta}}
   U(\dotwo - 2\delta, \dotwo-2\delta,4 \delta ) U(s-\delta ,\tfrac{d-s}{2} ,\tfrac{d-s}{2} + \delta)\\
   & \times U(s-2 \delta ,d-\tfrac{3 s}{2}+\delta,\tfrac{s}{2}+\delta) U(s-\delta , \tfrac{d-s}{2}-\delta ,\tfrac{d-s}{2} +2 \delta),
\end{align}
which is easily done using the \texttt{STR} package \cite{Preti:2018vog}.
But from \eqref{eq:requiredTaylorForm}, we see that these are also the values of our desired diagrams, up to an $O(\delta)$ correction term, because $
2\fdv{I_i}{\Sigma}(\delta,0) \times \eta^2 \propto \delta$.
After expanding the $U$s for small $\delta$, we find exactly \cref{eq:dItetdSigma,eq:dIprdSigma}.

\subsection{Evaluation of diagrams on spheres via Mellin-Barnes integrals}\label{app:sphereInts}

We use stereographic coordinates on the Euclidean sphere to evaluate the various sphere integrals.
For completeness, we give here two exactly solvable sphere integrals, \cite{Cardy:1988cwa}
\begin{align}\label{eq:CardyI2}
	I_2(\Delta) &\equiv \int_{x,y} \frac{1}{s(x,y)^{2\Delta}} %
	= \frac{\pi^d}{(2R)^{2 (\Delta-d)}} \frac{\Gamma(\dotwo)}{\Gamma(d)} \frac{\Gamma \left(\frac{d}{2}-\Delta \right)}{\Gamma (d-\Delta )} \\
	I_3(\Delta) &\equiv \int_{x,y,z} \frac{1}{(s(x,y) s(y,z) s(z,x))^{\Delta}} =\frac{8 \pi^{\frac{3(d+1)}{2}}}{\Gamma(d)} \frac{\Gamma(d-\frac{3\Delta}{2})}{\Gamma(\frac{d+1-\Delta}{2})^3} R^{3(d-\Delta)}.
\end{align}
We often use the fact that
\begin{equation}
I_2'(d) \equiv \odv{I_2}{\Delta} \Big\rvert_{\Delta=d} = \frac{2}{\sin \left(\frac{\pi  d}{2}\right)} \frac{\pi ^{d+1}}{\Gamma (d+1)}.
\end{equation}

\subsection{Mellin-Barnes notes}

We use the Mellin-Barnes procedures precisely as described in \cite{Fei:2015oha}, using \texttt{MBresolve} \cite{Smirnov:2009up} from the \texttt{MBtools} collection \cite{Belitsky:2022gba,belitskyFeynmanIntegralsMBGitLab2024}.
In each case, we always fix one of the points, $w\to \infty$, which reduces the number of integrals that we are required to do.
As mentioned above, this had the additional advantage of immediately giving a valid Mellin-Barnes representation; that is, we could avoid needing to perform the extra analytic shifts that were required in \cref{sec:applyingUniqueness}. 
The reason for this is not clear.

\subsection{The tetrahedron and prism diagrams}

To leading order in the shifts $\Delta_\phi = \tfrac{d-s}{2}+ \gamma_\phi$, $\Delta_\sigma = s+\gamma_\sigma$, we find:
\begin{equation}\begin{aligned}
I_{\mathrm{tet}} = \frac{1}{8N} 6 C_\sigma^2 C_\phi^4   \pi^{2d}  \frac{\Gamma(-\tfrac{d}{2})}{\Gamma(d)} \frac{\Gamma \left(\frac{d}{2}-s\right)}{\Gamma (s)} \left(\frac{\Gamma \left(\frac{s}{2}\right)}{\Gamma \left(\frac{d-s}{2}\right)}\right)^2 (1 + I_{\mathrm{tet},\mathrm{corr}}/N),
\end{aligned}\end{equation}
if each $\gamma$ is $\sim 1/N$. %
The next order in $N$ is easily calculated as well:
\begin{equation}\begin{aligned}
\frac{I_{\mathrm{tet},\text{corr}}^{(1)}}{N}= (&-2 (\gamma _{\sigma }+2 \gamma _{\phi })(\psi ^{(0)}(-\tfrac{d}{2})+\gamma+ 2 \log (2R))\\
& +\tfrac{2}{3} (\gamma _{\phi }-\gamma _{\sigma }) (\psi ^{(0)}(\tfrac{d}{2}-s)-\psi ^{(0)}(\tfrac{d-s}{2})-\psi ^{(0)}(\tfrac{s}{2})+\psi ^{(0)}(s))).
\end{aligned}\end{equation}
Though other ways of combining these terms exist, this can be written most compactly using \eqref{eq:polygamma}. 
With the usual $\varkappa = d-2\Delta_\phi-\Delta_\sigma =-2\gamma_\phi-\gamma_\sigma$:
\begin{equation}\begin{aligned}
\frac{I_{\mathrm{tet}}}{C_\sigma^2 C_\phi^4} &= \frac{3}{8N} \frac{2\pi ^{2 d}}{\Gamma(d)} e^{2 \gamma  \varkappa } (2R)^{4 \varkappa } \Gamma \left(2 \varkappa -\tfrac{d}{2}\right) \frac{\Gamma \left(-\frac{d}{2}+\frac{2 \varkappa }{3}+2 \Delta _{\phi }\right)}{\Gamma \left(d-2 \Delta _{\phi }-\frac{2 \varkappa }{3}\right)} \frac{\Gamma \left(\frac{d}{2}-\Delta _{\phi }-\frac{\varkappa }{3}\right){}^2}{\Gamma \left(\Delta _{\phi }+ \frac{\varkappa }{3}\right){}^2} + O(1/N^2)\\
&= \frac{3}{8N} \frac{2\pi ^{2 d}}{\Gamma(d)} e^{2 \gamma  \varkappa } (2R)^{4 \varkappa } \Gamma \left(2 \varkappa -\tfrac{d}{2}\right) \frac{\Gamma \left(\frac{d}{2}-\Delta_\sigma -\frac{\varkappa }{3}\right)}{\Gamma \left(\Delta_\sigma +\frac{\varkappa }{3}\right)} \frac{\Gamma \left(\frac{d}{2}-\Delta _{\phi }-\frac{\varkappa }{3}\right){}^2}{\Gamma \left(\Delta _{\phi }+\frac{\varkappa }{3}\right){}^2} + O(1/N^2)\\
&= \frac{3}{8N} \frac{2\pi ^{2 d}}{\Gamma(d)} e^{2 \gamma  \varkappa } (2R)^{4 \varkappa } \Gamma (2 \varkappa -\tfrac{d}{2})  A(\Delta_\sigma +\tfrac{\varkappa }{3}) A(\Delta _{\phi }+\tfrac{\varkappa}{3})^2 + O(1/N^2).
\end{aligned}\end{equation}

For our next diagram, we find 
\begin{equation}\begin{aligned}
\frac{I_\mathrm{pr}}{ C_\sigma^3 C_\phi^6}= \frac{5}{12N} \pi ^{3 d}\frac{ \Gamma(-\dotwo)  }{\Gamma (d)} \frac{\Gamma \left(\frac{3 s}{2}-\frac{d}{2}\right)}{\Gamma \left(d-\frac{3 s}{2}\right)}
\left(\frac{\Gamma(\dotwo-s)}{\Gamma (s)} \frac{\Gamma (\frac{s}{2})}{\Gamma(\frac{d-s}{2})}\right)^3 (1 + I_{\mathrm{pr},\mathrm{corr}}^{(1)}/N)
\end{aligned}\end{equation}
The subleading correction to this is compactly encoded by
\begin{equation}\begin{aligned}
\frac{I_\mathrm{pr}}{C_\sigma^3 C_\phi^6} &= \frac{1}{12N} \frac{5}{2} \frac{2\pi^{3d}}{\Gamma(d)} e^{3\gamma \varkappa} (2R)^{6\varkappa} \Gamma(3 \varkappa -\tfrac{d}{2})A(\Delta_\sigma +\tfrac{\varkappa }{5})^3 A(\Delta_\phi +\tfrac{3 \varkappa }{5})^3   A(-\tfrac{d}{2} +3\Delta_\phi +\tfrac{3 \varkappa }{5}) + O(1/N^2).
\end{aligned}\end{equation}
Of course, we can move the subleading terms around, and we note that up to $\gamma_i^2$ corrections that we did not calculate,
\begin{equation}\begin{aligned}
A(\tfrac{d}{2}-\Delta_\sigma +\Delta_\phi -\tfrac{2 \varkappa }{5}) &=A(d-\tfrac{3}{2}\Delta_\sigma-\tfrac{9\varkappa }{10}) = A(-\tfrac{d}{2} +3\Delta_\phi +\tfrac{3 \varkappa }{5}).
\end{aligned}\end{equation}

\subsection{The pillow diagram for arbitrary scaling dimensions}

Let us now evaluate in full generality the pillow integral for arbitrary $\Delta_i$ and $C_i$.
We use first \texttt{MBresolve}, then \texttt{MBsums} \cite{Ochman:2015fho}, and then the Mathematica \texttt{Sum} routine. 
Let us parametrise, for compactness, $d= 2\mu$, $\Delta_\sigma+2\Delta_\phi +\varkappa= d$, and $\Delta_\phi=\Delta$. Then $I_{\mathrm{pil}}(\mu, \varkappa,\Delta)$:
\begin{equation}\begin{aligned}
&\frac{I_\text{pil}}{C_\sigma^2 C_\phi^4} = \frac{1}{4N} \frac{\pi ^{4 \mu } (2 R)^{4 \varkappa }}{\Gamma (2 \mu )} \frac{\Gamma (\mu -\Delta )}{\Gamma (\Delta )} \frac{\Gamma (\mu -\varkappa )}{\Gamma (\varkappa )} \frac{\Gamma (\Delta +\varkappa -\mu )}{\Gamma (2 \mu -(\Delta +\varkappa ))} \Gamma (\varkappa -\mu +1)
\\
&\quad \quad \quad \left[
\Gamma (\mu -\Delta ) \Gamma (2 \varkappa -\mu ) \Gamma (\Delta +\varkappa -\mu ) \,\right.\\ 
& \quad \quad \quad \left.\times\, {}_4\tilde{F}_3(\varkappa ,\Delta +\varkappa -\mu ,2 \varkappa -\mu ,\mu -\Delta ;\varkappa -\mu +1,\Delta +2 \varkappa -\mu ,\mu+\varkappa -\Delta ;1)
\right.\\
&\left.
\, -\Gamma (\Delta ) \Gamma (\mu ) \Gamma (2 \mu-\Delta -\varkappa ) \, _4\tilde{F}_3(\Delta ,\varkappa ,\mu ,2 \mu -\Delta -\varkappa;\Delta +\varkappa , \mu-\varkappa +1,2 \mu -\Delta ;1)
\right] \label{eq:IpillowTotallyAnalytic}
\end{aligned}\end{equation}
The prefactor here can be simplified using $A(x)\equiv\Gamma(\tfrac{d}{2}-x)/\Gamma(x)$ to
\begin{equation}\begin{aligned}
= \frac{1}{4N} \frac{\pi ^{4 \mu } (2 R)^{4 \varkappa }}{\Gamma (2 \mu ) }\frac{A(\Delta ) A(\varkappa )}{A(\Delta +\varkappa -\mu )}\Gamma (\varkappa -\mu +1) \times \left[\cdots\right]
\end{aligned}\end{equation}
For generic $s$, $I_\mathrm{pil}$ vanishes $\propto \varkappa$, as given in \eqref{eq:ItetEvals}, or alternatively by
\begin{equation}\begin{aligned}
I_{\mathrm{pil}} = \frac{\pi^d}{2N}\frac{\Gamma \left(\frac{d}{2}\right) \Gamma \left(-\frac{d}{2}\right)}{\Gamma (d)}  \cN_b(\tfrac{d-s}{2}) \varkappa + O(\varkappa^2).
\end{aligned}\end{equation}
However, since  $\varkappa=\delta$, the existence of the counterterm means that this diagram may contribute at order $1/N^2$.
In the case of $s$ an even integer, we need to be a little more careful. 
Mathematica can only take this limit numerically, but for small $\varkappa=\delta$ we obtain from \eqref{eq:IpillowTotallyAnalytic} the finite result shown in \eqref{eq:ItetEvals}.

\subsection{Gross-Neveu for variable \texorpdfstring{$s$}{s}}\label{app:GNFdetails}

We can also calculate the arbitrary-$s$ contributions to the free energy for the GN CFT. 
They are diagrammatically identical to those of \eqref{eq:ONFcorrections} save for the absence of the prism diagram and differing symmetry factors (which here include the minus sign from the fermion loop):
\begin{equation}\label{eq:GNFcorrections}
-\delta F_{\mathrm{GN},1/N} \equiv
\begin{tikzpicture}[scale=1.0, line cap=round, line join=round,baseline={(0,-\MathAxis pt)}]
\begin{scope}[xshift=-2.5cm]
    \coordinate (C) at (0,0); %
    \draw[sigma] (C) circle (0.5);
    \coordinate (X) at (0,0.5);
    \draw[thick] ($(X)+(-0.15,-0.15)$) -- ($(X)+(0.15,0.15)$);
    \draw[thick] ($(X)+(-0.15,0.15)$) -- ($(X)+(0.15,-0.15)$);
    
    \node[below] at (0,-0.75) {$I_{\mathrm{ct}}^\mathrm{GN}: \half$};
    \node[right] at (0.85,0) {$+$};
\end{scope}
\tikzset{
  midarrow/.style={
    postaction={
      decorate,
      decoration={
        markings,
        mark=at position 0.5 with {\arrow{Stealth}}
}}}}

\begin{scope}
\coordinate (TL) at (-0.5, 0.5); 
\coordinate (BL) at (-0.5,-0.5); 
\coordinate (TR) at ( 0.5, 0.5);
\coordinate (BR) at ( 0.5,-0.5);

\draw[thick,midarrow] (TL) -- (TR);
\draw[thick,midarrow] (TR) -- (BR);
\draw[thick,midarrow] (BR) -- (BL);
\draw[thick,midarrow] (BL) -- (TL);
\draw[sigma]
    (TL) .. controls ($(TL)+(-0.3,0)$) and ($(BL)+(-0.3,0)$) .. (BL);
\draw[sigma] (TR) .. controls ($(TR)+(0.3,0)$) and ($(BR)+(0.3,0)$) .. (BR);
\node[below] at (0,-0.75) {$I_{\mathrm{pil}}^\mathrm{GN}: (-1)\frac{1}{2} $};
\node[right] at (1.05,0) {$+$};
\end{scope}

\begin{scope}[xshift=2.5cm]
    \coordinate (TL) at (-0.5, 0.5); %
    \coordinate (BL) at (-0.5,-0.5); %
    \coordinate (TR) at ( 0.5, 0.5); %
    \coordinate (BR) at ( 0.5,-0.5); %
  \draw[thick,midarrow] (TL) -- (TR);
\draw[thick,midarrow] (TR) -- (BR);
\draw[thick,midarrow] (BR) -- (BL);
\draw[thick,midarrow] (BL) -- (TL);
  \draw[sigma] (BL) -- (TR);
  \draw[sigma] (TL) -- (BR);
  \node[below] at (0,-0.75) {$I_{\mathrm{tet}}^\mathrm{GN}: (-1)\frac{1}{4}$};
\end{scope}
\end{tikzpicture}\; ,
\end{equation}
The counterterm diagram computation is identical to that of the $\gO(N)$ model.
To confirm these symmetry factors, we just expand
\begin{equation}\begin{aligned}
   &\log \int_{\sigma,\psi} e^{-\frac{1}{\sqrt{N}} \int \sigma \bar\psi_i \psi_i} \supset \frac{N}{4! (\sqrt{N})^4} \expval{(\int \sigma \bar\psi \psi)^4}_0^c\\
   = &\frac{1}{4! N} \expval{\sigma_1 \sigma_2 \sigma_3 \sigma_4}_0^c \underbrace{\expval{(\bar\psi \psi)_1 (\bar\psi \psi)_2 (\bar\psi \psi)_3 (\bar\psi \psi)_4}_0^c}_{\equiv (-1) 6 L(x_i)}.
\end{aligned}
\end{equation}
To evaluate the fermion loop $L(x_i)$ on the sphere\footnote{This happens to be the same as the flat-space fermion loop, mapped from $\abs{x-y} \to s(x,y)$: but in theories with more complex spin structure this is the correct method.}, we use the embedding space propagator $C_{12} = C(\eta_1(x_1), \eta_2(x_2))$ defined in \eqref{eq:fermionPropSphereEmbedding}; evaluating the traces\footnote{The traces are over $d_\text{embed}=d+1$-dimensional gamma matrices $\alpha_i$, but we keep $\Tr_{S_{d+1}} \alpha_i = T$; however, for example, $\alpha_i \alpha_j \alpha^i = (2-(d+1)) \alpha_j$.}, using $\eta_i \cdot \eta_j = -\half[(\eta_i-\eta_j)^2 -\eta_i^2 -\eta_j^2$ and $\eta_i^2=R^2$, and then mapping $\abs{\eta_1(x_1)-\eta_2(x_2)}=s(x_1,x_2)$, we find
\begin{equation}\begin{aligned}
    L(x_i) &\equiv \frac{1}{T}\Tr_{S_{d+1}}[C_{12}C_{23}C_{34} C_{41}]\\
    &= \half C_\psi^4\frac{s_{12}^2 s_{34}^2 -s_{13}^2 s_{24}^2+s_{14}^2 s_{23}^2}{(s_{12} s_{23} s_{34} s_{41})^{2\Delta_\psi}},
\end{aligned} \end{equation}
Thus our diagrams are
\begin{subequations}\label{eq:IdefsGN}
\begin{align}
I_{\mathrm{ct}}^\mathrm{GN}  &= \half \left(-\mu^{2\delta}\frac{2 Z_{g,1}}{N\delta}\right) \tr K^{-1} K_\delta\\
I_{\mathrm{pil}}^\mathrm{GN}  &= \frac{1}{N} \frac{-1}{2}\int_{x_{1,2,3,4}} \frac{C_\sigma^2}{(s_{12} s_{34})^{2\Delta_\sigma}} L(x_i) \\
I_{\mathrm{tet}}^\mathrm{GN}  &= \frac{1}{N} \frac{-1}{4} \int_{x_{1,2,3,4}} \frac{C_\sigma^2}{(s_{13} s_{24})^{2\Delta_\sigma}} L(x_i)
\end{align}
\end{subequations}
They can be calculated in the same way, though we note that ensuring the existence of the Mellin-Barnes representation of the pillow diagram for $s=2k, k -\half \in \mathbb{Z}$ requires a modification of the powers in the numerator of $L$ from 2 to $2 + \lambda$. 
Taking
\begin{equation}\begin{aligned}
\varkappa = d- \Delta_\sigma - 2\Delta_\psi,
\end{aligned}\end{equation}
we find once again that the pillow-type diagram vanishes $\propto \varkappa$, and hence in our case $\propto \delta$.
Specifically, for $s\neq 2k$, we find
\begin{equation}\begin{aligned}
F_{1/N,\mathrm{pil}}^\mathrm{GN} &= \frac{C_\sigma^2 C_{\psi }^4}{N} (2 R)^{4 \varkappa }  \pi^{2d} \frac{\Gamma(\tfrac{d}{2})\Gamma(-\tfrac{d}{2})}{\Gamma(d)} \frac{(d+s^2-1)}{s^2-1}
\tfrac{ \Gamma \left(\tfrac{1-s}{2}\right) \Gamma \left(\tfrac{1+s}{2}\right)}{\Gamma \left(\tfrac{d-s+1}{2}\right) \Gamma\left(\tfrac{d+s+1}{2}\right)} \varkappa +O(\varkappa^2)\\
&= -\frac{C_\sigma^2 C_{\psi }^4}{N} (2 R)^{4 \varkappa }  \frac{\Gamma(\tfrac{d}{2})\Gamma(-\tfrac{d}{2})}{\Gamma(d)} \frac{(d+s^2-1)}{s^2-1} \pi^{d}\cN_\psi(\tfrac{d-s}{2})\varkappa +O(\varkappa^2),
\end{aligned}\end{equation}
meaning that, just like the bosonic case, we have
\begin{equation}
    I_{\mathrm{pil}}^\mathrm{GN} = \begin{cases}
        O(\delta) & s \neq 2k\\
        3\times \frac{1}{N} F_b'(2k) \hat{\gamma}_{\psi,1} +O(\delta) & s = 2k
    \end{cases}.
\end{equation} 
Once again, due to $Z_{g,1} \supset 1/\delta$, the pillow may contribute at order $1/N^2$ in the arbitrary-$s$ case.

The contribution from $I_{\mathrm{tet}}^{\mathrm{GN}}$ is, for any value of $s$,
\begin{equation}\begin{aligned}
F_{1/N,\text{tet}}^\mathrm{GN} = -I_\mathrm{tet}^\mathrm{GN} =\frac{3 }{2 N}C_\sigma^2 C_{\psi }^4\pi ^{2 d} \frac{\Gamma \left(-\frac{d}{2}\right)}{\Gamma (d)} \frac{\Gamma \left(\frac{d}{2}-s\right)}{\Gamma (s)} \frac{\Gamma \left(\frac{s+1}{2}\right)^2}{\Gamma \left(\frac{d-s+1}{2}\right)^2},
\end{aligned}\end{equation}

which we write compactly with its $1/N$ correction as 
\begin{equation}\begin{aligned}
F_{1/N, \text{tet}}^{\mathrm{GN}}= \frac{3}{2N} C_\sigma^2 C_{\psi }^4 \frac{\pi ^{2 d}}{\Gamma(d)} e^{2 \gamma  \varkappa } (2R)^{4 \varkappa } \Gamma (2 \varkappa -\tfrac{d}{2})  A(\Delta_\sigma +\tfrac{\varkappa }{3}) A_f(\Delta _{\psi }+\tfrac{\varkappa}{3})^2 + O(1/N^2)\\
\end{aligned}\end{equation}
with the usual $A_f(\Delta) = \Gamma(\tfrac{d}{2}-\Delta+\tfrac{1}{2})/\Gamma(\Delta + \thalf)$.

Finally, we note that under an exchange of $\psi \to \phi$, with the extra factor of $i$ noted in \cref{sec:synchronicity} -- i.e.
\begin{equation}
i A_f\to A_b, \quad C_\psi \to C_\phi,\quad N_b = N \mapsto N_f = 2N,
\end{equation}
we find $F_{1/N, \text{tet}}^{\mathrm{GN}} \to F_{1/N, \text{tet}}^{\gO(N)}$.

\bibliographystyle{JHEP}
{\raggedright  %
\bibliography{references-SFE}    %

\providecommand{\href}[2]{#2}\begingroup\raggedright\begin{thebibliography}{10}

\bibitem{Tarnopolsky:2016vvd}
Grigory Tarnopolsky, \emph{On {{Large N Expansion}} of the {{Sphere Free Energy}}}, \href{http://dx.doi.org/10.1103/PhysRevD.96.025017}{\emph{Phys. Rev. D} {\bf 96} (July, 2017) 025017}, [\href{https://arxiv.org/abs/1609.09113}{{\tt arXiv:1609.09113}}\href{https://arxiv.org/pdf/1609.09113.pdf}{$^{\textsc{pdf}}$}], [\href{https://inspirehep.net/literature?q=Tarnopolsky:2016vvd}{\small \textsc{Inspire}}].

\bibitem{Giombi:2014xxa}
Simone Giombi and Igor~R. Klebanov, \emph{Interpolating between {$a$} and {$F$}}, \href{http://dx.doi.org/10.1007/JHEP03(2015)117}{\emph{JHEP} {\bf 03} (Mar., 2015) 117}, [\href{https://arxiv.org/abs/1409.1937}{{\tt arXiv:1409.1937}}\href{https://arxiv.org/pdf/1409.1937.pdf}{$^{\textsc{pdf}}$}], [\href{https://inspirehep.net/literature?q=Giombi:2014xxa}{\small \textsc{Inspire}}].

\bibitem{Fei:2015oha}
Lin Fei, Simone Giombi, Igor~R. Klebanov and Grigory Tarnopolsky, \emph{Generalized {{F-Theorem}} and the {$\epsilon$} {{Expansion}}}, \href{http://dx.doi.org/10.1007/JHEP12(2015)155}{\emph{JHEP} {\bf 12} (Dec., 2015) 155}, [\href{https://arxiv.org/abs/1507.01960}{{\tt arXiv:1507.01960}}\href{https://arxiv.org/pdf/1507.01960.pdf}{$^{\textsc{pdf}}$}], [\href{https://inspirehep.net/literature?q=Fei:2015oha}{\small \textsc{Inspire}}].

\bibitem{Giombi:2024zrt}
Simone Giombi, Elizabeth Himwich, Andrei Katsevich, Igor~R. Klebanov and Zimo Sun, \emph{Sphere free energy of scalar field theories with cubic interactions},  [\href{https://arxiv.org/abs/2412.14086}{{\tt arXiv:2412.14086}}\href{https://arxiv.org/pdf/2412.14086.pdf}{$^{\textsc{pdf}}$}], [\href{https://inspirehep.net/literature?q=Giombi:2024zrt}{\small \textsc{Inspire}}].

\bibitem{Zamolodchikov:1986c}
Alexander~B. Zamolodchikov, \emph{Irreversibility of the {{Flux}} of the {{Renormalization Group}} in a {{2D Field Theory}}}, \href{https://web.archive.org/web/20240405181444/http://jetpletters.ru/ps/1413/article_21504.pdf}{\emph{JETP Lett.} {\bf 43} (1986) 730--732}, [\href{https://inspirehep.net/literature/240292}{\small \textsc{Inspire}}]. \href{https://web.archive.org/web/20240405181444/http://jetpletters.ru/ps/1413/article_21504.pdf}{Archive}.

\bibitem{Cardy:1988cwa}
John~L. Cardy, \emph{Is {{There}} a c {{Theorem}} in {{Four-Dimensions}}?}, \href{http://dx.doi.org/10.1016/0370-2693(88)90054-8}{\emph{Phys. Lett. B} {\bf 215} (1988) 749--752}, [\href{https://inspirehep.net/literature?q=Cardy:1988cwa}{\small \textsc{Inspire}}].

\bibitem{Fraser-Taliente:2024hzv}
Ludo Fraser-Taliente and John Wheater, \emph{F-extremization determines certain large-{{N CFTs}}}, \href{http://dx.doi.org/10.1007/JHEP04(2025)085}{\emph{JHEP} {\bf 04} (Apr., 2025) 085}, [\href{https://arxiv.org/abs/2412.10499}{{\tt arXiv:2412.10499}}\href{https://arxiv.org/pdf/2412.10499.pdf}{$^{\textsc{pdf}}$}], [\href{https://inspirehep.net/literature?q=Fraser-Taliente:2024hzv}{\small \textsc{Inspire}}].

\bibitem{Fraser-Taliente2026prep}
Ludo {Fraser-Taliente}, \emph{In preparation},  2026.

\bibitem{Fei:2014yja}
Lin Fei, Simone Giombi and Igor~R. Klebanov, \emph{Critical {$O(N)$} {{Models}} in {$6-\epsilon$} {{Dimensions}}}, \href{http://dx.doi.org/10.1103/PhysRevD.90.025018}{\emph{Phys. Rev. D} {\bf 90} (July, 2014) 025018}, [\href{https://arxiv.org/abs/1404.1094}{{\tt arXiv:1404.1094}}\href{https://arxiv.org/pdf/1404.1094.pdf}{$^{\textsc{pdf}}$}], [\href{https://inspirehep.net/literature?q=Fei:2014yja}{\small \textsc{Inspire}}].

\bibitem{Gubser:2017vgc}
Steven~S. Gubser, Christian Jepsen, Sarthak Parikh and Brian Trundy, \emph{O({{N}}) and {{O}}({{N}}) and {{O}}({{N}})}, \href{http://dx.doi.org/10.1007/JHEP11(2017)107}{\emph{JHEP} {\bf 11} (Nov., 2017) 107}, [\href{https://arxiv.org/abs/1703.04202}{{\tt arXiv:1703.04202}}\href{https://arxiv.org/pdf/1703.04202.pdf}{$^{\textsc{pdf}}$}], [\href{https://inspirehep.net/literature?q=Gubser:2017vgc}{\small \textsc{Inspire}}].

\bibitem{Goykhman:2019kcj}
Mikhail Goykhman and Michael Smolkin, \emph{Vector model in various dimensions}, \href{http://dx.doi.org/10.1103/PhysRevD.102.025003}{\emph{Phys. Rev. D} {\bf 102} (July, 2020) 025003}, [\href{https://arxiv.org/abs/1911.08298}{{\tt arXiv:1911.08298}}\href{https://arxiv.org/pdf/1911.08298.pdf}{$^{\textsc{pdf}}$}], [\href{https://inspirehep.net/literature?q=Goykhman:2019kcj}{\small \textsc{Inspire}}].

\bibitem{Chai:2021wac}
Noam Chai, Soumangsu Chakraborty, Mikhail Goykhman and Ritam Sinha, \emph{Long-range fermions and critical dualities}, \href{http://dx.doi.org/10.1007/JHEP01(2022)172}{\emph{JHEP} {\bf 01} (Jan., 2022) 172}, [\href{https://arxiv.org/abs/2110.00020}{{\tt arXiv:2110.00020}}\href{https://arxiv.org/pdf/2110.00020.pdf}{$^{\textsc{pdf}}$}], [\href{https://inspirehep.net/literature?q=Chai:2021wac}{\small \textsc{Inspire}}].

\bibitem{Gracey:2017erc}
John~A. Gracey and Rebecca~M. Simms, \emph{Higher dimensional higher derivative {$\phi^4$} theory}, \href{http://dx.doi.org/10.1103/PhysRevD.96.025022}{\emph{Physical Review D} {\bf 96} (July, 2017) 025022}, [\href{https://arxiv.org/abs/1705.06983}{{\tt arXiv:1705.06983}}\href{https://arxiv.org/pdf/1705.06983.pdf}{$^{\textsc{pdf}}$}], [\href{https://inspirehep.net/literature?q=Gracey:2017erc}{\small \textsc{Inspire}}].

\bibitem{Fisher:1972zz}
Michael~E. Fisher, Shang-keng Ma and Bernhard~G. Nickel, \emph{Critical {{Exponents}} for {{Long-Range Interactions}}}, \href{http://dx.doi.org/10.1103/PhysRevLett.29.917}{\emph{Phys. Rev. Lett.} {\bf 29} (1972) 917--920}, [\href{https://inspirehep.net/literature?q=Fisher:1972zz}{\small \textsc{Inspire}}].

\bibitem{Giombi:2019enr}
Simone Giombi and Himanshu Khanchandani, \emph{{$O(N)$} models with {{Boundary Interactions}} and their {{Long Range Generalizations}}}, \href{http://dx.doi.org/10.1007/JHEP08(2020)010}{\emph{JHEP} {\bf 08} (Aug., 2020) 010}, [\href{https://arxiv.org/abs/1912.08169}{{\tt arXiv:1912.08169}}\href{https://arxiv.org/pdf/1912.08169.pdf}{$^{\textsc{pdf}}$}], [\href{https://inspirehep.net/literature?q=Giombi:2019enr}{\small \textsc{Inspire}}].

\bibitem{Paulos:2015jfa}
Miguel~F. Paulos, Slava Rychkov, Balt~C. van Rees and Bernardo Zan, \emph{Conformal {{Invariance}} in the {{Long-Range Ising Model}}}, \href{http://dx.doi.org/10.1016/j.nuclphysb.2015.10.018}{\emph{Nuclear Physics B} {\bf 902} (Jan., 2016) 246--291}, [\href{https://arxiv.org/abs/1509.00008}{{\tt arXiv:1509.00008}}\href{https://arxiv.org/pdf/1509.00008.pdf}{$^{\textsc{pdf}}$}], [\href{https://inspirehep.net/literature?q=Paulos:2015jfa}{\small \textsc{Inspire}}].

\bibitem{Behan:2017dwr}
Connor Behan, Leonardo Rastelli, Slava Rychkov and Bernardo Zan, \emph{Long-range critical exponents near the short-range crossover}, \href{http://dx.doi.org/10.1103/PhysRevLett.118.241601}{\emph{Physical Review Letters} {\bf 118} (June, 2017) 241601}, [\href{https://arxiv.org/abs/1703.03430}{{\tt arXiv:1703.03430}}\href{https://arxiv.org/pdf/1703.03430.pdf}{$^{\textsc{pdf}}$}], [\href{https://inspirehep.net/literature?q=Behan:2017dwr}{\small \textsc{Inspire}}].

\bibitem{Slade:2016yer}
Gordon Slade, \emph{Critical exponents for long-range {{O}}(n) models below the upper critical dimension}, \href{http://dx.doi.org/10.1007/s00220-017-3024-5}{\emph{Communications in Mathematical Physics} {\bf 358} (Nov., 2017) 343--436}, [\href{https://arxiv.org/abs/1611.06169}{{\tt arXiv:1611.06169}}\href{https://arxiv.org/pdf/1611.06169.pdf}{$^{\textsc{pdf}}$}], [\href{https://inspirehep.net/literature?q=Slade:2016yer}{\small \textsc{Inspire}}].

\bibitem{Behan:2018hfx}
Connor Behan, \emph{Bootstrapping the long-range {{Ising}} model in three dimensions}, \href{http://dx.doi.org/10.1088/1751-8121/aafd1b}{\emph{Journal of Physics A: Mathematical and Theoretical} {\bf 52} (Jan., 2019) 075401}, [\href{https://arxiv.org/abs/1810.07199}{{\tt arXiv:1810.07199}}\href{https://arxiv.org/pdf/1810.07199.pdf}{$^{\textsc{pdf}}$}], [\href{https://inspirehep.net/literature?q=Behan:2018hfx}{\small \textsc{Inspire}}].

\bibitem{Benedetti:2020rrq}
Dario Benedetti, Razvan Gurau, Sabine Harribey and Kenta Suzuki, \emph{Long-range multi-scalar models at three loops}, \href{http://dx.doi.org/10.1088/1751-8121/abb6ae}{\emph{J. Phys. A} {\bf 53} (Oct., 2020) 445008}, [\href{https://arxiv.org/abs/2007.04603}{{\tt arXiv:2007.04603}}\href{https://arxiv.org/pdf/2007.04603.pdf}{$^{\textsc{pdf}}$}], [\href{https://inspirehep.net/literature?q=Benedetti:2020rrq}{\small \textsc{Inspire}}]. Corrigendum and addendum in \cite{Benedetti:2024mqx}.

\bibitem{Benedetti:2024mqx}
Dario Benedetti, Razvan Gurau and Sabine Harribey, \emph{Corrigendum and addendum: {{Long-range}} multi-scalar models at three loops},  [\href{https://arxiv.org/abs/2411.00805}{{\tt arXiv:2411.00805}}\href{https://arxiv.org/pdf/2411.00805.pdf}{$^{\textsc{pdf}}$}], [\href{https://inspirehep.net/literature?q=Benedetti:2024mqx}{\small \textsc{Inspire}}].

\bibitem{Chai:2021arp}
Noam Chai, Mikhail Goykhman and Ritam Sinha, \emph{Long-{{Range Vector Models}} at {{Large N}}}, \href{http://dx.doi.org/10.1007/JHEP09(2021)194}{\emph{JHEP} {\bf 09} (Sept., 2021) 194}, [\href{https://arxiv.org/abs/2107.08052}{{\tt arXiv:2107.08052}}\href{https://arxiv.org/pdf/2107.08052.pdf}{$^{\textsc{pdf}}$}], [\href{https://inspirehep.net/literature?q=Chai:2021arp}{\small \textsc{Inspire}}].

\bibitem{Giombi:2022gjj}
Simone Giombi, Elizabeth Helfenberger and Himanshu Khanchandani, \emph{Long {{Range}}, {{Large Charge}}, {{Large}} {$N$}}, \href{http://dx.doi.org/10.1007/JHEP01(2023)166}{\emph{JHEP} {\bf 01} (Jan., 2023) 166}, [\href{https://arxiv.org/abs/2205.00500}{{\tt arXiv:2205.00500}}\href{https://arxiv.org/pdf/2205.00500.pdf}{$^{\textsc{pdf}}$}], [\href{https://inspirehep.net/literature?q=Giombi:2022gjj}{\small \textsc{Inspire}}].

\bibitem{Lohmann:2017qyq}
Martin Lohmann, Gordon Slade and Benjamin~C. Wallace, \emph{Critical two-point function for long-range {$O(n)$} models below the upper critical dimension}, \href{http://dx.doi.org/10.1007/s10955-017-1904-x}{\emph{Journal of Statistical Physics} {\bf 169} (Nov., 2017) 1132--1161}, [\href{https://arxiv.org/abs/1705.08540}{{\tt arXiv:1705.08540}}\href{https://arxiv.org/pdf/1705.08540.pdf}{$^{\textsc{pdf}}$}], [\href{https://inspirehep.net/literature?q=Lohmann:2017qyq}{\small \textsc{Inspire}}].

\bibitem{Behan:2023ile}
Connor Behan, Edoardo Lauria, Maria Nocchi and Philine {van Vliet}, \emph{Analytic and numerical bootstrap for the long-range {{Ising}} model}, \href{http://dx.doi.org/10.1007/JHEP03(2024)136}{\emph{JHEP} {\bf 03} (Mar., 2024) 136}, [\href{https://arxiv.org/abs/2311.02742}{{\tt arXiv:2311.02742}}\href{https://arxiv.org/pdf/2311.02742.pdf}{$^{\textsc{pdf}}$}], [\href{https://inspirehep.net/literature?q=Behan:2023ile}{\small \textsc{Inspire}}].

\bibitem{Behan:2017emf}
Connor Behan, Leonardo Rastelli, Slava Rychkov and Bernardo Zan, \emph{A scaling theory for the long-range to short-range crossover and an infrared duality}, \href{http://dx.doi.org/10.1088/1751-8121/aa8099}{\emph{Journal of Physics A: Mathematical and Theoretical} {\bf 50} (Aug., 2017) 354002}, [\href{https://arxiv.org/abs/1703.05325}{{\tt arXiv:1703.05325}}\href{https://arxiv.org/pdf/1703.05325.pdf}{$^{\textsc{pdf}}$}], [\href{https://inspirehep.net/literature?q=Behan:2017emf}{\small \textsc{Inspire}}].

\bibitem{Giombi:2019upv}
Simone Giombi, Richard Huang, Igor~R. Klebanov, Silviu~S. Pufu and Grigory Tarnopolsky, \emph{The {$O(N)$} {{Model}} in {$4<d<6$}: {{Instantons}} and {{Complex CFTs}}}, \href{http://dx.doi.org/10.1103/PhysRevD.101.045013}{\emph{Phys. Rev. D} {\bf 101} (Feb., 2020) 045013}, [\href{https://arxiv.org/abs/1910.02462}{{\tt arXiv:1910.02462}}\href{https://arxiv.org/pdf/1910.02462.pdf}{$^{\textsc{pdf}}$}], [\href{https://inspirehep.net/literature?q=Giombi:2019upv}{\small \textsc{Inspire}}].

\bibitem{Fraser-Taliente:2026gdh}
Kit {Fraser-Taliente} and Ludo {Fraser-Taliente}, \emph{The {$T^{\mu\nu}$} of the conformal scalars},  [\href{https://arxiv.org/abs/2601.05311}{{\tt arXiv:2601.05311}}\href{https://arxiv.org/pdf/2601.05311.pdf}{$^{\textsc{pdf}}$}], [\href{https://inspirehep.net/literature?q=Fraser-Taliente:2026gdh}{\small \textsc{Inspire}}].

\bibitem{Benedetti:2019ikb}
Dario Benedetti, Razvan Gurau, Sabine Harribey and Kenta Suzuki, \emph{Hints of unitarity at large {$N$} in the {$O(N)^3$} tensor field theory}, \href{http://dx.doi.org/10.1007/JHEP02(2020)072}{\emph{JHEP} {\bf 02} (Feb., 2020) 072}, [\href{https://arxiv.org/abs/1909.07767}{{\tt arXiv:1909.07767}}\href{https://arxiv.org/pdf/1909.07767.pdf}{$^{\textsc{pdf}}$}], [\href{https://inspirehep.net/literature?q=Benedetti:2019ikb}{\small \textsc{Inspire}}].

\bibitem{Benedetti:2019rja}
Dario Benedetti, Nicolas Delporte, Sabine Harribey and Ritam Sinha, \emph{Sextic tensor field theories in rank {$3$} and {$5$}}, \href{http://dx.doi.org/10.1007/JHEP06(2020)065}{\emph{JHEP} {\bf 06} (June, 2020) 065}, [\href{https://arxiv.org/abs/1912.06641}{{\tt arXiv:1912.06641}}\href{https://arxiv.org/pdf/1912.06641.pdf}{$^{\textsc{pdf}}$}], [\href{https://inspirehep.net/literature?q=Benedetti:2019rja}{\small \textsc{Inspire}}].

\bibitem{Fraser-Taliente:2025spectrum}
Ludo {Fraser-Taliente} and John Wheater, \emph{The spectrum in melonic {{CFTs}}, in preparation},  2026.

\bibitem{Gerchkovitz:2014gta}
Efrat Gerchkovitz, Jaume Gomis and Zohar Komargodski, \emph{Sphere {{Partition Functions}} and the {{Zamolodchikov Metric}}}, \href{http://dx.doi.org/10.1007/JHEP11(2014)001}{\emph{JHEP} {\bf 11} (Nov., 2014) 001}, [\href{https://arxiv.org/abs/1405.7271}{{\tt arXiv:1405.7271}}\href{https://arxiv.org/pdf/1405.7271.pdf}{$^{\textsc{pdf}}$}], [\href{https://inspirehep.net/literature?q=Gerchkovitz:2014gta}{\small \textsc{Inspire}}].

\bibitem{Jepsen:2023pzm}
Christian Jepsen and Yaron Oz, \emph{{{RG Flows}} and {{Fixed Points}} of {$O(N)^r$} {{Models}}}, \href{http://dx.doi.org/10.1007/JHEP02(2024)035}{\emph{JHEP} {\bf 02} (Feb., 2024) 035}, [\href{https://arxiv.org/abs/2311.09039}{{\tt arXiv:2311.09039}}\href{https://arxiv.org/pdf/2311.09039.pdf}{$^{\textsc{pdf}}$}], [\href{https://inspirehep.net/literature?q=Jepsen:2023pzm}{\small \textsc{Inspire}}].

\bibitem{Giombi:2015haa}
Simone Giombi, Igor~R. Klebanov and Grigory Tarnopolsky, \emph{Conformal {{QED}}{$_d$}, {$F$}-{{Theorem}} and the {$\epsilon$} {{Expansion}}}, \href{http://dx.doi.org/10.1088/1751-8113/49/13/135403}{\emph{J. Phys. A} {\bf 49} (Feb., 2016) 135403}, [\href{https://arxiv.org/abs/1508.06354}{{\tt arXiv:1508.06354}}\href{https://arxiv.org/pdf/1508.06354.pdf}{$^{\textsc{pdf}}$}], [\href{https://inspirehep.net/literature?q=Giombi:2015haa}{\small \textsc{Inspire}}].

\bibitem{DiPietro:2019hqe}
Lorenzo~Di Pietro, Davide Gaiotto, Edoardo Lauria and Jingxiang Wu, \emph{3d {{Abelian Gauge Theories}} at the {{Boundary}}}, \href{http://dx.doi.org/10.1007/JHEP05(2019)091}{\emph{JHEP} {\bf 05} (May, 2019) 091}, [\href{https://arxiv.org/abs/1902.09567}{{\tt arXiv:1902.09567}}\href{https://arxiv.org/pdf/1902.09567.pdf}{$^{\textsc{pdf}}$}], [\href{https://inspirehep.net/literature?q=DiPietro:2019hqe}{\small \textsc{Inspire}}].

\bibitem{Derkachov:1993uw}
Sergey~E. Derkachov, Nikolay~A. Kivel, Alexander~S. Stepanenko and Alexander~N. Vasil'ev, \emph{On {{Calculation}} of {$1/n$} {{Expansions}} of {{Critical Exponents}} in the {{Gross-Neveu Model}} with the {{Conformal Technique}}},  [\href{https://arxiv.org/abs/hep-th/9302034}{{\tt arXiv:hep-th/9302034}}\href{https://arxiv.org/pdf/hep-th/9302034.pdf}{$^{\textsc{pdf}}$}], [\href{https://inspirehep.net/literature?q=Derkachov:1993uw}{\small \textsc{Inspire}}].

\bibitem{Broadhurst:1996yc}
David~J. Broadhurst and A.~V. Kotikov, \emph{Compact analytical form for nonzeta terms in critical exponents at order {$1/N^3$}}, \href{http://dx.doi.org/10.1016/S0370-2693(98)01146-0}{\emph{Phys. Lett. B} {\bf 441} (1998) 345--353}, [\href{https://arxiv.org/abs/hep-th/9612013}{{\tt arXiv:hep-th/9612013}}\href{https://arxiv.org/pdf/hep-th/9612013.pdf}{$^{\textsc{pdf}}$}], [\href{https://inspirehep.net/literature?q=Broadhurst:1996yc}{\small \textsc{Inspire}}].

\bibitem{Farnsworth:2021ycg}
Kara Farnsworth, Kurt Hinterbichler and Ondrej Hulik, \emph{On the {{Conformal Symmetry}} of {{Exceptional Scalar Theories}}}, \href{http://dx.doi.org/10.1007/JHEP07(2021)198}{\emph{JHEP} {\bf 07} (July, 2021) 198}, [\href{https://arxiv.org/abs/2102.12479}{{\tt arXiv:2102.12479}}\href{https://arxiv.org/pdf/2102.12479.pdf}{$^{\textsc{pdf}}$}], [\href{https://inspirehep.net/literature?q=Farnsworth:2021ycg}{\small \textsc{Inspire}}].

\bibitem{Benedetti:2021wzt}
Dario Benedetti, Razvan Gurau, Sabine Harribey and Davide Lettera, \emph{The {{F-theorem}} in the melonic limit}, \href{http://dx.doi.org/10.1007/JHEP02(2022)147}{\emph{JHEP} {\bf 02} (Feb., 2022) 147}, [\href{https://arxiv.org/abs/2111.11792}{{\tt arXiv:2111.11792}}\href{https://arxiv.org/pdf/2111.11792.pdf}{$^{\textsc{pdf}}$}], [\href{https://inspirehep.net/literature?q=Benedetti:2021wzt}{\small \textsc{Inspire}}].

\bibitem{DeCesare:2025ukl}
Fabiana~De Cesare and Slava Rychkov, \emph{Disturbing news about the {$d=2+\epsilon$} expansion},  [\href{https://arxiv.org/abs/2505.21611}{{\tt arXiv:2505.21611}}\href{https://arxiv.org/pdf/2505.21611.pdf}{$^{\textsc{pdf}}$}], [\href{https://inspirehep.net/literature?q=DeCesare:2025ukl}{\small \textsc{Inspire}}].

\bibitem{Zinn-Justin:1991ksq}
Jean {Zinn-Justin}, \emph{Four-fermion interaction near four dimensions}, \href{http://dx.doi.org/10.1016/0550-3213(91)90043-W}{\emph{Nuclear Physics B} {\bf 367} (1991) 105--122}, [\href{https://inspirehep.net/literature?q=Zinn-Justin:1991ksq}{\small \textsc{Inspire}}].

\bibitem{Ferreira:1997he}
Pedro~M. Ferreira and John~A. Gracey, \emph{The beta-function of the {{Wess-Zumino}} model at {$O(N^2)$}}, \href{http://dx.doi.org/10.1016/S0550-3213(98)00236-3}{\emph{Nuclear Physics B} {\bf 525} (1998) 435--456}, [\href{https://arxiv.org/abs/hep-th/9712138}{{\tt arXiv:hep-th/9712138}}\href{https://arxiv.org/pdf/hep-th/9712138.pdf}{$^{\textsc{pdf}}$}], [\href{https://inspirehep.net/literature?q=Ferreira:1997he}{\small \textsc{Inspire}}].

\bibitem{Fraser-Taliente:2024rql}
Ludo {Fraser-Taliente} and John Wheater, \emph{Melonic limits of the quartic {{Yukawa}} model and general features of melonic {{CFTs}}}, \href{http://dx.doi.org/10.1007/JHEP01(2025)187}{\emph{JHEP} {\bf 01} (Jan., 2025) 187}, [\href{https://arxiv.org/abs/2410.09152}{{\tt arXiv:2410.09152}}\href{https://arxiv.org/pdf/2410.09152.pdf}{$^{\textsc{pdf}}$}], [\href{https://inspirehep.net/literature?q=Fraser-Taliente:2024rql}{\small \textsc{Inspire}}].

\bibitem{Gracey:1990aw}
John~A. Gracey, \emph{Probing the supersymmetric {{O}}{$(N)$} sigma model to {$O(1/N^2)$}: {{Critical}} exponent {$\eta$}}, \href{http://dx.doi.org/10.1016/0550-3213(91)90212-G}{\emph{Nucl. Phys. B} {\bf 348} (1991) 737--756}, [\href{https://inspirehep.net/literature?q=Gracey:1990aw}{\small \textsc{Inspire}}].

\bibitem{Gracey:1991yz}
John~A. Gracey, \emph{Critical exponent {$\eta$} at {{O}}{$(1/N^3)$} for the supersymmetric {{O}}{$(N)$} sigma model}, \href{http://dx.doi.org/10.1016/0370-2693(91)90641-3}{\emph{Phys. Lett. B} {\bf 262} (1991) 49--53}, [\href{https://inspirehep.net/literature?q=Gracey:1991yz}{\small \textsc{Inspire}}].

\bibitem{Popovic:1977cq}
Dragan~S. Popovi{\'c}, \emph{Anomalous {{Dimensions}} in the {$(\bar\psi \bar\phi)\psi\phi$} {{Model Field Theory}} in 1/{{N Expansion}}}, \href{http://dx.doi.org/10.1143/PTP.58.300}{\emph{Prog. Theor. Phys.} {\bf 58} (1977) 300}, [\href{https://inspirehep.net/literature?q=Popovic:1977cq}{\small \textsc{Inspire}}].

\bibitem{zinn-justin_quantum_2002}
Jean {Zinn-Justin}, \emph{Quantum {{Field Theory}} and {{Critical Phenomena}}}.
\newblock Oxford University Press, 5th~ed., June, 2002, \href{http://dx.doi.org/10.1093/acprof:oso/9780198509233.001.0001}{\small \textsc{doi}:10.1093/acprof:oso/9780198509233.001.0001}.

\bibitem{Vasiliev:1975mq}
Alexander~N. Vasil'ev and Mikhail~Yu. Nalimov, \emph{Analog of dimensional regularization for calculation of the renormalization-group functions in the 1/n expansion for arbitrary dimension of space}, \href{http://dx.doi.org/10.1007/BF01015800}{\emph{Theor. Math. Phys.} {\bf 55} (1983) 423--431}, [\href{https://inspirehep.net/literature?q=Vasiliev:1975mq}{\small \textsc{Inspire}}].

\bibitem{Ciuchini:1999wy}
Massimiliano Ciuchini, Sergey~E. Derkachov, John~A. Gracey and Alexander~N. Manashov, \emph{Computation of quark mass anomalous dimension at {{O}}{$(1/N_f^2)$} in quantum chromodynamics}, \href{http://dx.doi.org/10.1016/S0550-3213(00)00209-1}{\emph{Nuclear Physics B} {\bf 579} (2000) 56--100}, [\href{https://arxiv.org/abs/hep-ph/9912221}{{\tt arXiv:hep-ph/9912221}}\href{https://arxiv.org/pdf/hep-ph/9912221.pdf}{$^{\textsc{pdf}}$}], [\href{https://inspirehep.net/literature?q=Ciuchini:1999wy}{\small \textsc{Inspire}}].

\bibitem{Vasiliev:1981dg}
Alexander~N. Vasil'ev, Yuri~M. Pis'mak and Juha~R. Khonkonen, \emph{{$1/n$} expansion: {{Calculation}} of the exponents {$\eta$} and {$\nu$} in the order {$1/n^2$} for arbitrary number of dimensions}, \href{http://dx.doi.org/10.1007/BF01019296}{\emph{Theor. Math. Phys.} {\bf 47} (1981) 465--475}, [\href{https://inspirehep.net/literature?q=Vasiliev:1981dg}{\small \textsc{Inspire}}].

\bibitem{Gracey:1993kc}
John~A. Gracey, \emph{Computation of {{Critical Exponent}} {$\eta$} at {{O}}{$(1/N^3)$} in the {{Four Fermi Model}} in {{Arbitrary Dimensions}}}, \href{http://dx.doi.org/10.1142/S0217751X94000340}{\emph{International Journal of Modern Physics A} {\bf 9} (1994) 727--744}, [\href{https://arxiv.org/abs/hep-th/9306107}{{\tt arXiv:hep-th/9306107}}\href{https://arxiv.org/pdf/hep-th/9306107.pdf}{$^{\textsc{pdf}}$}], [\href{https://inspirehep.net/literature?q=Gracey:1993kc}{\small \textsc{Inspire}}].

\bibitem{Sun:2020ame}
Zimo Sun, \emph{{{AdS}} one-loop partition functions from bulk and edge characters}, \href{http://dx.doi.org/10.1007/JHEP12(2021)064}{\emph{JHEP} {\bf 12} (Dec., 2021) 064}, [\href{https://arxiv.org/abs/2010.15826}{{\tt arXiv:2010.15826}}\href{https://arxiv.org/pdf/2010.15826.pdf}{$^{\textsc{pdf}}$}], [\href{https://inspirehep.net/literature?q=Sun:2020ame}{\small \textsc{Inspire}}].

\bibitem{Harribey:2022esw}
Sabine Harribey, \emph{Renormalization in tensor field theory and the melonic fixed point},  [\href{https://arxiv.org/abs/2207.05520}{{\tt arXiv:2207.05520}}\href{https://arxiv.org/pdf/2207.05520.pdf}{$^{\textsc{pdf}}$}], [\href{https://inspirehep.net/literature?q=Harribey:2022esw}{\small \textsc{Inspire}}].

\bibitem{Fraser-Taliente:2024lea}
Ludo {Fraser-Taliente}, Christopher~P. Herzog and Abhay Shrestha, \emph{A {{Nonlocal Schwinger Model}}}, \href{http://dx.doi.org/10.1007/JHEP06(2025)252}{\emph{JHEP} {\bf 06} (June, 2025) 252}, [\href{https://arxiv.org/abs/2412.02514}{{\tt arXiv:2412.02514}}\href{https://arxiv.org/pdf/2412.02514.pdf}{$^{\textsc{pdf}}$}], [\href{https://inspirehep.net/literature?q=Fraser-Taliente:2024lea}{\small \textsc{Inspire}}].

\bibitem{Karateev:2018oml}
Denis Karateev, Petr Kravchuk and David {Simmons-Duffin}, \emph{Harmonic {{Analysis}} and {{Mean Field Theory}}}, \href{http://dx.doi.org/10.1007/JHEP10(2019)217}{\emph{JHEP} {\bf 10} (Oct., 2019) 217}, [\href{https://arxiv.org/abs/1809.05111}{{\tt arXiv:1809.05111}}\href{https://arxiv.org/pdf/1809.05111.pdf}{$^{\textsc{pdf}}$}], [\href{https://inspirehep.net/literature?q=Karateev:2018oml}{\small \textsc{Inspire}}].

\bibitem{Moshe:2003xn}
Moshe Moshe and Jean {Zinn-Justin}, \emph{Quantum {{Field Theory}} in the {{Large N Limit}}: A review}, \href{http://dx.doi.org/10.1016/S0370-1573(03)00263-1}{\emph{Physics Reports} {\bf 385} (Oct., 2003) 69--228}, [\href{https://arxiv.org/abs/hep-th/0306133}{{\tt arXiv:hep-th/0306133}}\href{https://arxiv.org/pdf/hep-th/0306133.pdf}{$^{\textsc{pdf}}$}], [\href{https://inspirehep.net/literature?q=Moshe:2003xn}{\small \textsc{Inspire}}].

\bibitem{Gubser:2019uyf}
Steven~S. Gubser, Christian~B. Jepsen, Ziming Ji, Brian Trundy and Amos Yarom, \emph{Non-local non-linear sigma models}, \href{http://dx.doi.org/10.1007/JHEP09(2019)005}{\emph{JHEP} {\bf 09} (Sept., 2019) 005}, [\href{https://arxiv.org/abs/1906.10281}{{\tt arXiv:1906.10281}}\href{https://arxiv.org/pdf/1906.10281.pdf}{$^{\textsc{pdf}}$}], [\href{https://inspirehep.net/literature?q=Gubser:2019uyf}{\small \textsc{Inspire}}].

\bibitem{Diaz:2008hy}
Danilo~E. Diaz, \emph{Polyakov formulas for {{GJMS}} operators from {{AdS}}/{{CFT}}}, \href{http://dx.doi.org/10.1088/1126-6708/2008/07/103}{\emph{JHEP} {\bf 07} (2008) 103}, [\href{https://arxiv.org/abs/0803.0571}{{\tt arXiv:0803.0571}}\href{https://arxiv.org/pdf/0803.0571.pdf}{$^{\textsc{pdf}}$}], [\href{https://inspirehep.net/literature?q=Diaz:2008hy}{\small \textsc{Inspire}}].

\bibitem{maalaouiConformalFractionalDirac2025}
Ali Maalaoui, \emph{Conformal {{Fractional Dirac Operator}} and {{Spinorial Q-curvature}}},  [\href{https://arxiv.org/abs/2505.05706}{{\tt arXiv:2505.05706}}\href{https://arxiv.org/pdf/2505.05706.pdf}{$^{\textsc{pdf}}$}].

\bibitem{fischmannConformalPowersDirac2014}
Matthias Fischmann, Christian Krattenthaler and Petr Somberg, \emph{On {{Conformal Powers}} of the {{Dirac Operator}} on {{Einstein Manifolds}}},  [\href{https://arxiv.org/abs/1405.7304}{{\tt arXiv:1405.7304}}\href{https://arxiv.org/pdf/1405.7304.pdf}{$^{\textsc{pdf}}$}].

\bibitem{Camporesi:1995fb}
Roberto Camporesi and Atsushi Higuchi, \emph{On the eigenfunctions of the {{Dirac}} operator on spheres and real hyperbolic spaces}, \href{http://dx.doi.org/10.1016/0393-0440(95)00042-9}{\emph{Journal of Geometry and Physics} {\bf 20} (1996) 1--18}, [\href{https://arxiv.org/abs/gr-qc/9505009}{{\tt arXiv:gr-qc/9505009}}\href{https://arxiv.org/pdf/gr-qc/9505009.pdf}{$^{\textsc{pdf}}$}], [\href{https://inspirehep.net/literature?q=Camporesi:1995fb}{\small \textsc{Inspire}}].

\bibitem{Preti:2018vog}
Michelangelo Preti, \emph{{{STR}}: A {{Mathematica}} package for the method of uniqueness}, \href{http://dx.doi.org/10.1142/S0129183120501466}{\emph{International Journal of Modern Physics C} {\bf 31} (Sept., 2020) 2050146}, [\href{https://arxiv.org/abs/1811.04935}{{\tt arXiv:1811.04935}}\href{https://arxiv.org/pdf/1811.04935.pdf}{$^{\textsc{pdf}}$}], [\href{https://inspirehep.net/literature?q=Preti:2018vog}{\small \textsc{Inspire}}].

\bibitem{Gracey:2018ame}
John~A. Gracey, \emph{Large {$N_f$} quantum field theory}, \href{http://dx.doi.org/10.1142/S0217751X18300326}{\emph{International Journal of Modern Physics A} {\bf 33} (Jan., 2019) 1830032}, [\href{https://arxiv.org/abs/1812.05368}{{\tt arXiv:1812.05368}}\href{https://arxiv.org/pdf/1812.05368.pdf}{$^{\textsc{pdf}}$}], [\href{https://inspirehep.net/literature?q=Gracey:2018ame}{\small \textsc{Inspire}}].

\bibitem{Benedetti:2018goh}
Dario Benedetti and Razvan Gurau, \emph{{{2PI}} effective action for the {{SYK}} model and tensor field theories}, \href{http://dx.doi.org/10.1007/JHEP05(2018)156}{\emph{JHEP} {\bf 05} (May, 2018) 156}, [\href{https://arxiv.org/abs/1802.05500}{{\tt arXiv:1802.05500}}\href{https://arxiv.org/pdf/1802.05500.pdf}{$^{\textsc{pdf}}$}], [\href{https://inspirehep.net/literature?q=Benedetti:2018goh}{\small \textsc{Inspire}}].

\bibitem{Vasiliev:1981yc}
Alexander~N. Vasil'ev, Yuri~M. Pis'mak and Juha~R. Khonkonen, \emph{Simple {{Method}} of {{Calculating}} the {{Critical Indices}} in the 1/{{N Expansion}}}, \href{http://dx.doi.org/10.1007/BF01030844}{\emph{Theor. Math. Phys.} {\bf 46} (1981) 104--113}, [\href{https://inspirehep.net/literature?q=Vasiliev:1981yc}{\small \textsc{Inspire}}].

\bibitem{Vasiliev:1982dc}
Alexander~N. Vasil'ev, Yuri~M. Pis'mak and Juha~R. Khonkonen, \emph{{$1/n$} expansion: Calculation of the exponent {$\eta$} in the order {$1/n^3$} by the conformal bootstrap method}, \href{http://dx.doi.org/10.1007/BF01015292}{\emph{Theor. Math. Phys.} {\bf 50} (1982) 127--134}, [\href{https://inspirehep.net/literature?q=Vasiliev:1982dc}{\small \textsc{Inspire}}].

\bibitem{Dobrev:1977qv}
Vladimir~K. Dobrev, Gerhard Mack, Valentina~B. Petkova, Svetlana~G. Petrova and Ivan~T. Todorov, \emph{Harmonic {{Analysis}} on the {$n$}-{{Dimensional Lorentz Group}} and {{Its Application}} to {{Conformal Quantum Field Theory}}}, vol.~63 of \emph{Lect. {{Notes Phys}}.}
\newblock Springer Berlin, Heidelberg, 1977, \href{http://dx.doi.org/10.1007/BFb0009678}{\small \textsc{doi}:10.1007/BFb0009678}, [\href{https://inspirehep.net/literature?q=Dobrev:1977qv}{\small \textsc{Inspire}}].

\bibitem{Henriksson:2022rnm}
Johan Henriksson, \emph{The critical {{O}}({{N}}) {{CFT}}: {{Methods}} and conformal data}, \href{http://dx.doi.org/10.1016/j.physrep.2022.12.002}{\emph{Phys. Rept.} {\bf 1002} (Feb., 2023) 2250}, [\href{https://arxiv.org/abs/2201.09520}{{\tt arXiv:2201.09520}}\href{https://arxiv.org/pdf/2201.09520.pdf}{$^{\textsc{pdf}}$}], [\href{https://inspirehep.net/literature?q=Henriksson:2022rnm}{\small \textsc{Inspire}}].

\bibitem{Douglas:2010ic}
Michael~R. Douglas, \emph{Spaces of {{Quantum Field Theories}}}, \href{http://dx.doi.org/10.1088/1742-6596/462/1/012011}{\emph{J. Phys. Conf. Ser.} {\bf 462} (Dec., 2013) 012011}, [\href{https://arxiv.org/abs/1005.2779}{{\tt arXiv:1005.2779}}\href{https://arxiv.org/pdf/1005.2779.pdf}{$^{\textsc{pdf}}$}], [\href{https://inspirehep.net/literature?q=Douglas:2010ic}{\small \textsc{Inspire}}].

\bibitem{Smirnov:2009up}
Alexander~V. Smirnov and Vladimir~A. Smirnov, \emph{On the {{Resolution}} of {{Singularities}} of {{Multiple Mellin-Barnes Integrals}}}, \href{http://dx.doi.org/10.1140/epjc/s10052-009-1039-6}{\emph{The European Physical Journal C} {\bf 62} (May, 2009) 445--449}, [\href{https://arxiv.org/abs/0901.0386}{{\tt arXiv:0901.0386}}\href{https://arxiv.org/pdf/0901.0386.pdf}{$^{\textsc{pdf}}$}], [\href{https://inspirehep.net/literature?q=Smirnov:2009up}{\small \textsc{Inspire}}].

\bibitem{Belitsky:2022gba}
Andrei~V. Belitsky, Alexander~V. Smirnov and Vladimir~A. Smirnov, \emph{{{MB Tools}} reloaded}, \href{http://dx.doi.org/10.1016/j.nuclphysb.2022.116067}{\emph{Nuclear Physics B} {\bf 986} (Jan., 2023) 116067}, [\href{https://arxiv.org/abs/2211.00009}{{\tt arXiv:2211.00009}}\href{https://arxiv.org/pdf/2211.00009.pdf}{$^{\textsc{pdf}}$}], [\href{https://inspirehep.net/literature?q=Belitsky:2022gba}{\small \textsc{Inspire}}].

\bibitem{belitskyFeynmanIntegralsMBGitLab2024}
Andrei~V. Belitsky, Alexander~V. Smirnov and Vladimir~A. Smirnov, \emph{{feynmanIntegrals/MB} {$\cdot$} {{GitLab}}},  Mar., 2024.
\newblock \href{https://gitlab.com/feynmanintegrals/MB}{\small \textsc{Url}:https://gitlab.com/feynmanintegrals/MB}.

\bibitem{Ochman:2015fho}
Michal Ochman and Tord Riemann, \emph{{{MBsums}} - a {{Mathematica}} package for the representation of {{Mellin-Barnes}} integrals by multiple sums}, \href{http://dx.doi.org/10.5506/APhysPolB.46.2117}{\emph{Acta Phys. Polon. B} {\bf 46} (2015) 2117}, [\href{https://arxiv.org/abs/1511.01323}{{\tt arXiv:1511.01323}}\href{https://arxiv.org/pdf/1511.01323.pdf}{$^{\textsc{pdf}}$}], [\href{https://inspirehep.net/literature?q=Ochman:2015fho}{\small \textsc{Inspire}}].

\end{thebibliography}\endgroup
}              %
\end{document}